\documentclass[a4paper,twocolumn,11pt,accepted=2024-04-09]{quantumarticle}
\pdfoutput = 1
\usepackage{graphicx}
\usepackage[colorlinks=true,citecolor=blue]{hyperref}
\usepackage{amsmath}
\usepackage{multirow}
\usepackage{amssymb}
\usepackage{bbm}
\usepackage{color}
\usepackage{dcolumn}
\usepackage{bm}
\usepackage{float}
\usepackage[normalem]{ulem}
\usepackage{braket}
\usepackage{changepage}
\usepackage[subrefformat=parens,labelformat=parens,caption=false]{subfig}
\usepackage{bbm}
\usepackage[dvipsnames]{xcolor}
\usepackage{esint}
\usepackage{lineno}
\usepackage{verbatim}
\usepackage{bm}
\usepackage{multicol}
\usepackage{bbold}
\usepackage{placeins}
\usepackage{mwe}
\usepackage[numbers,sort&compress]{natbib}
\usepackage{amsmath,stmaryrd,graphicx}

\newcommand{\ceil}[1]{\lceil #1 \rceil}

\usepackage[subrefformat=parens,labelformat=parens,caption=false]{subfig}

\usepackage{amsthm}
\usepackage[export]{adjustbox}

\usepackage{soul}

\usepackage{tikz,xcolor,hyperref}

\definecolor{lime}{HTML}{A6CE39}
\DeclareRobustCommand{\orcidicon}{%
	\begin{tikzpicture}
	\draw[lime, fill=lime] (0,0) 
	circle [radius=0.16] 
	node[white] {{\fontfamily{qag}\selectfont \tiny ID}};
	\draw[white, fill=white] (-0.0625,0.095) 
	circle [radius=0.007];
	\end{tikzpicture}
	\hspace{-2mm}
}

\foreach \x in {A, ..., Z}{%
	\expandafter\xdef\csname orcid\x\endcsname{\noexpand\href{https://orcid.org/\csname orcidauthor\x\endcsname}{\noexpand\orcidicon}}
}
\makeatletter
\newcommand{\fixed@sra}{$\vrule height 2\fontdimen22\textfont2 width 0pt\shortrightarrow$}
\newcommand{\shortarrow}[1]{%
  \mathrel{\text{\rotatebox[origin=c]{\numexpr#1*45}{\fixed@sra}}}
}

\makeatother

\begin{document}
\title{Low-depth simulations of fermionic systems on square-grid quantum hardware}

\author{Manuel G. Algaba}
\thanks{manuel.algaba@meetiqm.com}
\affiliation{IQM, Nymphenburgerstr. 86, 80636 Munich, Germany}
\author{P.V. Sriluckshmy}
\affiliation{IQM, Nymphenburgerstr. 86, 80636 Munich, Germany}
\author{Martin Leib}
\affiliation{IQM, Nymphenburgerstr. 86, 80636 Munich, Germany}
\author{Fedor \v{S}imkovic IV}
\thanks{fedor.simkovic@meetiqm.com}
\affiliation{IQM, Nymphenburgerstr. 86, 80636 Munich, Germany}

\begin{abstract}

We present a general strategy for mapping fermionic systems to NISQ hardware with square qubit connectivity which yields low-depth quantum circuits, counted in the number of native two-qubit fSIM gates. We achieve this by leveraging novel operator decomposition and circuit compression techniques paired with specifically chosen \emph{low-depth fermion-to-qubit mappings} and allow for a high degree of gate cancellations and parallelism. Our mappings retain the flexibility to simultaneously optimize for qubit counts or qubit operator weights and can be used to investigate arbitrary fermionic lattice geometries. We showcase our approach by investigating the tight-binding model, the Fermi-Hubbard model as well as the multi-orbital Hubbard-Kanamori model. We report unprecedentedly low circuit depths per single Trotter layer with up to a $70 \%$ improvement upon previous state-of-the-art. Our compression technique also results in significant reduction of two-qubit gates. We find the lowest gate-counts when applying the XYZ-formalism to the DK mapping. Additionally, we show that our decomposition and compression formalism produces favourable circuits even when no native parameterized two-qubit gates are available.

\end{abstract}

% \date{\today}
\maketitle

\section{Introduction}
\label{sec:introduction}

One of the most promising applications of future quantum computers is the simulation of fermionic quantum systems. These systems are of major relevance to various scientific fields, such as quantum chemistry, condensed matter physics, lattice gauge theories and nuclear physics. Simultaneously, they are related to many unsolved scientific problems with potential industrial applications, such as high-temperature superconductivity~\cite{McArdle2020Quantum}, battery design~\cite{Delgado2022}, chemical reaction rates~\cite{Andersson2022}, and nitrogen fixation in fertilizers~\cite{reiher2017elucidating}. 

Classically, despite extensive scientific efforts in developing and benchmarking computational tools~\cite{leblanc2015, Motta2017, Schaefer2021}, most such systems remain too difficult to study with current state-of-the-art techniques. This is mainly due to their exponentially growing computational space~\cite{Georgescu2014} as well as due to the anti-commuting nature of fermions, which leads to the infamous sign problem  for many computational methods based on quantum Monte Carlo~\cite{troyer2005computational, rossi2017polynomial}. 

Whilst it is generally believed that quantum computers will eventually be able to circumvent these difficulties and provide at least polynomial speedups~\cite{lee2022there} for problems of relevance, finding suitable quantum, or hybrid quantum-classical, algorithms is an active field of research. The need for developing highly optimized quantum algorithms is further amplified by the shortcomings of currently available noisy quantum hardware, which is severely limited in their qubits counts, coherence times and gate fidelities~\cite{Koch2020}. Small scale proof-of-principle implementations of algorithms on such NISQ hardware have nevertheless been performed,  encouraging further research in the field \cite{,Arute2020, Tazhingulov2022, Stanisic2022}. 

A necessary first step for any quantum computing approach to fermionic systems is to find an adequate transformation between the fermionic Fock space and the computational space of a multi-qubit quantum device. This transformation is not unique and \emph{fermion-to-qubit mappings} can have vastly different properties from one another. The right choice of mappings is thus fundamental to successfully simulating fermionic systems. 

The Jordan-Wigner transformation (JW)~\cite{Jordan1928} is not only  historically the first fermion-to-qubit mapping, but also the most popular choice today, especially due to its simplicity and favourable performance for small systems. It maps fermionic modes one-to-one to qubits arranged on a line at the expense of generating qubit operators whose weights scale linearly with the total system size. 

Numerous mappings have been since proposed that improve the JW transformation in terms of specific optimisation criteria, especially in cases where one can exploit some known underlying structure in the fermionic Hamiltonian to be simulated. In particular, some mappings aim at reducing the total number of qubits needed to encode a fermionic system by using symmetries~\cite{Bravyi2017, Steudtner2018, Chiew2021, Chien2022, Kirby2022, Harrison2022}, by encoding fermionic modes through tree-graph structures~\cite{Havlicek2017,Vlasov2019}, or by considering specific hardware topology and connectivity structures~\cite{Chien2022,Aaron2022}. Some mappings are designed with partial quantum error correction in mind~\cite{Haah2017}, and allow to either identify~\cite{Bausch2020, Chien2022} or correct~\cite{Setia2019, Tantivasadakarn2020, Chen2022b} errors that occur during circuit execution. All the previous techniques can effectively be viewed as being forms of exact bosonization   \cite{Bochniak2020,Arkadiusz2020} tailored to quantum computers.

Recently, the focus in the study of fermion-to-qubit mappings has increasingly shifted towards efficient implementation of mappings on near-term (NISQ) quantum devices~\cite{Clinton2021, Clinton2022, Nys2022}. Indeed, current NISQ systems are mostly limited by gate count and coherence times. This is reflected in the fact that even though processors with hundreds of qubits are currently available, the largest reported fermionic simulations to date typically use less than thirty~\cite{Huggins2022}.

It is often implicitly assumed that the mappings which yield the lowest circuit depths are the ones with the lowest operator weights, corresponding to the number of qubits they act upon. These weights can be reduced by introducing additional ancilla qubits used to resolve fermionic commutation relations~\cite{Bravyi2002, Ball2005, Verstraete2005, Chen2018, Setia2018, Derby2020, Chen2022}. In particular, many (geometrically) \emph{local} mappings, with circuit depths which are independent of fermionic lattice size, have been proposed. Alternatively, one can dynamically alter the association of fermionic modes with hardware qubits using fermionic swap (fSWAP) networks~\cite{Bravyi2002, Kivlichan2018, Gorman2019, Cade2020, Clinton2022}. 

For fermionic systems with some degree of sparsity it turns out that a combination of local mappings and fSWAP networks yields the lowest-depth circuits to date~\cite{Clinton2021, Clinton2022}. This approach has reduced the necessary circuit depths per Trotter layer by multiple orders of magnitude to the range of roughly $10^2-10^4$, depending on the fermionic model to be simulated. The lingering question is whether further improvements to those depths are possible, especially since reducing the circuit depths by even a small constant factor can potentially accelerate the advent of useful fermionic simulations on quantum computers by years. Another issue with many of the existing fermion-to-qubit mappings is that they assume qubit layouts that are tailored towards one singular model or demand connectivity graphs that are not compatible with the restricted qubit topologies provided by some of the leading quantum computing hardware platforms. It is therefore important to investigate mappings that bring future fermionic simulation away from the hypothetical and closer to the practical. 

In this work we focus on designing a universal strategy of simulating multi-orbital fermionic systems on an arbitrary two-dimensional lattice by means of a quantum processor with a square qubit connectivity graph, which is one of the standard topologies found in the superconducting quantum computer industry~\cite{Arute2019,Valery2022}. We restrict our analysis to NISQ applications as fault-tolerant approaches may involve other steps not considered here. Our particular focus is on generating minimal-depth quantum circuits, counted in the number of parallelizable native two-qubit gate layers per single Trotter step. To this end, we present operator decomposition and compression techniques which we apply to a number of fermion-to-qubit mappings which we introduce in this paper. We show that our formalism generally improves the circuit depths of Trotterized time evolution for the investigated models and that mappings with the lowest operator weights are not necessarily optimal in terms of circuit depth because of their respective lack in parallelism.

The paper is structured as follows: In Section \ref{sec:xyzdecomposition} we introduce our decomposition and compression techniques (which we collectively name XYZ). In Section \ref{sec:models} we introduce the fermionic models of interest and in Section \ref{sec:themapping} we discuss the general formalism we use to map them to square qubit layouts. In Section \ref{sec:fermionicsimulation} we present our analysis of circuit depths obtained by using the XYZ formalism in combination with various local fermion-to-qubit mappings for various fermionic lattice geometries. We inspect the tight-binding model (TB) in Sec.~\ref{subsec:spinlessmodels}, the Fermi-Hubbard model (FH) in Sec.~\ref{subsec:spinfulmodels}, and the Hubbard-Kanamori model (HK) in Sec.~\ref{subsec:multiorbital}. In Sec.~\ref{subsec:comparison} we investigate how the advantage of using the XYZ formalism depends on the type of available native gates. We finish with a discussion of our main findings as well as future directions in Section \ref{sec:discussion}. 

\section{XYZ operator decomposition and circuit compression techniques}
\label{sec:xyzdecomposition}

Most quantum algorithms involve multi-qubit operators, yet current quantum processors only allow for one and two-qubit gates to be executed at any given point. As a consequence, it is necessary to decompose such multi-qubit operators into circuits involving executable native gates. Depending on which gates are considered native, there are multiple ways to decompose operators which yield different levels of efficiency in terms of resulting gate counts and circuit depths. Let us consider as example the operator $e^{i\alpha Z_1 Z_2 Z_3 Z_4}$ where $\alpha \in \mathbb{R}$ and $Z_i$ is a Pauli operator acting on qubit $q_i$ \footnote{In what follows we label circuit qubits with ascending integers from top to bottom.}: 
\begin{equation}\label{eq:cnotlad}
    \raisebox{-.45\height}{\includegraphics[scale=0.08]{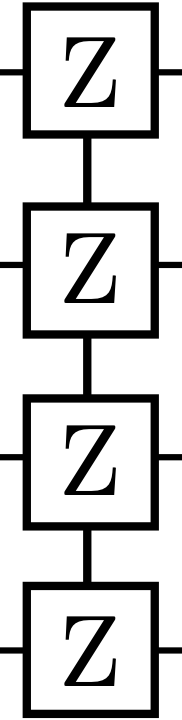}} =
    \raisebox{-.42\height}{\includegraphics[scale=0.08]{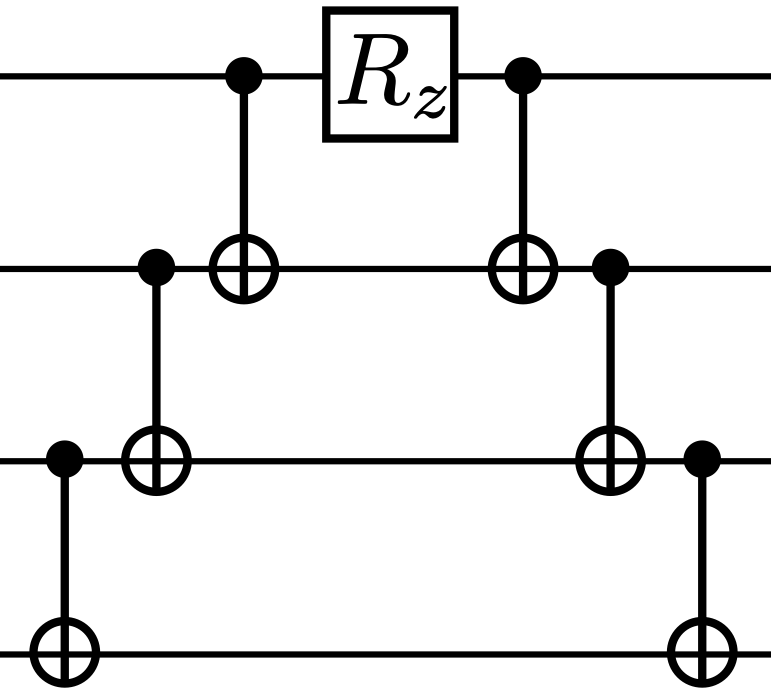}} =
    \raisebox{-.42\height}{\includegraphics[scale=0.08]{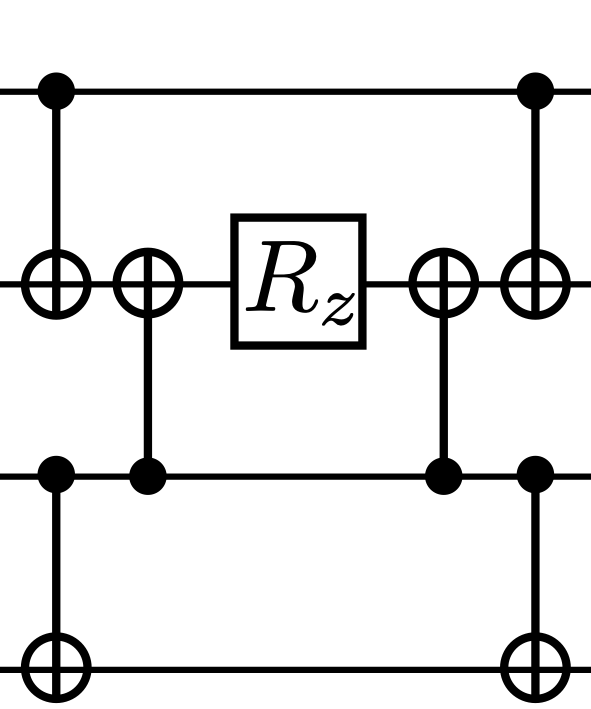}}
\end{equation}
We can decompose the operator (l.h.s. of Eq.~\ref{eq:cnotlad}) using the standard approach in the  field~\cite{nielsen2002quantum}, which generates a \emph{V-shaped} circuit involving CNOT gates and single qubit rotations (left circuit in Eq.~\ref{eq:cnotlad}). However, it is more efficient in terms of circuit depth to decompose the operator into an \emph{X-shaped} circuit instead (right circuit in Eq.~\ref{eq:cnotlad}) ~\cite{Cowtan2020}. For multiple operators acting on overlapping sets of qubits one can further compress the resulting circuits by combining their outermost CNOT gates using known identities (we list these in Appendix~\ref{appendix:app_CNOT}).

Here, we use an alternative, recently introduced technique, named XYZ-decomposition~\cite{xyzdecomposition} which, as we will proceed to show, yields considerably shallower circuits compared to the decomposition into CNOTs, especially when parameterized two-qubit gates are available. The XYZ technique is based on the decomposition of a multi-qubit gate of the form $e^{i\alpha \mathcal{O}}$, where $\mathcal{O}$ is a tensor product of Pauli operators on a given number of qubits, into three operators of lower or equal weight:
\begin{equation}
\label{eq:xyzdecomp}
    e^{i\alpha \mathcal{O}}=e^{i\frac{\pi}{4}\mathcal{O}_1}e^{i\alpha \mathcal{O}_2}e^{-i\frac{\pi}{4}\mathcal{O}_1}
\end{equation}
where $\mathcal{O},\mathcal{O}_1,\mathcal{O}_2$ are unitary and fulfil the relations $\mathcal{O}^2=\mathbb{1}$ and $\mathcal{O}=\frac{i}{2}[\mathcal{O}_1,\mathcal{O}_2]$. All three operators must act on a connected set of qubits and $\mathcal{O}_1$, $\mathcal{O}_2$ are acting on a subset of the qubits that $\mathcal{O}$ is acting on. For simplicity we assume a linear connectivity, meaning qubits can interact with up to two of their neighbors. This approach can be recursively applied to obtain a complete decomposition of the original multi-qubit operator into two-qubit gates.  As an example, let us consider the following operator acting on four qubits:
\begin{align}
&e^{i\alpha X_1 Z_2 Z_3 Y_4}=
e^{i\frac{\pi}{4}X_1 Y_2}
e^{i\alpha X_2 Z_3 Y_4}
e^{-i\frac{\pi}{4}X_1 Y_2}  \\
&= e^{i\frac{\pi}{4}X_1 Y_2}e^{i\frac{\pi}{4}X_3 Y_4}e^{-i\alpha X_2 Y_3}e^{-i\frac{\pi}{4}X_1 Y_2}e^{-i\frac{\pi}{4}X_3 Y_4} \nonumber
\end{align}
Here, we first decompose the operator at the \emph{central} qubit $q_2$ to generate two- and three-qubit operators. Subsequently, we further decomposed the three-qubit operators by breaking it up at $q_3$. Depending on the prefactor in the exponential we can introduce a graphical notation for the three different types of two-qubit gates occurring in such decompositions:
\begin{equation}\label{eq:xyzgraphical}
    \raisebox{-.45\height}{\includegraphics[scale=0.08]{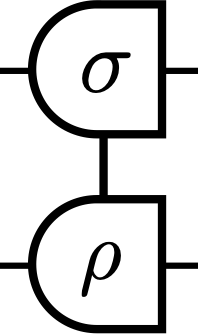}} \equiv e^{i\frac{\pi}{4}\sigma_1 \rho_2}
,\ 
    \raisebox{-.44\height}{\includegraphics[scale=0.08]{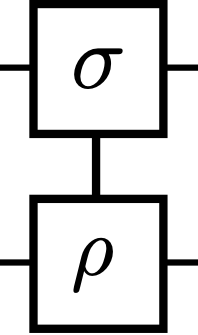}} \equiv e^{i\alpha\sigma_1 \rho_2}
,\ 
    \raisebox{-.44\height}{\includegraphics[scale=0.08]{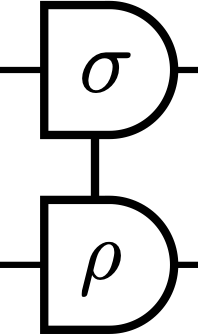}} \equiv e^{-i\frac{\pi}{4}\sigma_1 \rho_2}
\end{equation}
These are the only three types of two-qubit gates generated by the XYZ-decomposition (single-qubit gate equivalents of the above can also appear). The two gates with prefactors $\pm \pi/4$ are simply special cases of the gate with prefactor $\alpha$. Here, $\sigma$ and $\rho$ are Pauli operators acting on qubits $q_1$ and $q_2$, respectively. Note that the square two-qubit operator does not indicate the value of $\alpha$, which has to be tracked separately. Only a single such two-qubit gate is generated per multi-qubit operator. If parameterized two-qubit gates are not available it can be further decomposed into a parameterized single-qubit gate and two two-qubit gates using Eq.~\ref{eq:xyzdecomp}. 

Generally, one has a choice of which qubit to designate as central for the next decomposition as well as the choice of how to distribute the two new Paulis of the new operators acting on this qubit. Let's consider the operator $e^{i\alpha Y_1 Z_2 Z_3 X_4}$ for which we show below four of the possible decompositions:
\begin{equation}\label{eq:legs}
\begin{matrix}
    \raisebox{-.46\height}{\includegraphics[scale=0.08]{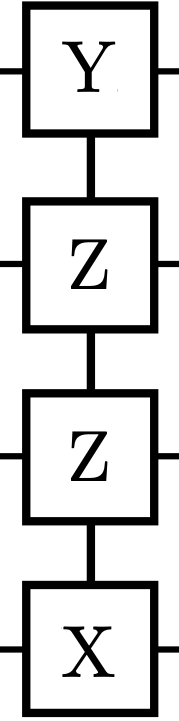}}\ =\
    \raisebox{-.46\height}{\includegraphics[scale=0.08]{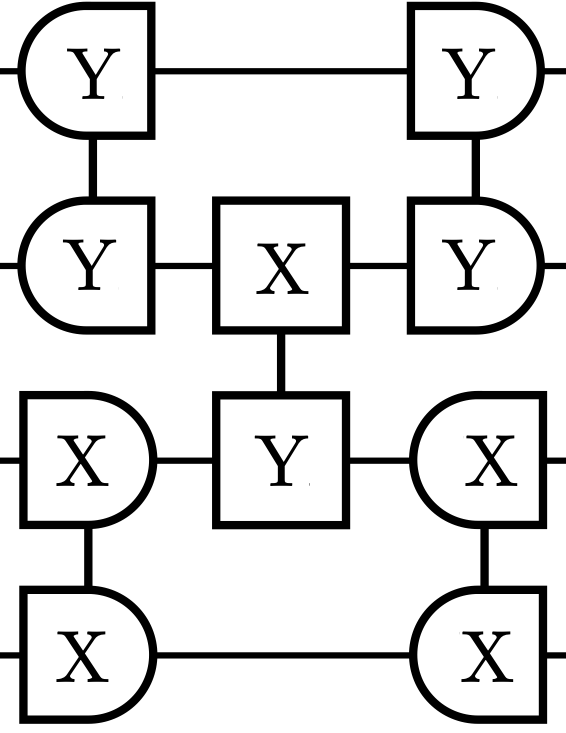}}\ =\
    \raisebox{-.46\height}{\includegraphics[scale=0.08]{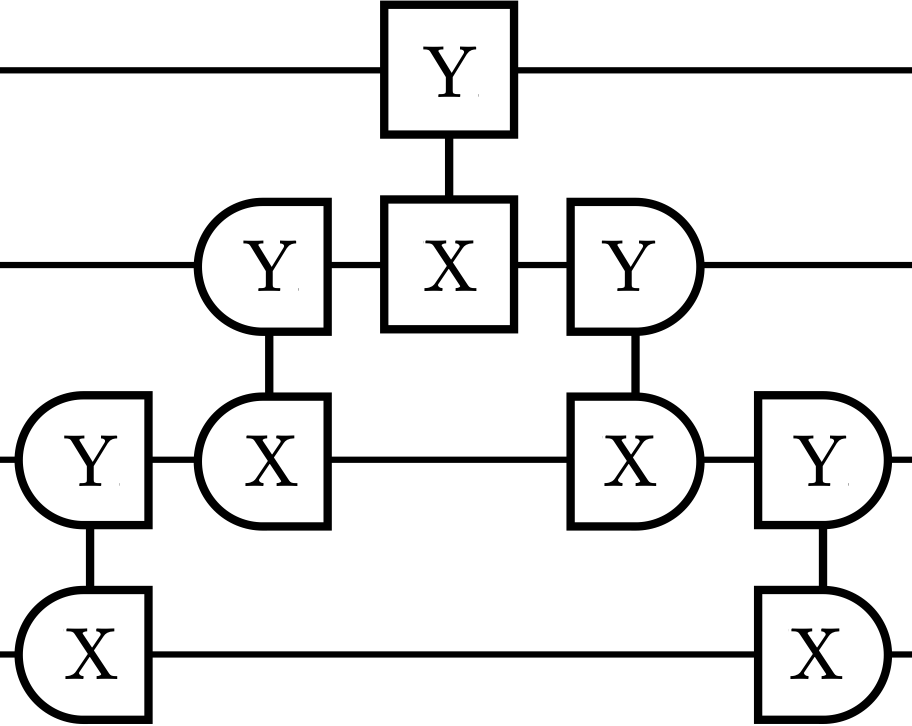}}\ = \\ \\  \hspace*{-0.55cm}\quad \quad   =\  \raisebox{-.46\height}{\includegraphics[scale=0.08]{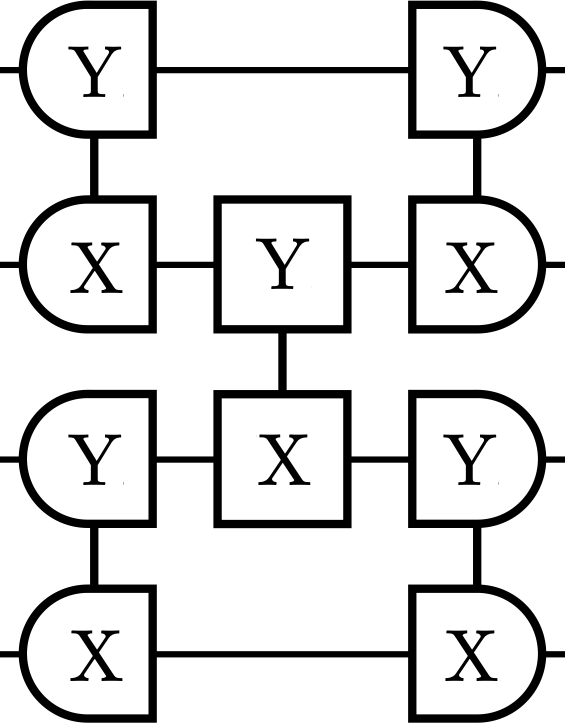}}\ =\  \raisebox{-.46\height}{\includegraphics[scale=0.08]{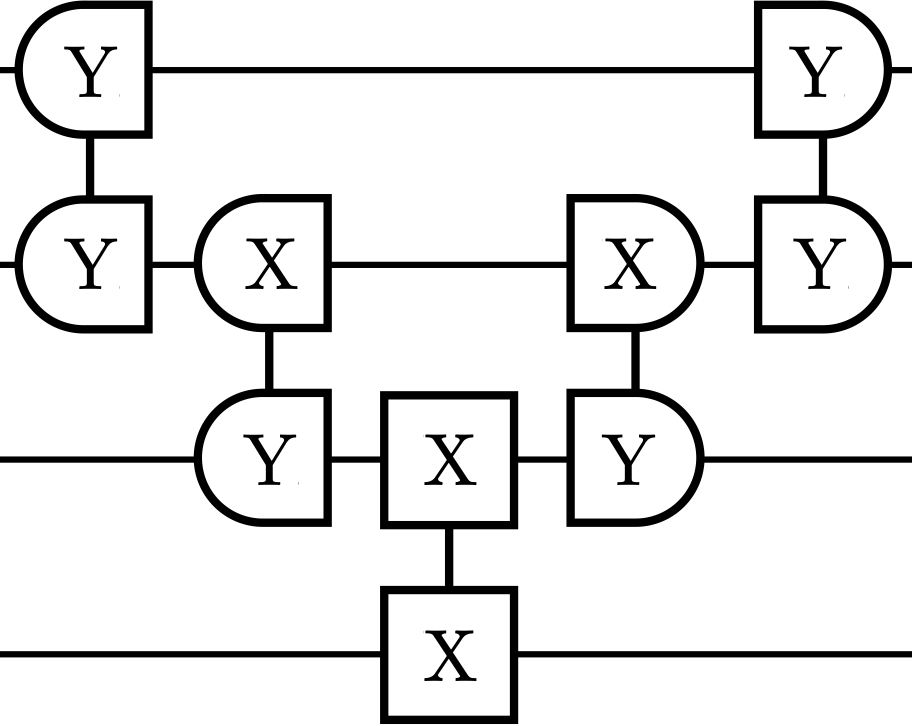}}
\end{matrix}
\end{equation}
Starting with four-qubit operators one has the choice of decomposing operators into a V-shaped or an X-shaped circuit. As shown in Ref.~\cite{xyzdecomposition}, the latter always produces lower-depth circuits and we found that this statement generally holds true even when multiple operators are considered simultaneously. It is also possible to construct \emph{asymmetric} X-shaped circuits for operators acting on five or more qubits, but this rarely leads to improvements in overall circuit depth. One can also always turn around the shape of the decomposition, vertically, in some sense reminiscent of using the Yang-Baxter equation to decrease the depths of standard circuits \cite{Peng2022}.

Let us now turn to compressing circuits resulting from applying multiple multi-qubit operators acting on overlapping sets of qubits. Our main focus is on treating operations between consecutive two-qubit gates, as all single-qubit gates appearing in between them can be transported through to the side as long as they have a $\pi/4$ prefactor. Consider an arbitrary two qubit gate of the form $e^{i\alpha\sigma^{a}_1 \rho_2}$ and a single qubit gate of the form $e^{\pm i\frac{\pi}{4}\sigma^{b}}$. Then we have (see Appendix \ref{appendix:xyz_derivations}):
\begin{equation}\label{eq:sqgcompression}
    \raisebox{-.40\height}{\includegraphics[scale=0.08]{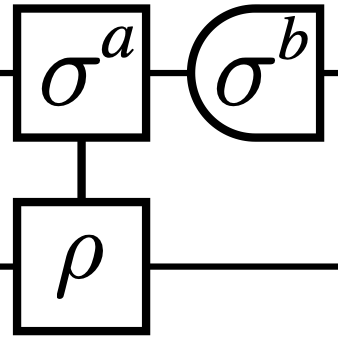}} =
    \raisebox{-.40\height}{\includegraphics[scale=0.08]{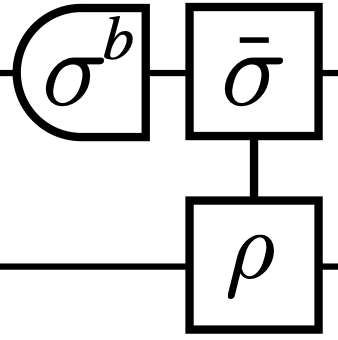}}
\end{equation}
where $\bar{\sigma}=\pm \frac{i}{2}[\sigma^a,\sigma^b]+\delta_{ab}\sigma^{a}$. This works for any single qubit gate with a prefactor of $\pm \pi/4$ but in the negative case we have to reverse the appropriate signs for $\bar{\sigma}$. If there is an arbitrary prefactor $\alpha$, this transportation identity does not work. However, in practice, such situations will not appear in our circuits.
 
Given a commuting pair of two-qubit gates acting on the same qubits they can be either commuted through, annihilated if they differ only in their sign prefactor in the exponential or combined into a single two-qubit gate if this gate is considered native. Graphically we have:
\begin{equation}\label{eq:xyzdecomp4}
\begin{matrix}
    \ \ \ \ \  \raisebox{-.40\height}{\includegraphics[scale=0.08]{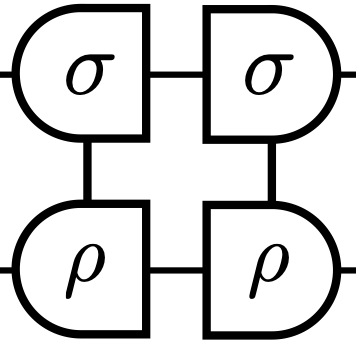}} =
    \raisebox{-.40\height}{\includegraphics[scale=0.08]{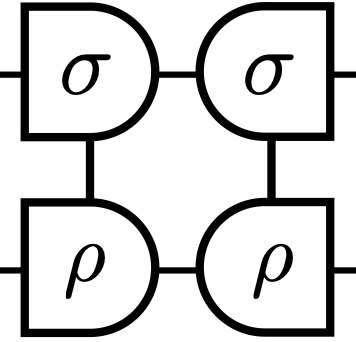}} =
    \raisebox{-.40\height}{\includegraphics[scale=0.08]{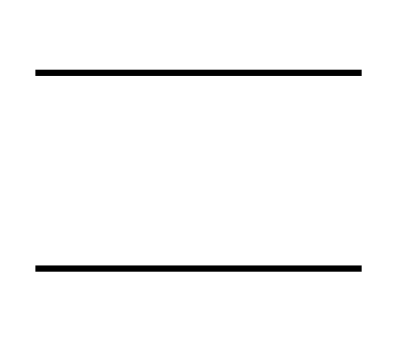}} ,
    \\ \\
    \ \ \, \, \raisebox{-.36\height}{\includegraphics[scale=0.08]{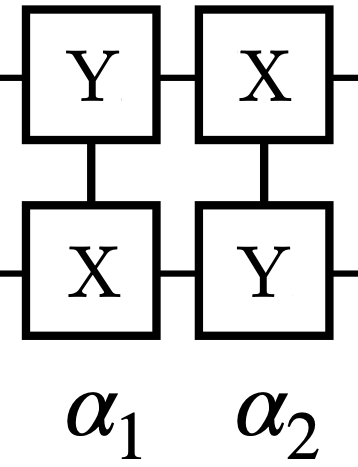}}  =
    \raisebox{-.36\height}{\includegraphics[scale=0.08]{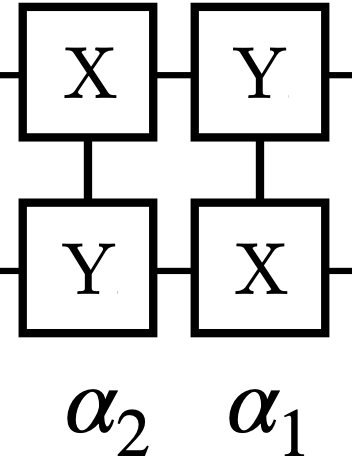}}\   = \,    \ \ 
    \raisebox{-.36\height}{\includegraphics[scale=0.08]{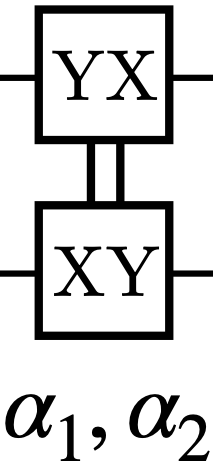}}
\end{matrix}
\end{equation}
where we grouped the two last gates into $e^{i\alpha_1X_1Y_2+\alpha_2Y_1X_2}$, with $\alpha_1=\pm \alpha_2$, which up to single qubit rotations corresponds to a parameterised iSWAP gate ($e^{i\alpha(X_1X_2+Y_1Y_2)}$). This way, two commuting Pauli strings with the same prefactor (independently of their sign) can be implemented using one two-qubit and some single-qubit gates. We thus assume that the squared 'native' two-qubit gate on the lower r.h.s. of Eq.~\ref{eq:xyzdecomp4} can admit any combination of signs. All commuting operators fulfilling this can be combined into one two-qubit gate if a set of three native gates is available, i.e. the set $\{e^{i\phi Z_1Z_2}, e^{i\phi(X_1Y_2+Y_1X_2)}, e^{i\phi(X_1Y_2+Z_1Z_2)}\}$. More generally, all such gates can be implemented, up to single qubit rotations, as a fermionic simulation (fSIM) gate \cite{Kivlichan2018}:
\begin{equation}
    \text{fSIM}_{ij}(\theta,\phi) = e^{i\frac{\theta}{2}(X_iX_j+Y_iY_j)+i\frac{\phi}{4}(Z_i+Z_j-Z_iZ_j)}
\end{equation}
This gate also allows to merge three commuting Pauli strings if two of them have a similar absolute prefactor. Such a two-qubit gate has been natively implemented on existing superconducting platforms~\cite{foxen2020demonstrating} and has been shown to be useful for compressing fermionic operators in the context of the JW mapping combined with a fermionic swap (fSWAP) gate network~\cite{Kivlichan2018}. In principle, we found that nearly all two-qubit gate pairs appearing within our circuits can be transformed into $e^{i\alpha(X_1Y_2+Y_1X_2)}$ (with gates coming from some quartic fermionic interaction terms being notable exceptions which generate $e^{i\alpha Z_1 Z_2}$ type gates). This is why we will use its graphical form from Eq.~\ref{eq:xyzdecomp4} throughout later sections instead of fSIM gates.

For anti-commuting two-qubit gates acting on the same qubits it is possible to combine them into a single two-qubit gate if at least one of them has a $\pm \pi/4$ prefactor in the exponential, which corresponds to the common case encountered (see Appendix \ref{appendix:xyz_derivations}):
\begin{equation}\label{eq:xyzdecomp8}
    \raisebox{-.42\height}{\includegraphics[scale=0.08]{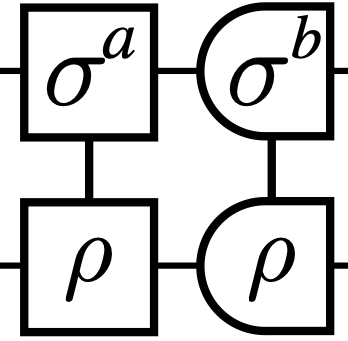}} =
    \raisebox{-.42\height}{\includegraphics[scale=0.08]{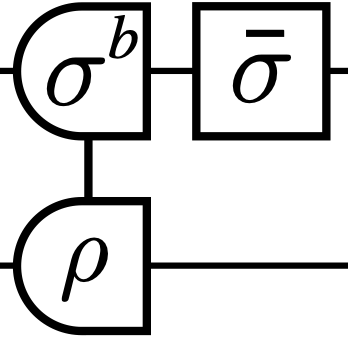}} 
\end{equation}
where $\sigma^a=\frac{i}{2}[\sigma^b,\bar{\sigma}]$. Let us now show how one can use these circuit compression identities in practice. 
Consider the two operators $e^{i\alpha Y_1 Z_2 Z_3 X_4}e^{i\alpha X_1 Z_2 Z_3 Y_4}$:\vspace{0.1cm}
\begin{equation}\label{eq:exhc}
\begin{matrix}
    \raisebox{-.46\height}{\includegraphics[scale=0.08]{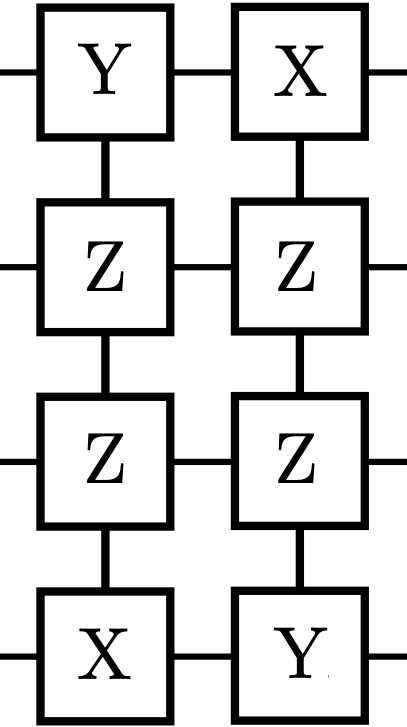}} =
    \raisebox{-.46\height}{\includegraphics[scale=0.08]{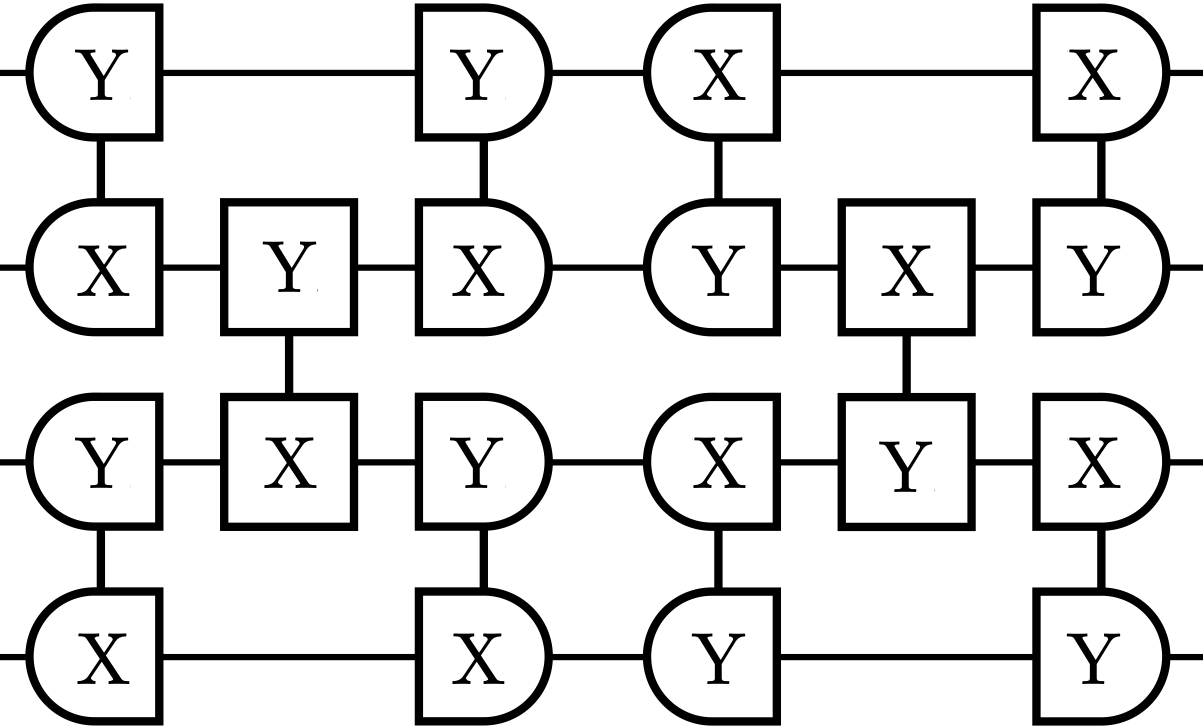}} =
    \raisebox{-.46\height}{\includegraphics[scale=0.08]{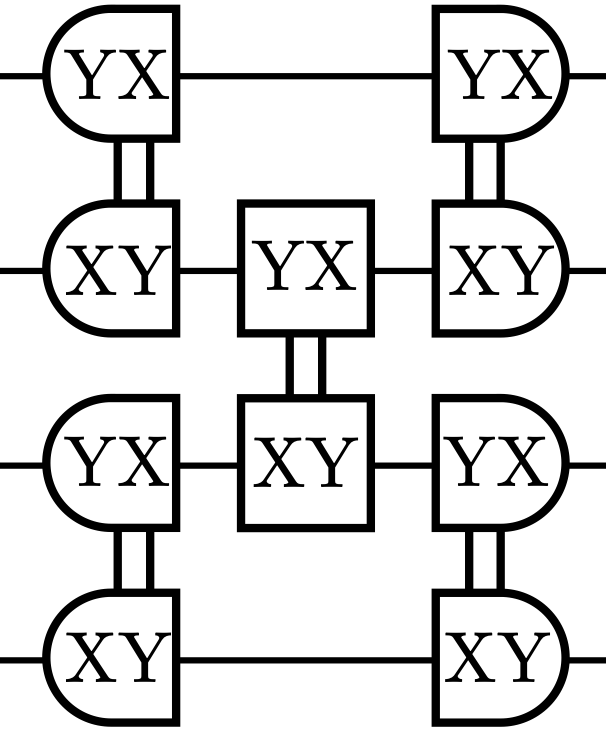}}
\end{matrix}
\end{equation}\vspace{0.1cm}
We have decomposed the two four-qubit operators (l.h.s.) in such a way that most gates commute (center) and subsequently paired them into native two-qubit gates (r.h.s.), which has reduced the depth by half. As we will see, this is a common theme for fermionic Hamiltonians and can be proven to be optimal. Specifically, for $\alpha=\pi/8$:
\begin{align}
    e^{i\frac{\pi}{8}(X_1 X_2...X_{n-1} Y_n+}& ^{Y_1 X_2...X_{n-1}X_n)}|0\rangle^{ \otimes n}= \\ &\frac{|0\rangle^{\otimes n}+|1\rangle^{\otimes n}}{\sqrt{2}} \nonumber
\end{align}
The implementation of two such operators together therefore gives rise to a maximally entangled GHZ state. Following the steps in \cite{xyzdecomposition}, this implies that for every pair of qubits there has to be a chain of non-commuting two-qubit gates sequentially connecting them, so a minimal depth of $n-1$ is needed for operators with even weight and $n$ for operators with odd weight, proving the XYZ decomposition for fermionic hoppings to be optimal in terms of depth and in the number of two-qubit gates.

Now let us add another two operators to the example of Eq.~\ref{eq:exhc}, which have a partial qubit overlap with the first two: $e^{i\alpha Y_1 Z_2 Z_3 X_4}e^{i\alpha X_1 Z_2 Z_3 Y_4}e^{i\alpha Y_3 Z_4 Z_5 X_6}e^{i\alpha X_3 Z_4 Z_5 Y_6}$. We obtain the circuit:
\begin{equation}\label{eq:exdisj}
\begin{matrix}
    \raisebox{-.47\height}{\includegraphics[scale=0.08]{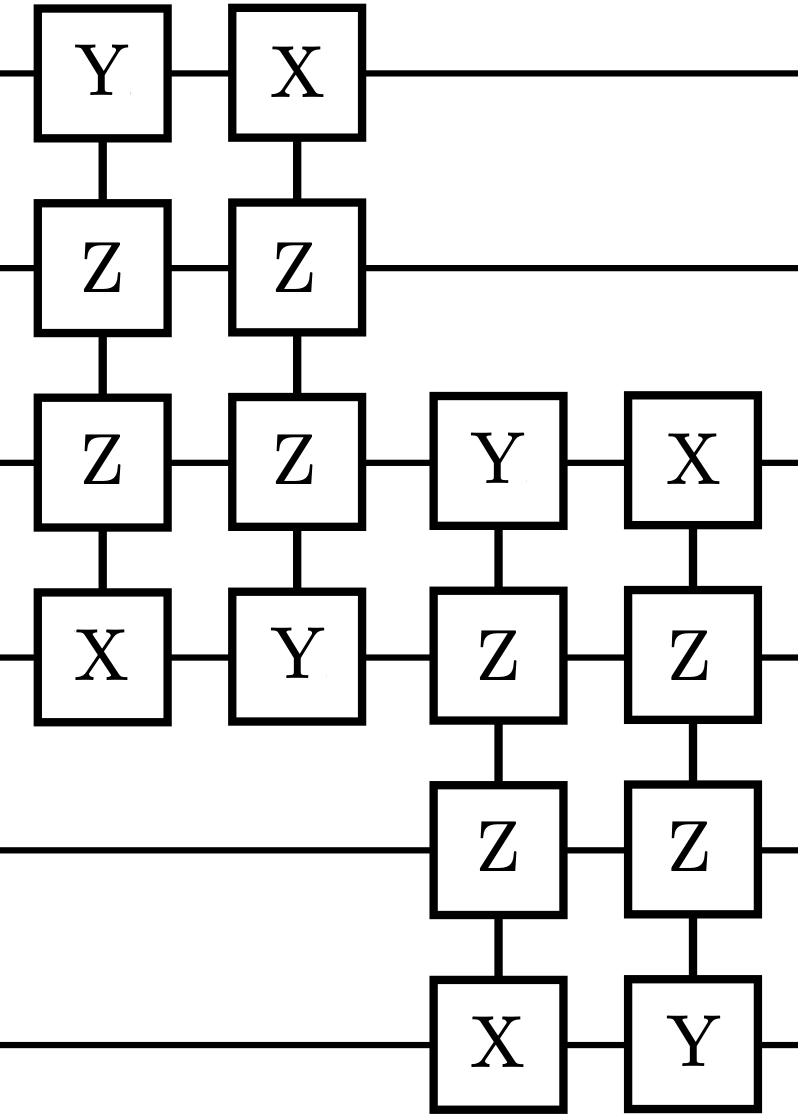}} =
    \raisebox{-.47\height}{\includegraphics[scale=0.08]{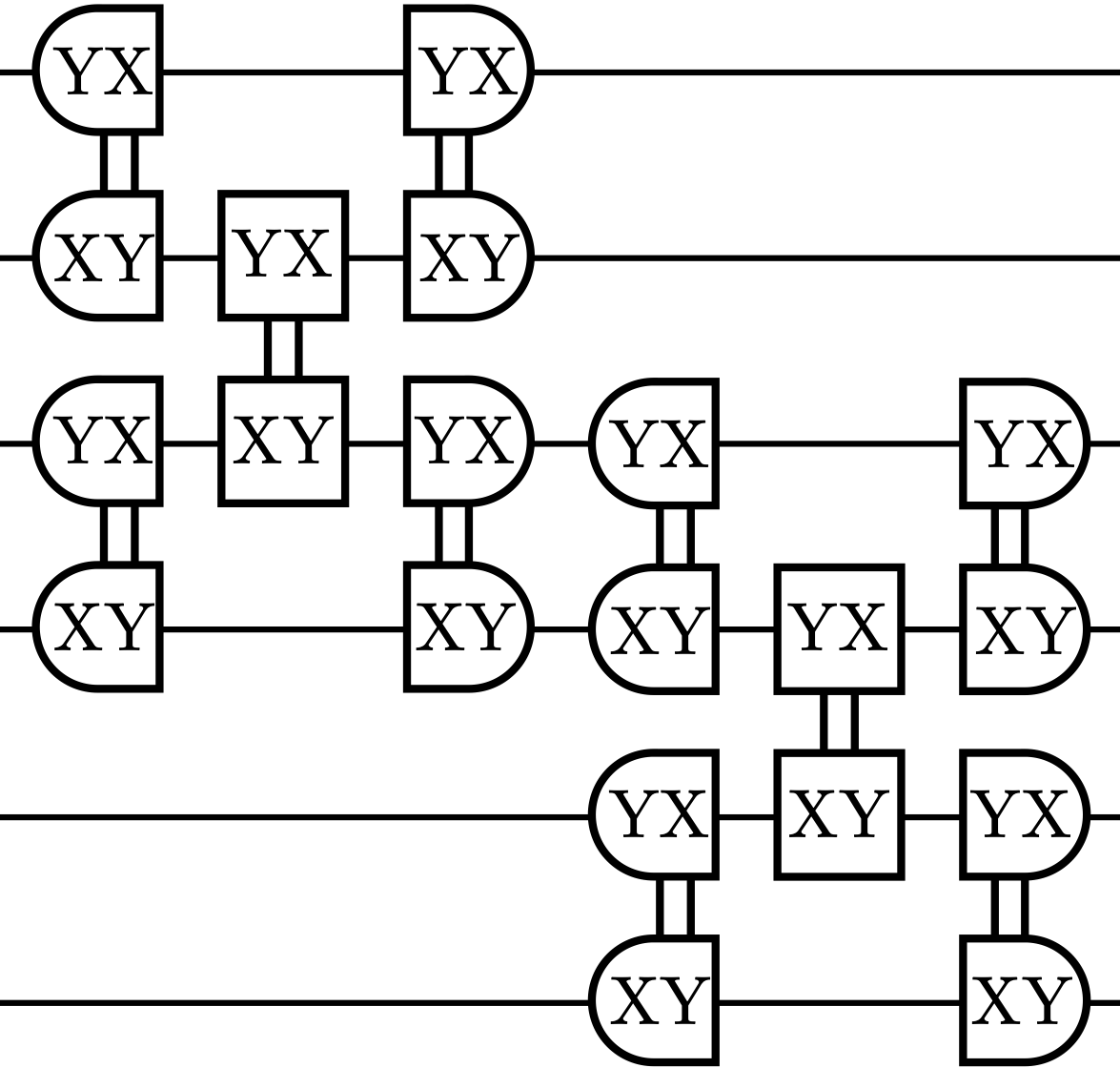}} =
    \raisebox{-.47\height}{\includegraphics[scale=0.08]{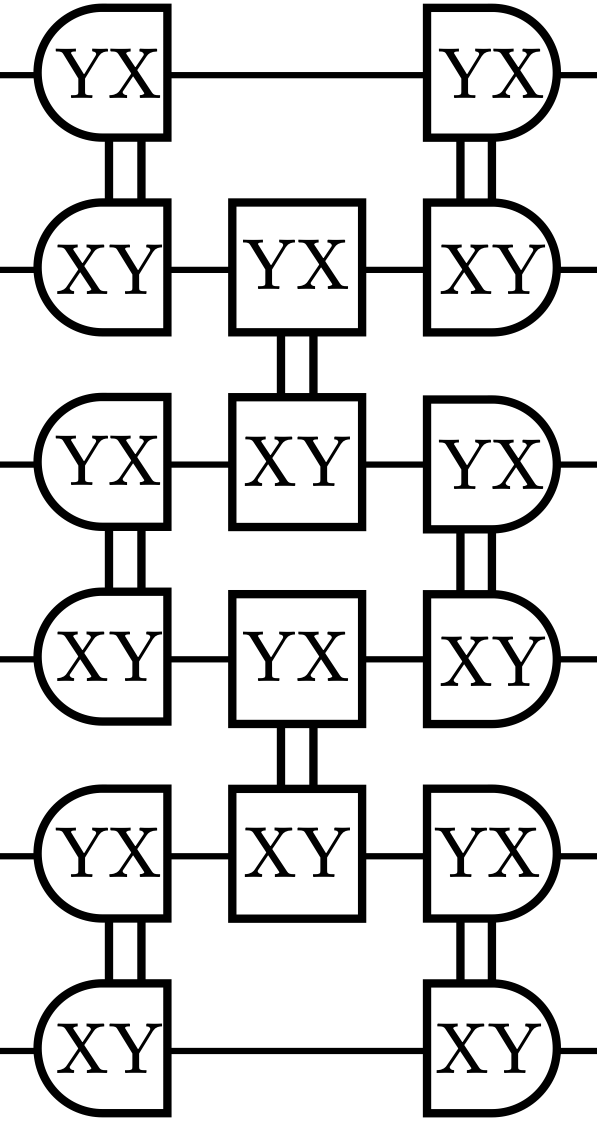}}
\end{matrix}
\end{equation}
where we have been able to mutually annihilate two-qubit gates between the two pairs, which allowed us to further parallelize  gates and reduce the total circuit depth by a further factor two. 

These examples clearly demonstrate the power of our approach. 
However, the overall efficiency of the formalism will differ for different fermion-to-qubit mappings. In the following section, we will present a general strategy for generating mappings for a square qubit layout. These mappings will then be compared against each other in the context of these XYZ decomposition and compression techniques.

\section{Fermionic models on regular lattices}
\label{sec:models}
The most general model used to describe fermionic systems in quantum chemistry and material science is the electronic structure Hamiltonian. Its second-quantized form is given by:
\begin{equation}
    \mathcal{H}_{\text{ES}} = \sum_{pq}h^{pq} c^\dagger_pc_q^{\phantom{\dagger}} + \sum_{pqrs}h^{pqrs} c^\dagger_p c^\dagger_q c_r^{\phantom{\dagger}} c_s^{\phantom{\dagger}}
\end{equation}
where $p,q,r,s$ are indices of a given basis set, $c^\dagger$ and $c$ are the fermionic creation and annihilation operators and $h^{pq}$, $h^{pqrs}$ are constants called the one- and two-electron integrals, respectively. In general, the coefficient tensor of electronic structure Hamiltonians is maximally dense, thus containing all $N^4$ possible quadratic and quartic terms, where $N$ is the total number of fermionic modes in the system. For such dense Hamiltonian, it has been shown that that the optimal strategy is to use the Jordan-Wigner mapping in combination with fSWAP networks~\cite{Kivlichan2018, Gorman2019}. However, this optimality no longer holds true for structured Hamiltonians with some degree of sparsity, like those defined on regular lattices. Indeed, these are the systems for which we expect our approach to be the most fruitful.  

One of the most complex models falling into this class is the Hubbard-Kanamori model (HK), designed to accurately describe the competing spin- and orbital degrees of freedom in d- and f- electron materials. 
Notably, the strongly correlated transition-metal oxides~\cite{Kanamori1963, Georges2013, Hao2019}, which are of fundamental relevance to the design of  batteries~\cite{Delgado2022} fall into this category. The HK Hamiltonian reads:
\begin{align}\label{eq:hkham}
&\mathcal{H}_{\text{HK}} = \nonumber \sum_{i,j, m, \sigma} t^{ijm\sigma}c^\dagger_{i m \sigma} \; c_{jm\sigma}^{\phantom{\dagger}}\\& + \sum_{i,m} U^{im} \; n_{im\uparrow}^{\phantom{\dagger}}n_{im\downarrow}^{\phantom{\dagger}}\\&+ \!\!\! \sum_{i,m < \bar{m}} \!\!\! U_1^{im\bar{m}}\left(n_{im\uparrow}^{\phantom{\dagger}}n_{i\bar{m}\downarrow}^{\phantom{\dagger}}+n_{im\downarrow}^{\phantom{\dagger}}n_{i\bar{m}\uparrow}^{\phantom{\dagger}}\right) \nonumber \\&+ \!\!\! \sum_{i,m < \bar{m}} \!\!\! U_2^{im\bar{m}}  \left(n_{im\uparrow}^{\phantom{\dagger}}n_{i\bar{m}\uparrow}^{\phantom{\dagger}}+n_{im\downarrow}^{\phantom{\dagger}}n_{i\bar{m}\downarrow}^{\phantom{\dagger}}\right) \nonumber \\& + \!\!\! \sum_{i,m < \bar{m}} \!\!\! J^{im\bar{m}} \left(c^\dagger_{im\uparrow}c^\dagger_{im\downarrow}c_{i\bar{m}\downarrow}^{\phantom{\dagger}}c_{i\bar{m}\uparrow}^{\phantom{\dagger}} +c^\dagger_{i\bar{m}\uparrow}c^\dagger_{i\bar{m}\downarrow}c_{im\downarrow}^{\phantom{\dagger}}c_{im\uparrow}^{\phantom{\dagger}} \right. \nonumber \\& \left. \quad\quad\quad\quad\quad \!
+c^\dagger_{im\uparrow}c^\dagger_{i\bar{m}\downarrow}c_{im\downarrow}^{\phantom{\dagger}}c_{i\bar{m}\uparrow}^{\phantom{\dagger}} +c^\dagger_{i\bar{m}\uparrow}c^\dagger_{im\downarrow}c_{i\bar{m}\downarrow}^{\phantom{\dagger}}c_{im\uparrow}^{\phantom{\dagger}} \right) \nonumber   
\end{align}
where  $n$ is the number operator, the indices $i$, $j$ denote lattice sites ($i\neq j$), the indices  $m$, $\bar{m}$ label orbitals and there are two spins $\sigma \in \{\uparrow,\downarrow \}$ per orbital. The terms with prefactor $t$ correspond to fermionic hopping, $U$, $U_1$ and $U_2$ terms are intra- and inter-band density-density interactions, and $J$ terms represent pair-hopping and spin-exchange interaction terms. Following common practice~\cite{Hao2019}, we only allow for hopping to occur between neighboring lattice sites. 

One can trivially reduce this Hamiltonian to the single-orbital Fermi-Hubbard model (FH) which is believed to capture the physics of copper-based high-temperature superconductors. The model is obtained from Eq.~\ref{eq:hkham} by setting the number of orbitals to one and therefore suppressing all inter-orbital interaction terms, i.e. $U_1^{im\bar{m}}=U_2^{im\bar{m}}=J^{im\bar{m}}=0$. Finally, we set $t^{ijm\sigma}\equiv - t^{ij}$ with $t^{ij} \neq 0$ only if $i$,$j$ are considered neighbors on the fermionic lattice. The resulting Hamiltonian has the form:
\begin{align}\label{eq:fhham}
\mathcal{H}_{\text{FH}} & =  -  \sum_{i,j, \sigma} t^{ij} c^\dagger_{i \sigma}c_{j\sigma}^{\phantom{\dagger}}+U \sum_{i} n_{i\uparrow}^{\phantom{\dagger}}n_{i\downarrow}^{\phantom{\dagger}}
\end{align}
Note, that in what follows we omit any chemical potential terms ($\mu^{\sigma} = t^{iim\sigma}$) of the Fermi-Hubbard Hamiltonian for simplicity reasons. Such terms could be trivially added to the Hamiltonian without requiring any changes to our approach. The FH model is extremely challenging to classical computational methods and, with the exception of some specific regimes, remains largely unsolved.

If we further allow for only one spin type, and consequently also set $U=0$, we recover the one-spin tight-binding model Hamiltonian (TB):
\begin{align}\label{eq:tbham}
\mathcal{H}_{\text{TB}} & =  - \sum_{i,j} t^{ij} c^\dagger_{i}c_{j}^{\phantom{\dagger}}
\end{align}
This model is relatively trivial to solve, but is nevertheless useful to investigate it for demonstration purposes, as will be done in section \ref{subsec:spinlessmodels}. 

\begin{figure}[h!btp]
    \centering
    \includegraphics[width=\columnwidth]{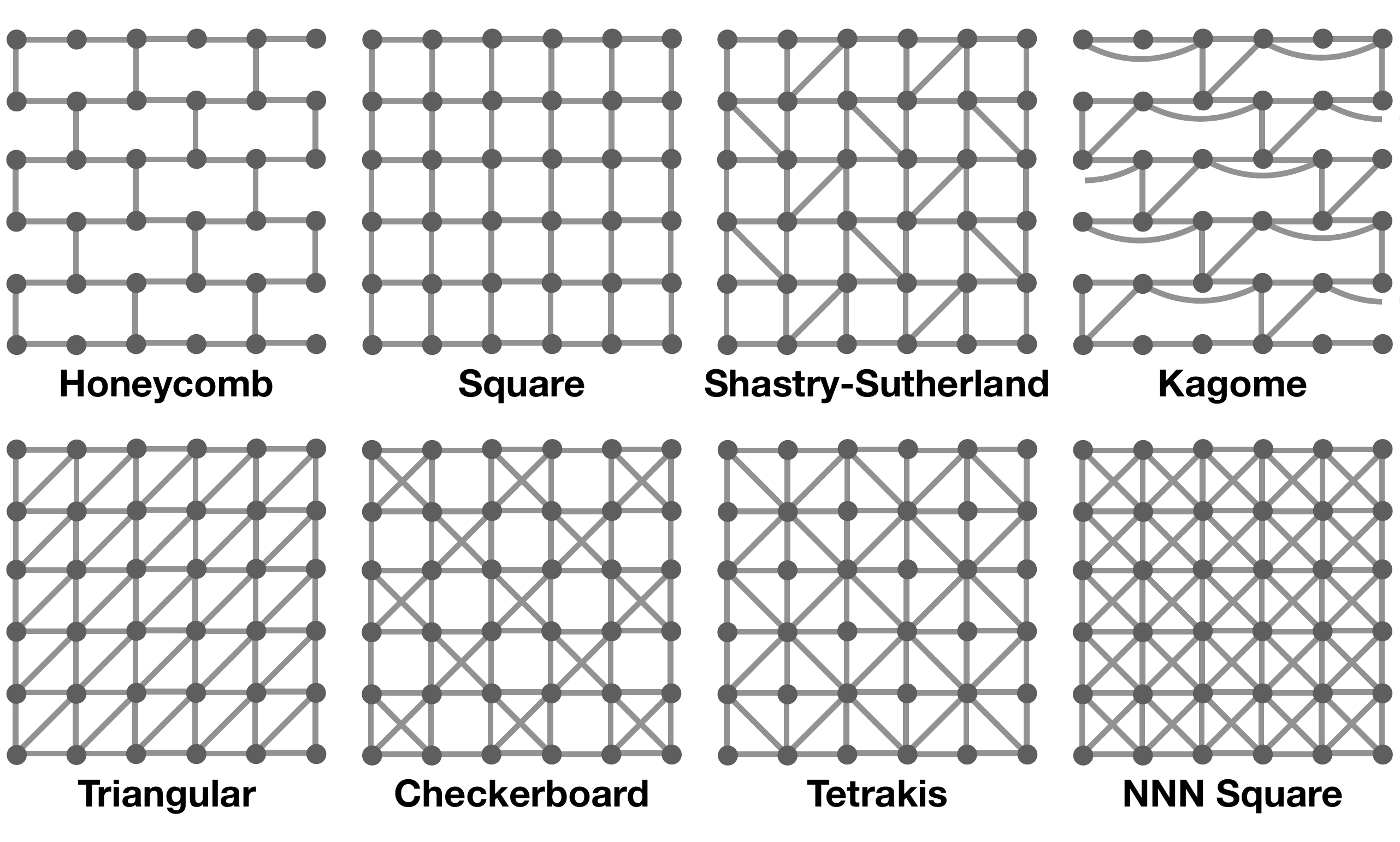} 
    \caption{Different fermionic lattices  embedded into a square lattice geometry by allowing for higher-neighbour connectivity. Nodes represent fermionic lattice sites and links indicate the existence of hopping terms between two sites. }\label{fig:lattices}
\end{figure}

In principle, one can define all of the aforementioned models on any fermionic connectivity graph, but here we only focus on regular, two-dimensional lattices. The most common choice is the square lattice with nearest-neighbor (NN) and optionally next-nearest-neighbor (NNN) connectivity. However, other geometries have also been extensively studied in literature, notably the triangular~\cite{Szasz2020, Wietek2021}, honeycomb~\cite{Raczkowski2020, Yang2021} and Kagome~\cite{Medeiros2023} lattices. In this paper, we consider eight different geometries as shown in Fig.~\ref{fig:lattices}. Rather than investigating optimal mappings for each of these lattices individually, we embed each of them into a square lattice layout whilst allowing for higher-neighbor connectivity (and the corresponding inter-site hopping terms). This way it is sufficient to only consider strategies of mapping a square fermionic lattice to a square qubit layout. We do not consider one- or three-dimensional fermionic models here, as for one-dimensional systems the Jordan-Wigner mapping is already optimal and in the case of three-dimensional systems one faces the additional difficulty of accommodating the third dimension on a two-dimensional lattice. The best approach in this case is to treat it on equal footing with orbital degrees of freedom \cite{Clinton2022}, as we will show in detail in further sections. 

\section{Local fermion-to-qubit mappings for square qubit layouts}
\label{sec:themapping}

In this section we will investigate how to efficiently transform fermionic Hamiltonians into spin Hamiltonians containing operators that act on a set of qubits $Q$ which can be implemented on a quantum device: 
\begin{align}
\label{eq:spin_ham}
    \mathcal{H}_{\text{Q}}  = \sum_i a_i  \bigotimes_{j\in Q}\, \tilde{\sigma}_i^j,
\end{align}
where, $a_i$ are constants and $\tilde{\sigma}_i^j \in \{ I, X, Y, Z\}$. The antisymmetric nature of fermionic systems is encoded in the state wavefunction when working in first quantization. In second quantization this is already included in the fermionic operators used to describe the Hamiltonian of the system. This poses a challenge for transforming the antisymmetric operators of fermionic Hamiltonians into Pauli qubit operators that have no native antisymmetry. This transformation should preserve the locality of the fermionic interactions to avoid increases in the scaling of the number of gates or the circuit depth.

The most convenient approach for creating local fermion-to-qubit mappings (i.e. those where the operator weight is constant with respect to the system size) consists of defining the vertex $V_{i}$ and edge $E_{ij}$ operators for every fermionic mode $i$ and pair of modes $(i,j)$. To facilitate the conversion between the two types of operators, we can additionally introduce the Majorana fermionic operators $\gamma_i=c_i^{\phantom{\dagger}}+c_i^\dagger$ and $\Bar{\gamma}_i=\frac{1}{i}(c_i^{\phantom{\dagger}}-c_i^\dagger)$~\cite{Bravyi2002}. This allows us to compose the edge and vertex operators as $E_{ij}=-i\gamma_i \gamma_j=-i(c_ic_j+c^\dagger_ic^\dagger_j+c_ic^\dagger_j+c^\dagger_ic_j)$, $V_i=-i\gamma_i\Bar{\gamma}_i=c_i,c^\dagger_i-c^\dagger_ic_i$. In order to be compatible with the fermionic anticommutation relations, the edge and vertex operators must themselves satisfy the following relations:
\begin{align}\label{eq:cond1} &\quad \quad \{E_{ij},V_i\}=\{E_{ij},E_{jk}\}=0 \\
 &[E_{ij},E_{kl}]=[E_{ij},V_{k}]=[V_i,V_j]=0 \nonumber
\end{align}
for indices $i\neq j\neq k \neq l$. This means that edges must anticommute with vertices which they are incident on, two edges must anticommute if they share a vertex, and all other combinations of two operators must commute. 
Additionally, since $\gamma_i \gamma_i = 1$, then $\gamma_i \gamma_j\gamma_j\gamma_i = 1$ and any cyclic product of Majorana operators must be equal to the identity: $\prod_j^{|p|-1}\gamma_{p_j} \gamma_{p_{j+1}}=\mathbb{1}$, where $p=\{p_1,p_2,...\}$ forms a closed path. By using $E_{ij} = -i \gamma_i \gamma_j$, this equation in terms of Majorana operators is translated into a condition that any valid mapping must fulfill:
\begin{equation}\label{eq:cond3} 
i^{(|p|-1)}\prod_j^{|p|-1} E_{p_j,p_{j+1}}= \mathbb{1} 
\end{equation}
 The condition in Eq.~\ref{eq:cond3} is not generally fulfilled since most combinations for edge and vertex operators will create a non-trivial operator $S_p$ instead of the identity $\mathbb{1}$. However, one can solve this issue by initializing the quantum device in the common eigenspace $U=\bigcap_p U_p$ of each product's $+1$ eigenspace $U_p$ and thus $S_p\sim \mathbb{1}$. A standard way to do this \cite{Derby2020,Derby2021} consists of measuring the stabilisers $S_p$ at the beginning of each run and transform the state if it is not in the +1 eigenspace by implementing a 1-depth layer of single qubit gates based on the outcome of the stabilisers measurement.

This measurement feedback loop is not always available in experiments. In that case, one can alternatively prepare the correct state in the +1 eigenspace of $S_p$ via unitary encoding using only quantum gates. We refer the reader to general strategies for implementing this approach for a given set of stabilisers whose depth is at worst $\mathcal{O}(n)$ for $n$ qubits  \cite{Cleve1997, higgott2021optimal}.

The most straightforward way to prepare the initial state \cite{hagge2023} when having access to measurement feedback consists of constructing the desired fermionic state in the physical qubits considering that qubit zero-states correspond to unoccupied fermionic sites and one-states to occupied ones. Then, since vertex operators commute with stabilisers, we are allowed to measure the stabilisers without changing the state of the fermions itself. We choose a pair of defective stabilisers and connect them through a path of stabilisers in such a way that each one shares an edge with the following one. We finally flip the sign of the edge operators of the shared edges so that the original pair of stabilisers is no longer defective and the additionally involved ones will not be affected since two of their edges have been flipped. We repeat this until no more than one defective stabilisers exists. If that is the case, we take the last defective stabiliser and construct a path of edge-sharing stabilisers to an outer stabiliser and flip all the edge operator signs of the shared edges. This will flip the sign of the outer stabiliser, but we can flip the edge of the outer stabiliser that is not shared with any other stabiliser to finally eliminate all the defective measurements. This procedure implies a change in the application of the evolution operators of hopping terms what should not be a problem if we can incorporate measurements feedback.

As an extra condition for the mapping to be correct, one has to define inverse edge operators as $E_{ji}=-E_{ij}$, which we do by assigning one sign prefactor to the ordering $(i,j)$ and the opposite sign to the inverse one $(j,i)$. Edge operators may be constructed from other existing edges by using the \emph{composite} rule: 
\begin{equation} \label{eq:composite}
E_{ik}=iE_{ij}E_{jk}
\end{equation}
This means that it is not necessary to define a unique edge for every connected fermionic mode pair, as long as a composite edge path exists between them. An alternative strategy to composing edge operators is 
to instead swap fermionic modes between adjacent vertices. This can be done using an fSWAP operator between two modes $(i,j)$, which is defined as \cite{Bravyi2002}:
\begin{equation}\label{eq:fswap}
    \operatorname{fSWAP}_{i,j}=e^{i\frac{\pi}{4}V_i}e^{i\frac{\pi}{4}V_j}e^{\frac{\pi}{4}(E_{ij}V_j+V_iE_{ij})}
\end{equation}
Given these definitions we can now translate the quadratic and quartic fermionic operators occurring in the models of Sec.~\ref{sec:models} into products of edge and vertex operators:
\begin{equation}
    n_j^{\phantom{\dagger}} \rightarrow \frac{1}{2}(1-V_j)
\end{equation}
\begin{equation}\label{eq:onsite}
    n_j^{\phantom{\dagger}} n_k^{\phantom{\dagger}} \rightarrow \frac{1}{4} (1-V_j-V_k+V_j V_k)
    % (1-V_j)(1-V_k)
\end{equation}
\begin{equation}\label{eq:hoppings}
    c^\dagger_j c_k^{\phantom{\dagger}} \rightarrow \frac{i}{4}(1+V_k-V_j-V_jV_k)E_{jk}
\end{equation}
\begin{align}\label{eq:hoppings2}
    &c^\dagger_jc^\dagger_kc_l^{\phantom{\dagger}}c_m^{\phantom{\dagger}} \to - \frac{1}{16}(1+V_m-V_k +V_l-V_j \\&-V_kV_m +V_lV_m+V_jV_k-V_kV_l-V_jV_l-V_jV_m\nonumber \\&-V_kV_lV_m+V_jV_kV_m-V_jV_lV_m+V_jV_kV_l\nonumber \\& \quad \quad \quad \quad +V_jV_kV_lV_m)E_{jl}E_{km} \nonumber
\end{align}
where $j\neq k \neq l \neq m$. One can notice that the resulting formulation involves significantly more terms than the original fermionic operator. However, most of these cancel out once multiple symmetry-related fermionic operators are combined:
\begin{equation}\label{eq:quadraticsumhc}
    c^\dagger_j c_k^{\phantom{\dagger}} + c^\dagger_k c_j^{\phantom{\dagger}} \rightarrow \frac{i}{2}(V_k-V_j)E_{jk}
\end{equation}
\begin{align}\label{eq:quarticsumhc}
 &c^\dagger_jc^\dagger_kc_l^{\phantom{\dagger}}c_m^{\phantom{\dagger}}+c^\dagger_jc^\dagger_lc_k^{\phantom{\dagger}}c_m^{\phantom{\dagger}}+c^\dagger_lc^\dagger_mc_j^{\phantom{\dagger}}c_k^{\phantom{\dagger}}+c^\dagger_mc^\dagger_kc_l^{\phantom{\dagger}}c_j^{\phantom{\dagger}} \to \nonumber \\& -\frac{1}{4}(V_jV_k-V_jV_l+V_lV_m-V_mV_k)E_{jl}E_{km}
\end{align}
which can be expressed using other edges since $E_{jl}E_{km}=-E_{jm}E_{kl}$. Given a set of edges and vertices from a fermionic connectivity graph, one has to define their matching (Pauli string) operators acting on hardware qubits which satisfy all the commutation relations from Eq.~\ref{eq:cond1}. The choice of these operators is not unique and will influence the performance of the underlying algorithm, quantified either in terms of the qubit-to-mode ratio, the gate count, the circuit depth or the error detecting and correcting properties of the mapping. 

\begin{figure}
  \centering
  \begin{minipage}[c]{.36\linewidth}
    \subfloat
      {\includegraphics[width=.9\linewidth]{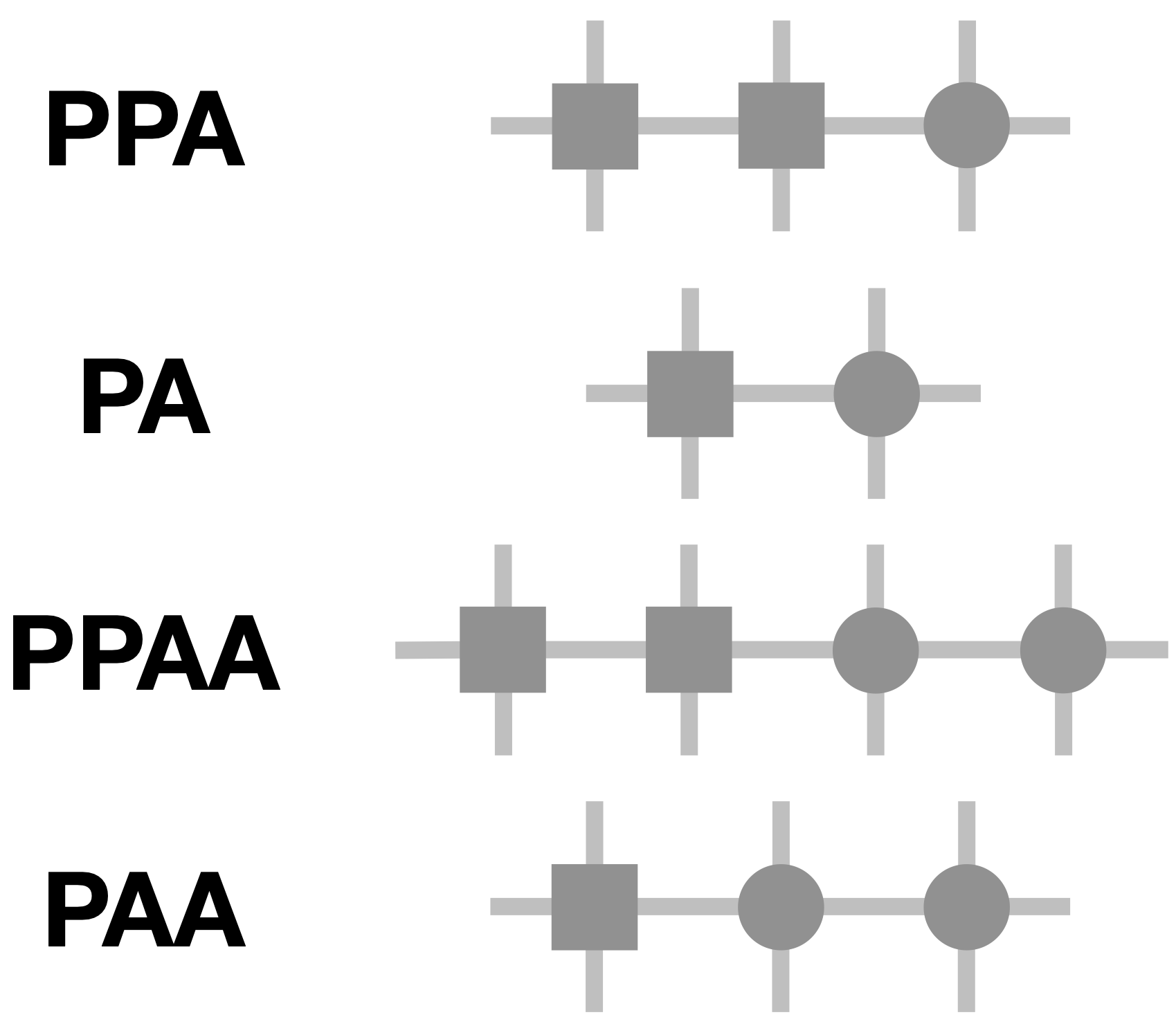}}
  \end{minipage}
  \hfill
  \begin{minipage}[c]{.62\linewidth}
    \subfloat
      {\includegraphics[width=.9\linewidth]{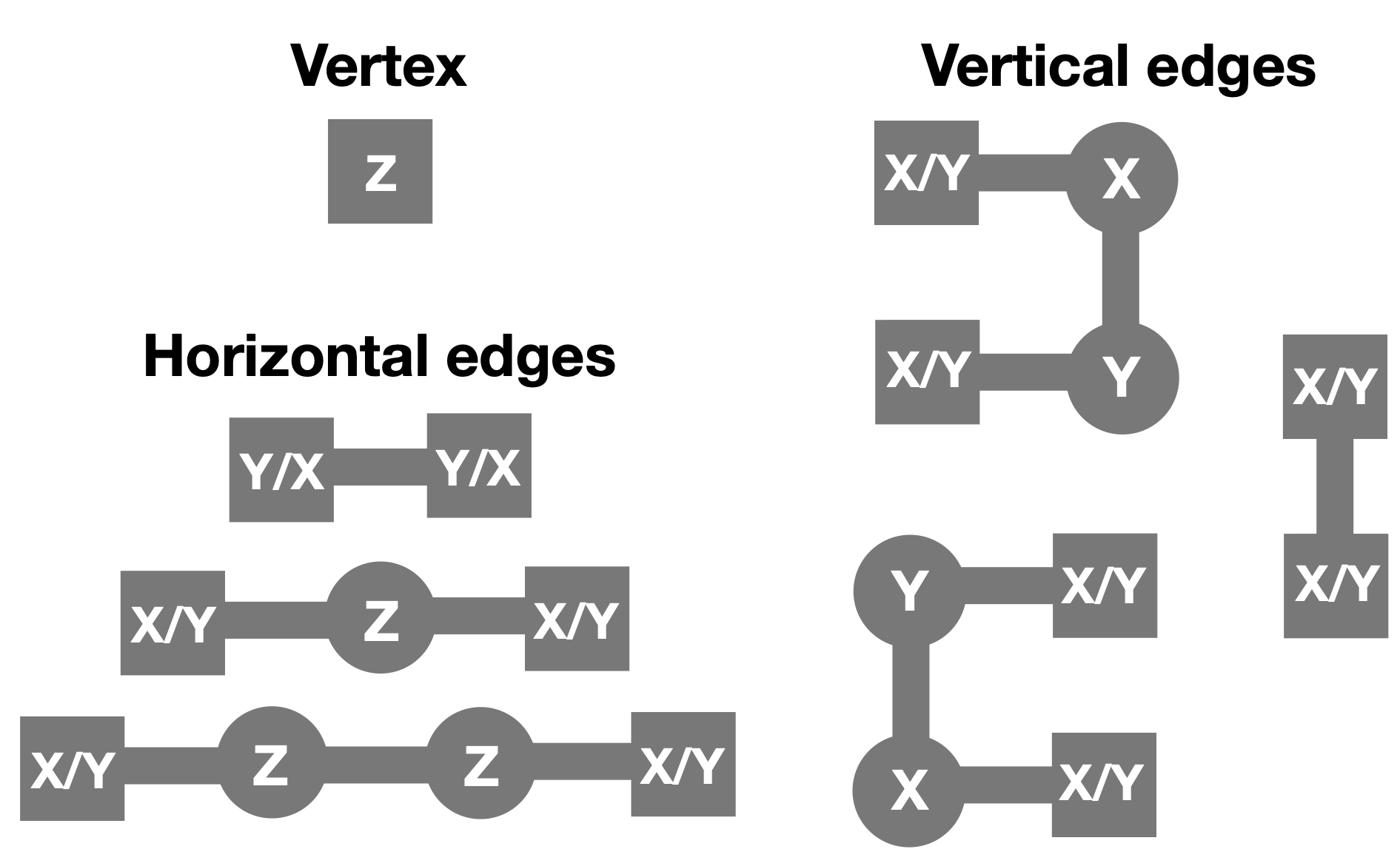}}
  \end{minipage}
  \caption
    {Fermion-to-qubit mappings for square qubit layout: \emph{left)} Unit cells showing different horizontal sequences of physical qubits (squares, P) and ancilla qubits (circles, A). \emph{right)} Vertex operators, horizontal and vertical edge operators defined in the horizontal and vertical directions. The X/Y symbols indicate two possible Pauli strings for a given edge, chosen depending on the particular mapping instance.
      \label{fig:Picture4}
    }
\end{figure}

The family of mappings proposed in this paper uses common structures for the edge and vertex operators (see Fig.~\ref{fig:Picture4} and Table~\ref{table:edge_ops}). For any given mapping we define a unit cell, which consists of a sequence of physical (P) and ancilla (A) qubits. Each unit cell is repeated horizontally and vertically to cover the qubit lattice. We define physical qubits as those on which a fermionic mode exists, meaning that a vertex operator will be acting on it with a Pauli Z operator. Additionally, we introduce ancilla qubits, whose purpose is to resolve (anti-)commutation relations between edge operators. 
In cases where the qubit lattice is incommensurate with the unit cell one can partially cover additional qubits with the given unit cell pattern. In what follows, we will not consider the effects of finite lattices as ultimately they will not influence the relative performance, in terms of circuit depth, of the mappings considered. 

\begin{table*}[]

\begin{center}
\begin{adjustwidth}{0.40cm}{}
\begin{tabular}{|c|c|c|c|}
\hline
\multicolumn{2}{|c|}{Mapping}                   & \multicolumn{2}{c|}{Operator}     \\ \hline \hline 
\multirow{22}{*}{Tight-binding}  &  \multirow{5}{*}{\raisebox{-.40\height}{\includegraphics[scale=0.350]{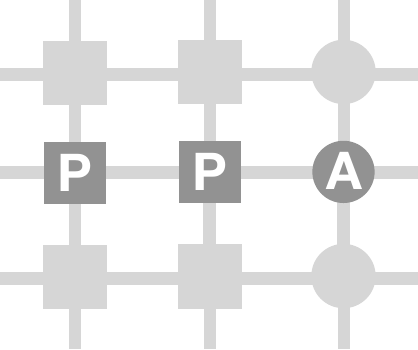}}} & \multicolumn{1}{c|}{$V^{1,1}_{x,y}$} & $Z_{x,y}$ \\ \cline{3-4} 
                 &  &  \multicolumn{1}{c|}{$E^{2,1}_{x,y\rightarrow x,y+1}$} &  $X_{x,y}Y_{a(x,y)}X_{a(x,y+1)}X_{x,y+1}$ \\ \cline{3-4} 
                 &  &  \multicolumn{1}{c|}{$E^{2,1}_{x+1,y\rightarrow x+1,y+1}$} & $X_{x+1,y}X_{a(x,y)}Y_{a(x,y+1)}X_{x+1,y+1}$ \\ \cline{3-4} 
                 &  & \multicolumn{1}{c|}{$E^{2,1}_{x,y\rightarrow x+1,y}$} & $X_{x,y}Z_{a(x,y)}X_{x+1,y}$ \\ \cline{3-4} 
                 &  & \multicolumn{1}{c|}{$E^{2,1}_{x+1,y\rightarrow x+2,y}$} & $Y_{x+1,y}Y_{x+2,y}$ \\ \cline{2-4}
& \multirow{5}{*}{\raisebox{-.40\height}{\includegraphics[scale=0.350]{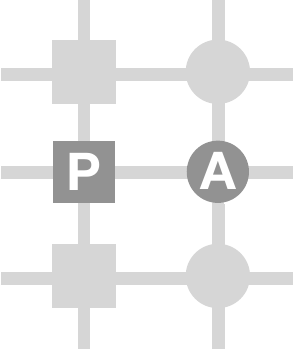}}} & \multicolumn{1}{c|}{$V^{1,1}_{x,y}$} & $Z_{x,y}$ \\ \cline{3-4} 
                 &  & \multicolumn{1}{c|}{$E^{2,1}_{x,y\rightarrow x,y+1}$} &   $X_{x,y}Y_{a(x,y)}X_{a(x,y+1)}X_{x,y+1}$  \\ \cline{3-4} 
                 &  & \multicolumn{1}{c|}{$E^{2,1}_{x+1,y\rightarrow x+1,y+1}$} & $X_{x+1,y}Y_{a(x+1,y)}X_{a(x+1,y+1)}Y_{x+1,y+1}$ \\ \cline{3-4} 
                 &  & \multicolumn{1}{c|}{$E^{2,1}_{x,y\rightarrow x+1,y}$} & $X_{x,y}Z_{a(x,y)}X_{x+1,y}$ \\ \cline{3-4} 
                 &  & \multicolumn{1}{c|}{$E^{2,1}_{x+1,y\rightarrow x+2,y}$} & $Y_{x+1,y}Z_{a(x+1,y)}Y_{x+2,y}$ \\ \cline{2-4}
& \multirow{5}{*}{\raisebox{-.40\height}{\includegraphics[scale=0.350]{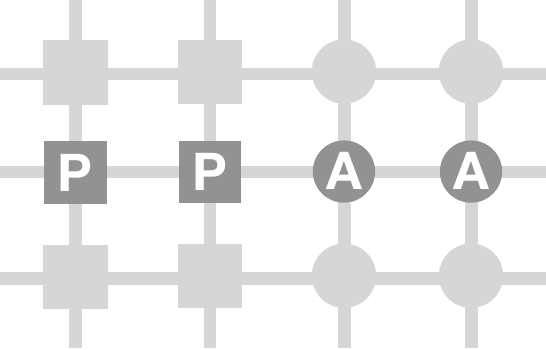}}} & \multicolumn{1}{c|}{$V^{1,1}_{x,y}$} & $Z_{x,y}$ \\ \cline{3-4} 
                 &  & \multicolumn{1}{c|}{$E^{2,1}_{x,y\rightarrow x,y+1}$} & $X_{x,y}Y_{b(x-1,y)}X_{b(x-1,y+1)}X_{x,y+1}$ \\ \cline{3-4} 
                 &  & \multicolumn{1}{c|}{$E^{2,1}_{x+1,y\rightarrow x+1,y+1}$} & $X_{x+1,y}Y_{a(x+1,y)}X_{a(x+1,y+1)}X_{x+1,y+1}$ \\ \cline{3-4} 
                 &  & \multicolumn{1}{c|}{$E^{2,1}_{x,y\rightarrow x+1,y}$} & $Y_{x,y}Y_{x+1,y}$ \\ \cline{3-4} 
                 & & \multicolumn{1}{c|}{$E^{2,1}_{x+1,y\rightarrow x+2,y}$} & $X_{x+1,y}Z_{a(x+1,y)}Z_{b(x+1,y)}X_{x+2,y}$ \\ \cline{2-4} 
& \multirow{7}{*}{\raisebox{-.40\height}{\includegraphics[scale=0.350]{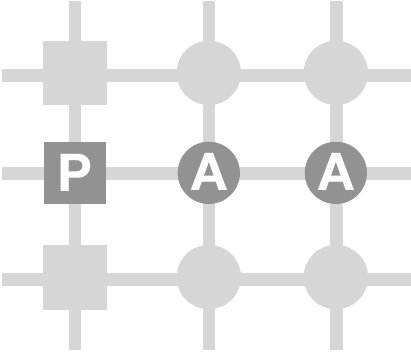}}} & \multicolumn{1}{c|}{$V^{1,1}_{x,y}$} & $Z_{x,y}$ \\ \cline{3-4} 
                 &  & \multicolumn{1}{c|}{$E^{2,1}_{x,y\rightarrow x,y+1}$} & $X_{x,y}X_{b(x-1,y)}Y_{b(x-1,y+1)}X_{x,y+1}$  \\ \cline{3-4} 
                 &  & \multirow{3}{*}{$E^{2,1}_{x+1,y\rightarrow x+1,y+1}$}     &  $X_{x+1,y}X_{a(x+1,y)}Y_{a(x+1,y+1)}X_{x+1,y+1}$ \\
                 &    &                            &     or \\
                 &    &                            &     $Y_{x+1,y}Y_{b(x,y)}X_{b(x,y+1)}Y_{x+1,y+1}$                 \\ \cline{3-4} 
                 &  & \multicolumn{1}{c|}{$E^{2,1}_{x,y\rightarrow x+1,y}$} & $Y_{x,y}Z_{a(x,y)}Z_{b(x,y)}Y_{x+1,y}$ \\ \cline{3-4} 
                 &  & \multicolumn{1}{c|}{$E^{2,1}_{x+1,y\rightarrow x+2,y}$} & $X_{x+1,y}Z_{a(x+1,y)}Z_{b(x+1,y)}X_{x+2,y}$ 
\\ \hline \hline
\multirow{16}{*}{Fermi-Hubbard} & \multirow{9}{*}{\raisebox{-.40\height}{\includegraphics[scale=0.350]{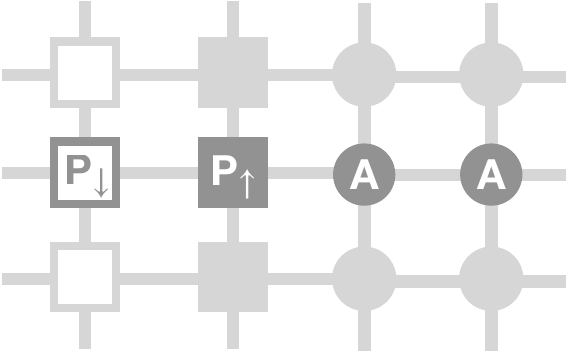}}} & \multicolumn{1}{c|}{$V^{1,1}_{x,y}$} & $Z_{x,y}$ \\  \cline{3-4} 
                 &  & \multicolumn{1}{c|}{$E^{4,1}_{x,y\rightarrow x,y+1}$} & $X_{x,y}X_{b(x-1,y)}Y_{b(x-1,y+1)}X_{x,y+1}$ \\ \cline{3-4} 
                 &  & \multicolumn{1}{c|}{$E^{4,1}_{x+1,y\rightarrow x+1,y+1}$} & $Y_{x+1,y}Y_{a(x+1,y)}X_{a(x+1,y+1)}Y_{x+1,y+1}$ \\ \cline{3-4} 
                 &  & \multicolumn{1}{c|}{$E^{4,1}_{x+2,y\rightarrow x+2,y+1}$} & $Y_{x+1,y}Y_{b(x+1,y)}X_{b(x+1,y+1)}Y_{x+2,y+1}$ \\ \cline{3-4} 
                 &  & \multicolumn{1}{c|}{$E^{4,1}_{x+3,y\rightarrow x+3,y+1}$} & $X_{x+3,y}X_{a(x+3,y)}Y_{a(x+3,y+1)}X_{x+3,y+1}$  \\ \cline{3-4} 
                 &  & \multicolumn{1}{c|}{$E^{4,1}_{x,y\rightarrow x+1,y}$} & $Y_{x,y}X_{x+1,y}$ \\ \cline{3-4} 
                 & 
                   & \multicolumn{1}{c|}{$E^{4,1}_{x+1,y\rightarrow x+2,y}$} & $Y_{x+1,y}Z_{a(x+1,y)}Z_{b(x+1,y)}Y_{(x+2,y)}$ \\ \cline{3-4} 
                 &  & \multicolumn{1}{c|}{$E^{4,1}_{x+2,y\rightarrow x+3,y}$} & $X_{x+2,y}Y_{x+3,y}$ \\ \cline{3-4} 
                 &  & \multicolumn{1}{c|}{$E^{4,1}_{x+3,y\rightarrow x+4,y}$} & $X_{x+3,y}Z_{a(x+3,y)}Z_{b(x+3,y)}X_{(x+4,y)}$  \\ \cline{2-4}
& \multirow{7}{*}{\raisebox{-.40\height}{\includegraphics[scale=0.350]{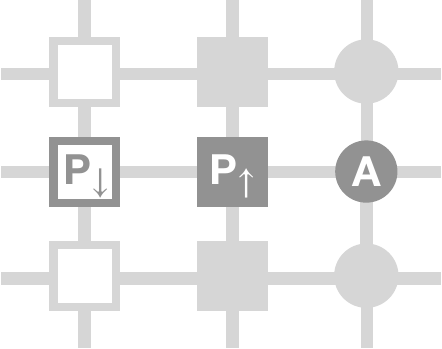}}} & \multicolumn{1}{c|}{$V^{1,1}_{x,y}$} & $Z_{x,y}$ \\ \cline{3-4} 
                 &  & \multicolumn{1}{c|}{$E^{4,1}_{x+1,y\rightarrow x+1,y+1}$} & $X_{x+1,y}Y_{a(x+1,y)}X_{a(x+1,y+1)}X_{x+1,y+1}$ \\ \cline{3-4} 
                 &  & \multicolumn{1}{c|}{$E^{4,1}_{x+3,y\rightarrow x+3,y+1}$} & $Y_{x+3,y}Y_{a(x+3,y)}X_{a(x+3,y+1)}Y_{x+3,y+1}$  \\ \cline{3-4} 
                 &  & \multicolumn{1}{c|}{$E^{4,1}_{x,y\rightarrow x+1,y}$} & $X_{x,y}Y_{x+1,y}$ \\ \cline{3-4} 
                 &  & \multicolumn{1}{c|}{$E^{4,1}_{x+1,y\rightarrow x+2,y}$} & $X_{x+1,y}Z_{a(x+1,y)}X_{(x+2,y)}$ \\ \cline{3-4} 
                 &  & \multicolumn{1}{c|}{$E^{4,1}_{x+2,y\rightarrow x+3,y}$} & $X_{x+2,y}Y_{x+3,y}$ \\ \cline{3-4} 
                 &  & \multicolumn{1}{c|}{$E^{4,1}_{x+3,y\rightarrow x+4,y}$} & $Y_{x+3,y}Z_{a(x+3,y)}Y_{(x+4,y)}$  \\ \hline
\end{tabular}
\end{adjustwidth}
\caption{Explicit definition of edge and vertex operators for each mapping, where $x,y$ represent the physical qubits associated to the site with Cartesian coordinates $x,y$ while $a(x,y)$ and $b(x,y)$ represent the first and second ancillas to the right of the $x,y$ physical qubit. The superindices $m,n$ indicate the regularity of each operator, so a given operator $A^{m,n}_{x,y\rightarrow x+p,y+q}$ will have to be applied every $m$ sites in the horizontal axis and every $n$ sites in the vertical axis. One of the edge operators in the TB-PAA mapping can be implemented through two different instances of Pauli strings. Even though they are slightly different, both of them yield very similar results up to a difference of 1 step in depth.  \label{table:edge_ops}}
\end{center}
\end{table*}

The qubit-to-mode ratio of a mapping is given by $r=(N_P+N_A)/N_P$, where $N_P$ and $N_A$ are the number of physical and ancilla qubits, respectively. We allow a unit cell to contain any number of physical qubits followed by up to two ancilla qubits. We found that whilst two consecutive ancilla qubits can improve the performance of the mapping by allowing for additional degrees of parallelism, a third ancilla does not yield further improvements in this respect but simultaneously worsens other metrics of the mapping. Similarly, we found that allowing for more than two consecutive physical qubits only leads to an improved qubit-to-mode ratio whilst increasing maximum weights of operators as well as the resulting circuit depths. Nevertheless, it is important to note that such mappings might be a viable option if the number of qubits is the limiting factor, as one can, in principle, come arbitrarily close to a ratio of $r=1$. In this limit we would recover the JW mapping embedded in a two-dimensional lattice using a snake pattern consisting solely of two-qubit horizontal edge string and two-qubit vertical edges to connecting them at the ends.

The above physical-ancillary pattern restrictions leave us with four possible choices of unit cells: PPA, PA, PPAA and PAA (Fig.~\ref{fig:Picture4}.a)), which we will consider for mapping particular models in the next sections. We can now define a common set of edge operators for these four unit cells (Fig.~\ref{fig:Picture4}.b)). This includes horizontal edges, which can act on between two, three and four qubits, depending on the number of ancilla qubits connecting two physical ones.  Vertical edges act on four qubits and can only be defined when a pair of ancilla qubits is adjacent to a pair of physical qubits.  In cases where a vertical edge exists on the boundary of the lattice and no other vertical edge shares physical qubits with it (i.e. in the Honeycomb lattice of Fig.~\ref{fig:lattices}), one can omit the ancilla qubits in the vertical edge operator and reduce it to the two-qubit form shown in Fig.~\ref{fig:Picture4}.b).
All additional higher-order edges occurring in the lattices of Fig.~\ref{fig:lattices} are generated using the composite rule of Eq.~\ref{eq:composite}. Further, consecutive edges in the horizontal direction must act on their common physical qubit with different Pauli operators, which means that we alternate between the set of operators \{XX, XZX, XZZX, XXYX, XYXX\} and the set \{YY, YZY, YZZY, YXYX, YYXY\}, resulting in a pool of operators of weights 2-4. This, together with the definitions of the edge operators in Fig.~\ref{fig:Picture4}, guarantees that all commutation relations of Eq.~\ref{eq:cond1} are satisfied.

Note that the PA mapping is equivalent to the Verstraete-Cirac mapping \cite{Verstraete2005} embedded in a square qubit topology (the authors of the original paper considered two superimposed square grids of qubits, whilst the mapping has been embedded into a non-square topology in Ref.~\cite{Clinton2021}). The remaining three mappings have, to our knowledge, not been previously considered.

\begin{figure*}
\centering
\includegraphics[width=0.75\linewidth]{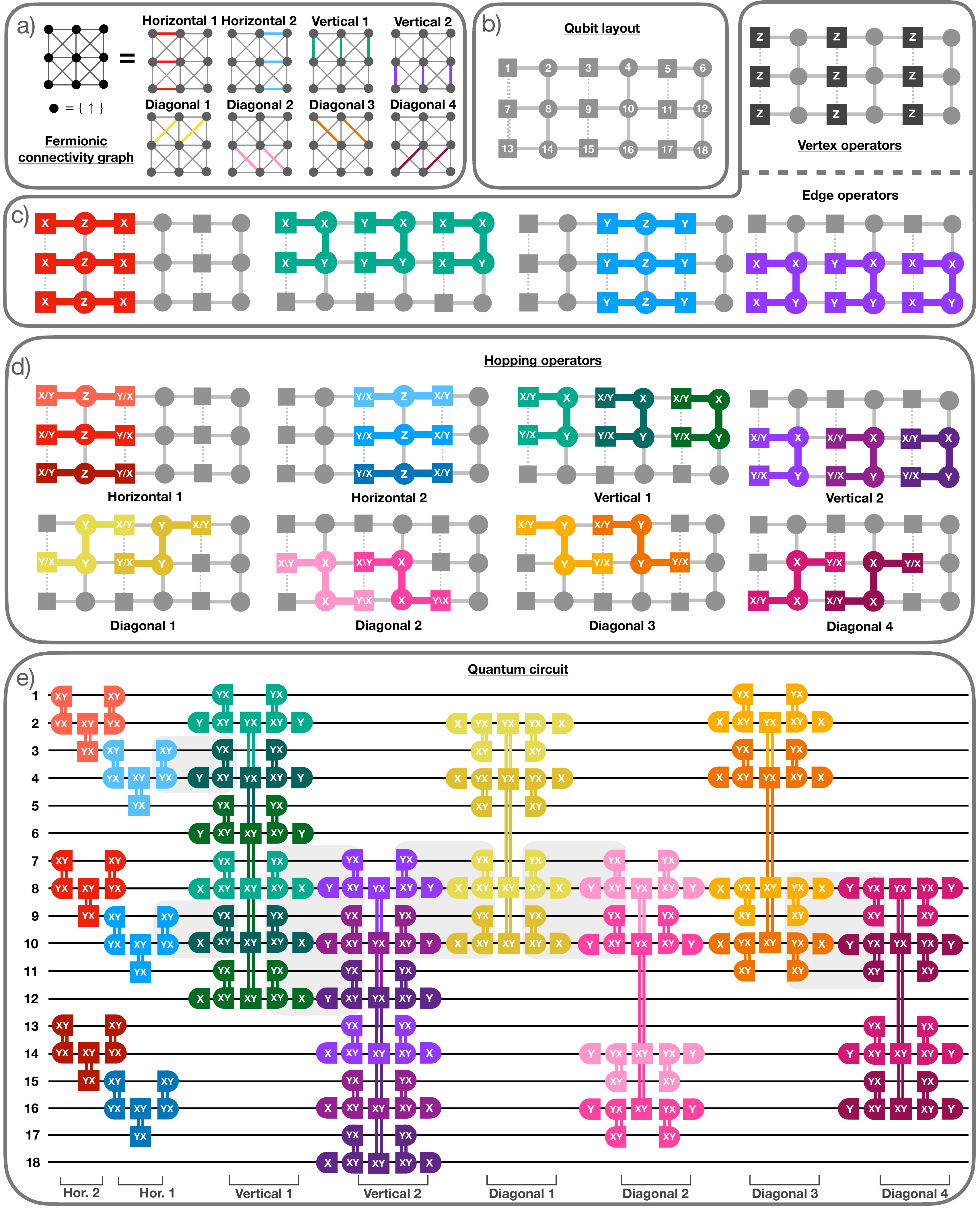}
\caption{a) Fermionic connectivity of the 3-by-3  NNN square lattice, arranged into parallel groups. b) The corresponding numbered qubits in a 6-by-3 square qubit layout using the PA mapping. Square shapes represent physical qubits, circles are ancilla qubits. c) Vertex (black) and edge operators for this mapping. Red and blue strings represent horizontal edge operators while green and purple strings depict vertical ones. The letters X,Y,Z indicate the Pauli acting on a given qubit. Dashed qubit connections are not required. d) All operators obtained from applying the mapping to the hopping terms of the tight-binding model, grouped according to a). $X/Y$ and $Y/X$ terms indicate two operators of the form $X\dots Y$ and $Y\dots X$. Diagonal terms are obtained from the composite edge rule. e) Resulting quantum circuit for the execution of a single Trotter step for the Hamiltonian, obtained using the XYZ-decomposition, and assuming the two-qubit gate $e^{i\alpha(XY+YX)}$ is native. Further circuit compression can be applied within grey shaded areas of the circuit. }\label{fig:Picture1}
\end{figure*}

As an example of how to use the XYZ formalism efficiently in conjunction with these mappings, let us consider PAA, which contains four-qubit edge operators both in the horizontal and vertical directions. The corresponding nearest-neighbor fermionic hopping terms can then be efficiently transformed in Hermitian conjugate pairs into four-qubit operators using Eq.~\ref{eq:quadraticsumhc}. Applying the tricks we introduced in the context of the XYZ formalism we can compress the circuit for the horizontal hopping operators, $t_h=e^{i\alpha(X_1Z_2Z_3Y_4-Y_1Z_2Z_3X_4)}$ and the vertical hopping operators $t_v=e^{i\alpha(X_1Y_2X_3Y_4-Y_1Y_2X_3X_4)}$, into depth three:
\begin{equation}\label{eq:HandV}
    \raisebox{-.47\height}{\includegraphics[scale=0.08]{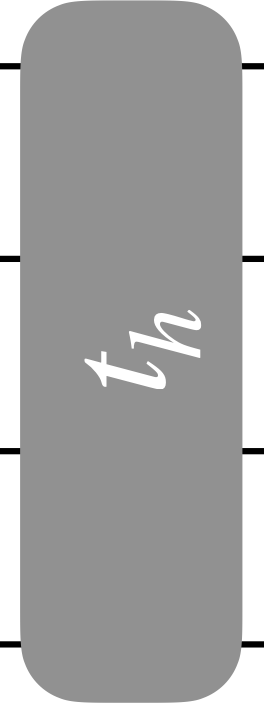}}=\raisebox{-.47\height}{\includegraphics[scale=0.08]{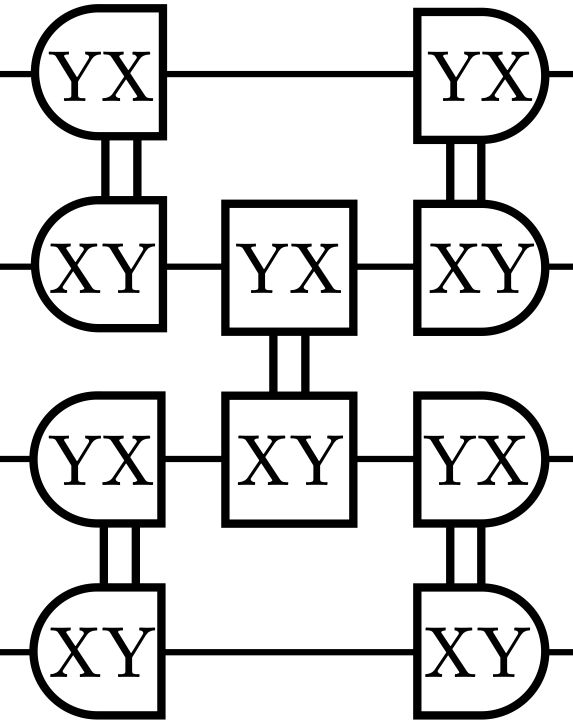}}\  ,\quad 
    \raisebox{-.47\height}{\includegraphics[scale=0.08]{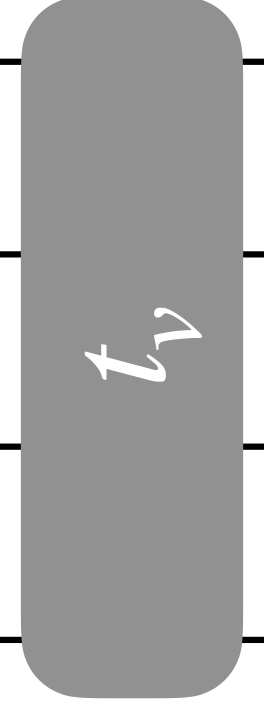}}=\raisebox{-.47\height}{\includegraphics[scale=0.08]{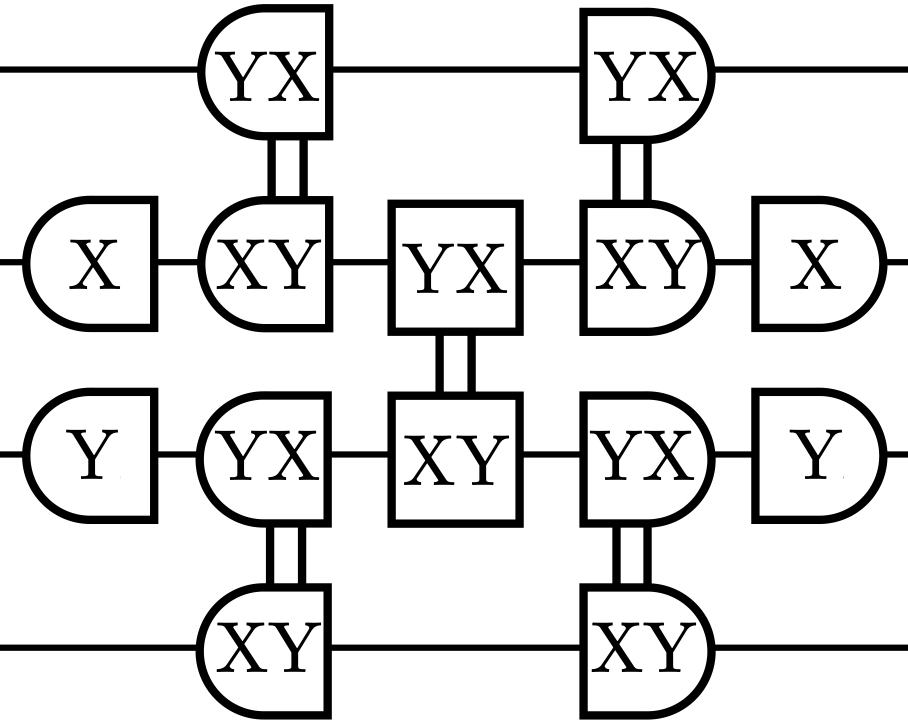}}
\end{equation}
where the two outside qubits are physical ones and the two inside qubits are ancillas. Similarly, all density-density fermionic terms of the form of Eq.~\ref{eq:onsite} can be straightforwardly transformed into circuits with a single two-qubit gate:
\begin{equation}\label{eq:Uterm}
    \raisebox{-.43\height}{\includegraphics[scale=0.085]{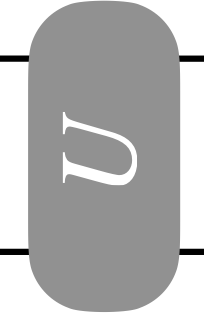}} = \raisebox{-.43\height}{\includegraphics[scale=0.085]{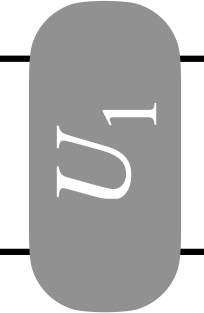}} = \raisebox{-.43\height}{\includegraphics[scale=0.085]{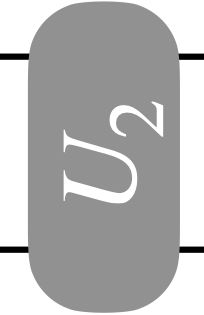}} = \raisebox{-.43\height}{\includegraphics[scale=0.085]{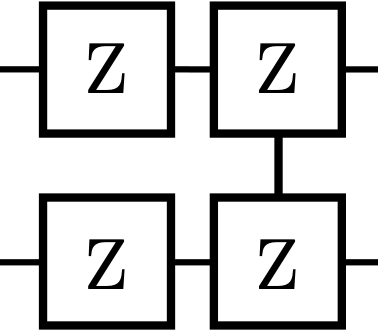}}
\end{equation}
The density-density term and the hopping terms between two modes can, in fact, be implemented as efficiently as the hopping terms alone, even if the modes are connected through a number of ancilla qubits:
\begin{equation}\label{eq:hopwithint}
\begin{matrix}
    \ \ \ \ \raisebox{-.43\height}{\includegraphics[scale=0.105]{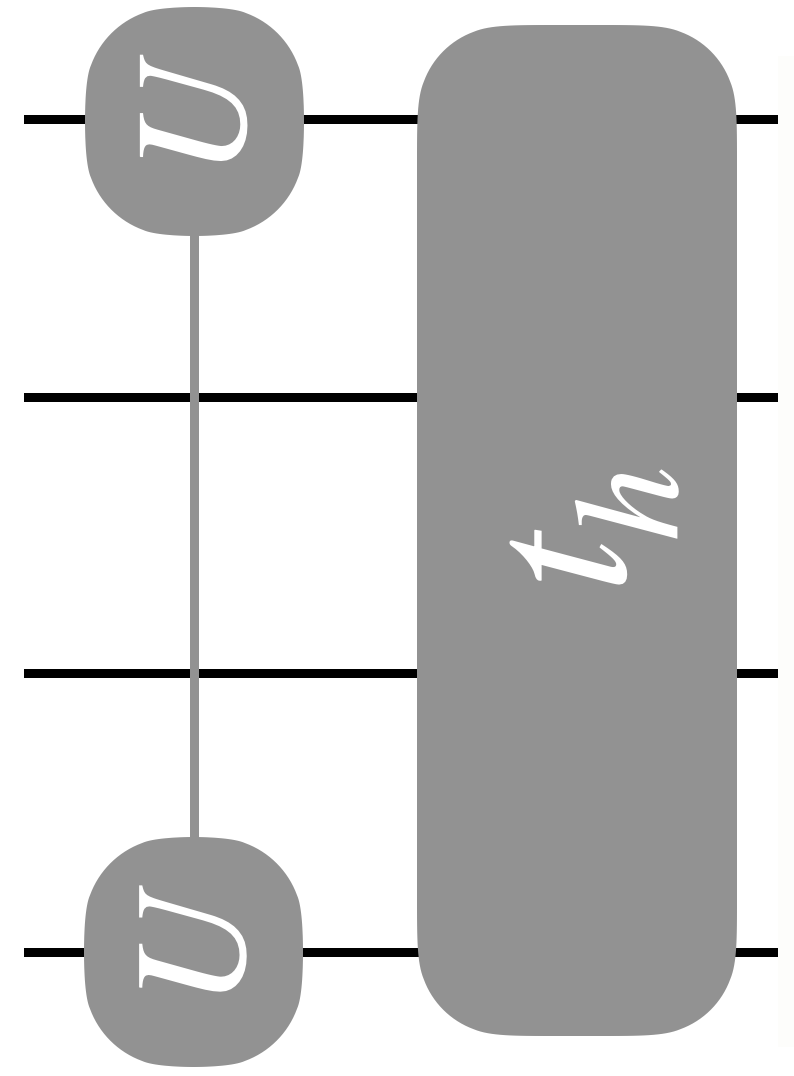}} =
    \raisebox{-.43\height}{\includegraphics[scale=0.105]{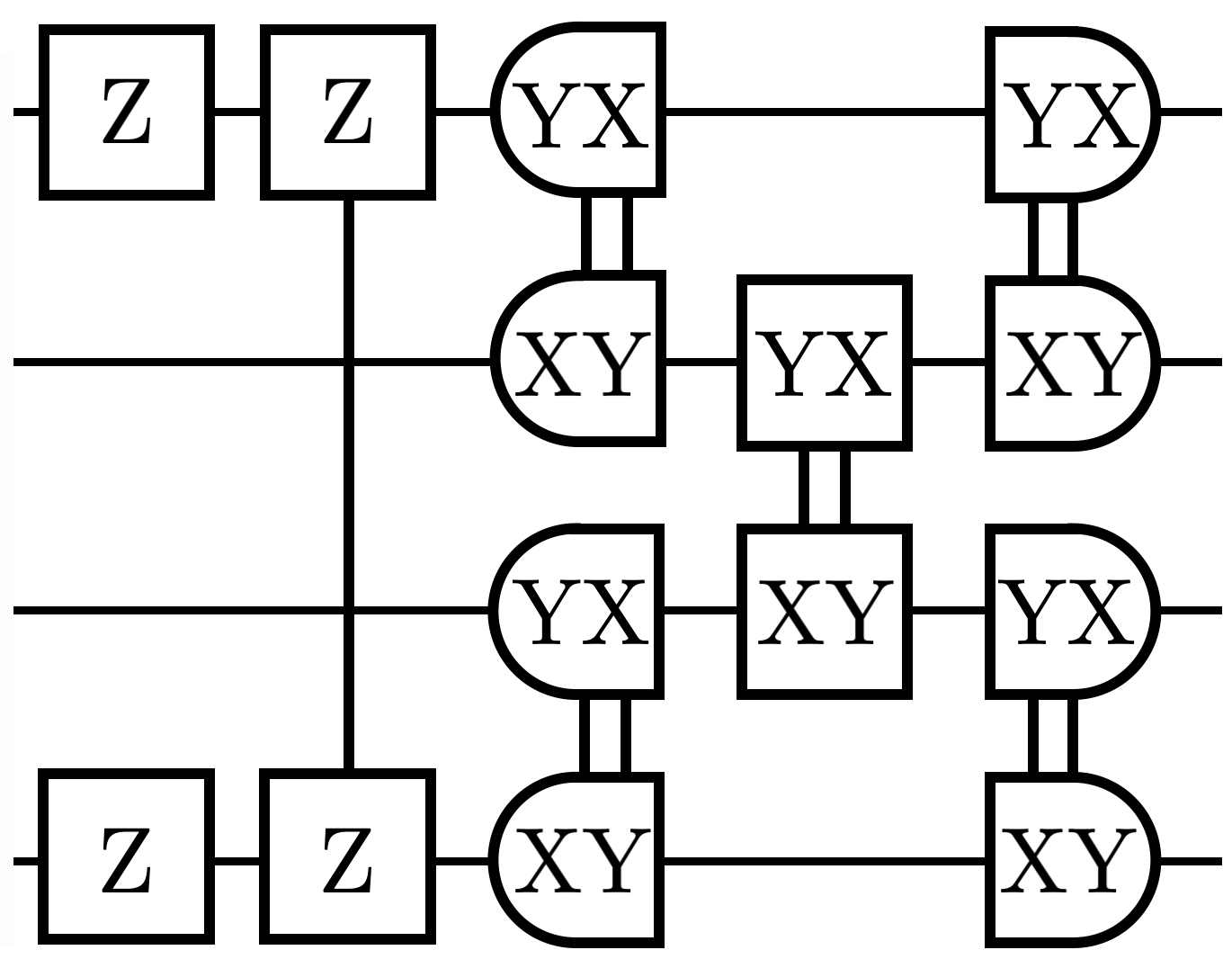}} = \quad \quad \\ \\ =
    \raisebox{-.43\height}{\includegraphics[scale=0.105]{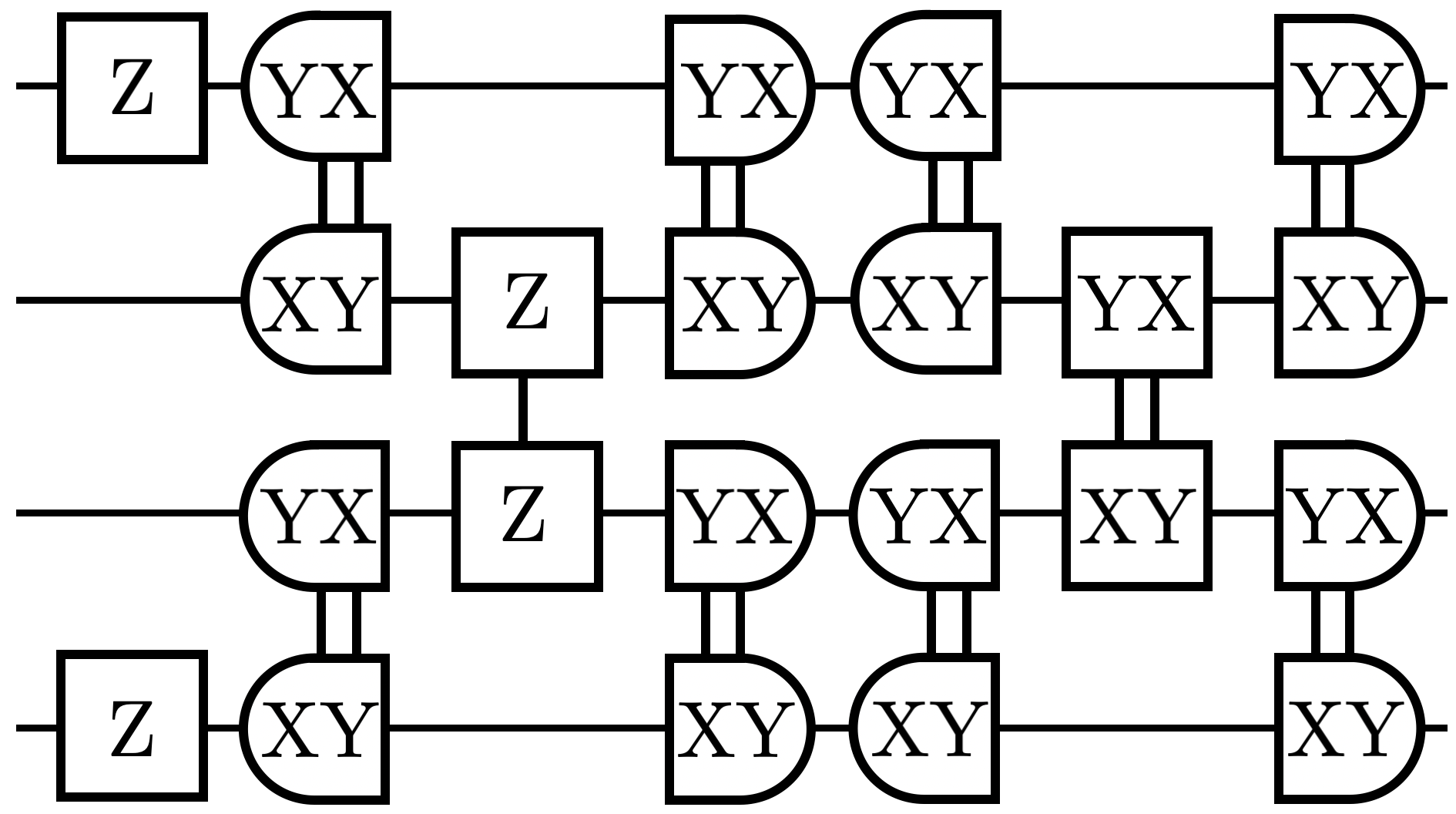}} =
    \raisebox{-.43\height}{\includegraphics[scale=0.105]{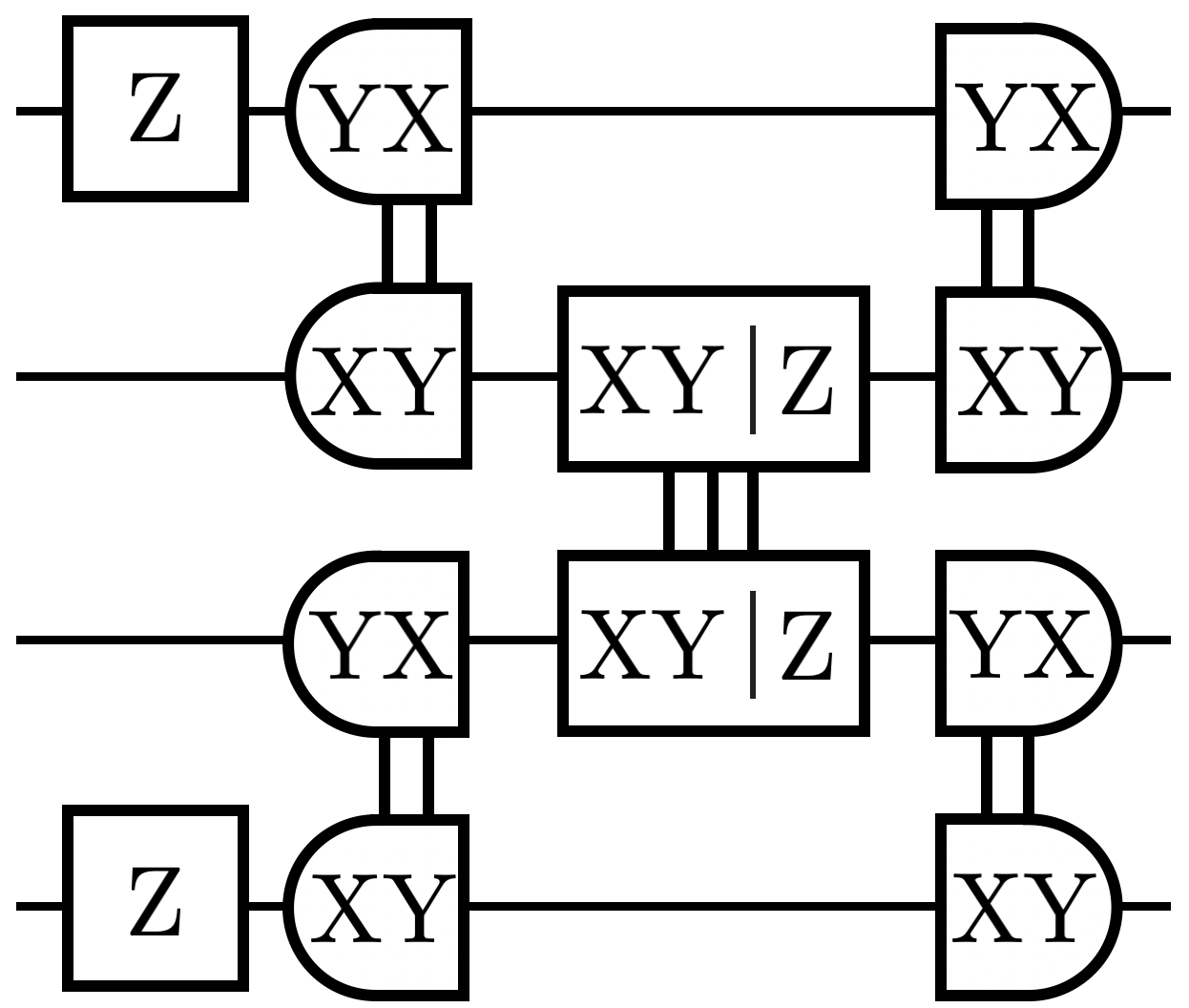}}
\end{matrix}
\end{equation}

Here, the central gate on the r.h.s. represents an interaction of the form $e^{i\theta(X_1Y_2+Y_1X_2)+i\phi Z_1Z_2}$, which is an fSIM gate up to single qubit rotations. This shows that using the XYZ decomposition for jointly implementing hopping and density-density operators between two physical modes is optimal.

One can also efficiently compress the four additional quartic pair-hopping and spin-exchange terms which appear in the HK model of Eq.~\ref{eq:hkham} into a depth-five circuit by using Eq.~\ref{eq:quarticsumhc}:
\begin{equation}\label{eq:Jterm}
    \raisebox{-.48\height}{\includegraphics[scale=0.08]{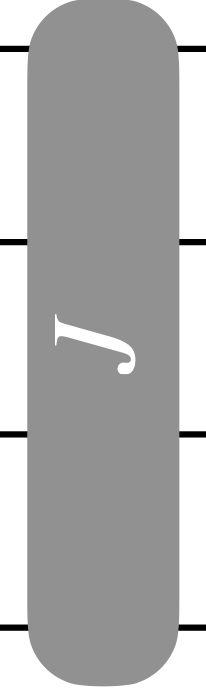}}=\raisebox{-.49\height}{\includegraphics[scale=0.08]{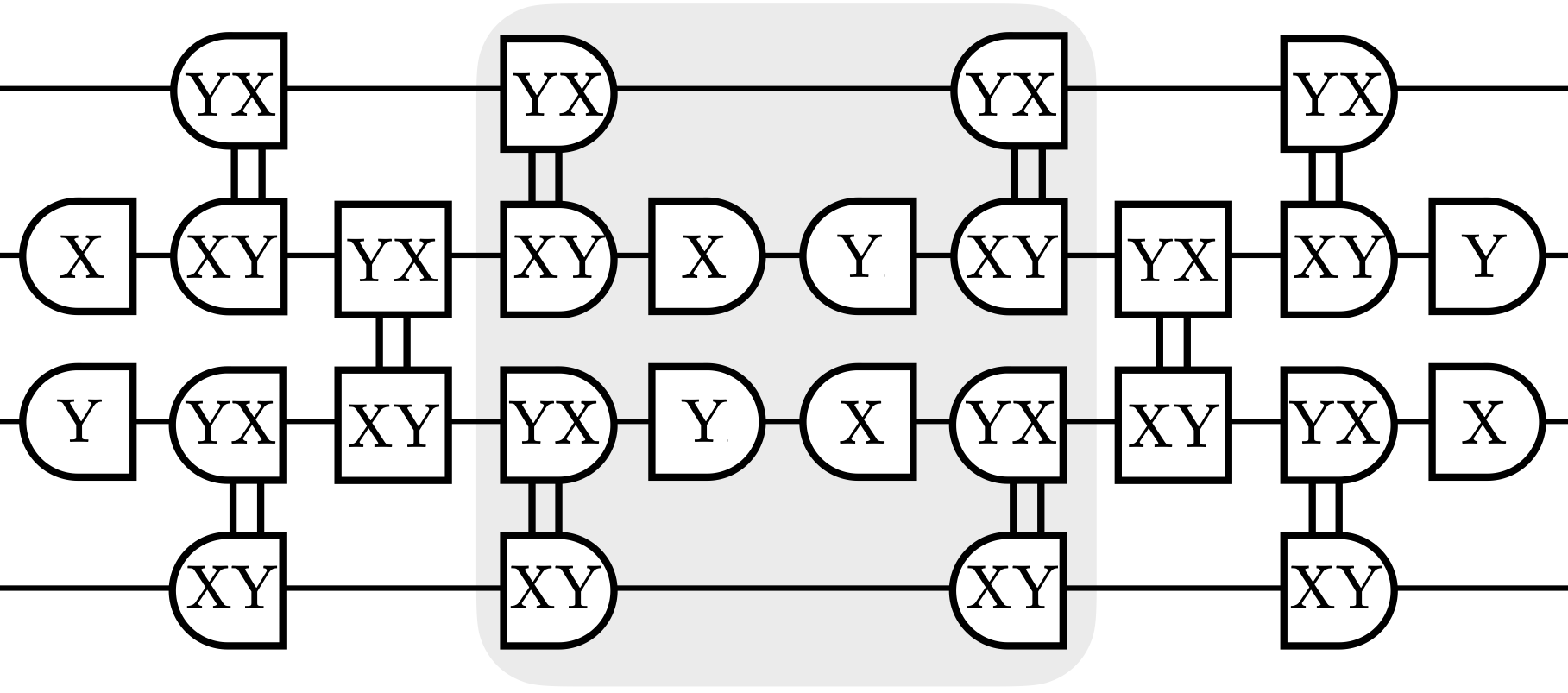}}
\end{equation}
Here, we assume that the four modes correspond to pairs of spins $\{\uparrow,\downarrow\}$ on two arbitrary orbitals $\{1,2\}$ within this operator are ordered (up to spin and orbital relabeling) as: $1\!\!\uparrow$, $2\!\!\downarrow$, $1\!\!\downarrow$, $2\!\!\uparrow$. 
The shaded area in Eq.~\ref{eq:Jterm} indicates that additional compression is possible within contained gates. Note that whilst for hopping terms it was more efficient to combine Hermitian conjugates, in the case of the quartic terms we first combine operators between the two groups (pair hopping and spin exchange) as these differ by exactly two Paulis acting on the first and last qubit, which leads to maximal cancellations. Beyond the HK model, the density-assisted hoppings operators \cite{Jiang2023} of the form $(c^\dagger_{i,\downarrow}c_{j,\downarrow}+c^\dagger_{j,\downarrow}c_{i,\downarrow})n_{i,\uparrow}$ can also be straightforwardly implemented with the presented techniques.

We can also consider the computational cost of implementing an instance of the fSWAP operator within our mappings. From Eq.~\ref{eq:fswap} it is evident that the first two exponential terms correspond to single-qubit gates and the last has similar structure to a hopping term together with its Hermitian conjugate (see Eq.~\ref{eq:quadraticsumhc}). This term can be implemented especially efficiently, in depth one, if we swap modes between vertices connected by a two-qubit edge (meaning two consecutive P qubits). If we explicitly use $V_i=Z_i$, $V_j=Z_j$ and $E_{ij}=X_iY_j$ we
obtain the following circuit by just applying Eq.~\ref{eq:fswap}:
\begin{equation}\label{eq:fswapexpression}
    \raisebox{-.42\height}{\includegraphics[scale=0.08]{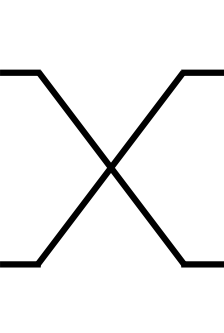}}=\raisebox{-.42\height}{\includegraphics[scale=0.08]{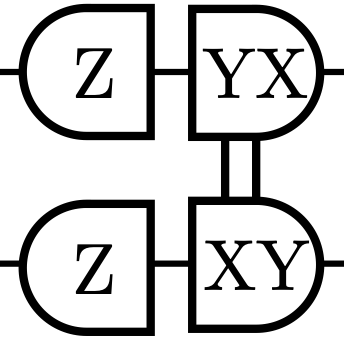}}=\raisebox{-.30\height}{\includegraphics[scale=0.12]{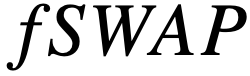}}
\end{equation}
Note that fSWAPs are reversible and sometimes this can be useful when combining it with other operators within a specific ordering. Fermionic swap operators can also be implemented between modes connected through ancilla qubits, but at a higher cost and we thus refrain from using them in this work. Nonetheless, it can be noted that through Eq.~\ref{eq:hopwithint}, they can be optimally combined with a pair of hopping operators together with the density-density operator by shifting the angle of the hoppings by $\pi/4$. This extends the work in \cite{Kivlichan2018}, where they used the fSIM gate for executing together these three types of operators between neighboring modes on JW chain.

fSWAPs are especially practical for models containing multiple modes per fermionic lattice site. They allow us to keep the same general unit cell structure of our mappings derived for one mode per site by replacing every physical qubit P by a chain of physical qubits corresponding to all modes defined on the given lattice site (this approach has been first introduced in Ref.\cite{Clinton2022}). This chain will locally have the properties of a JW mapping on each individual site and all modes except for those at the ends will only be connected by horizontal edges. These modes can then be transported through the chain using a fermionic SWAP network, which due to the JW nature of the chain leads to much lower circuit depths compared to applying the composite rule. 

As an example, consider the Fermi-Hubbard model with two spin-modes per site. As explained, we arrange the modes next to each other horizontally ($\textit{P}_{\uparrow}$,$\textit{P}_{\downarrow}$) and connect them through a two-qubit edge operator $E_{\uparrow \downarrow}=X_{\uparrow}X_{\downarrow}$ and the vertex operators of both modes remain single-qubit Z's. In cases where one has more than one mode per lattice site it is necessary to implement an fSWAP network, which alternates fSWAP operations between even ($2k$, $2k+1$) and odd ($2k+1$, $2k+2$) pairs of modes associated to the site, as done for normal qubit swap networks \cite{Algaba2022}. This ensures that every mode travels to both ends of the chain and then interacts with similar modes of neighboring lattice sites after applying only linear number of swap layers. 

In order to additionally ensure that all intra-site Hamiltonian terms from Eq.~\ref{eq:hkham} can be implemented at some point in the network, the relative position of modes is equally important. The described swap network only ensures that every pair of modes is adjacent at some point, but this requirement cannot be satisfied for every group of four modes. Luckily, the quartic spin-exchange and pair-hopping terms always act on two pairs of spin-modes within two orbitals. With this restriction it is indeed possible to make all such groups adjacent within the swap network if we ensure that pairs of spins of an orbital always move in the same direction. If we consider the example of four orbitals, then this can be ensured by arranging the modes in the pattern: $1\!\!\uparrow$, $2\!\!\downarrow$, $1\!\!\downarrow$, $2\!\!\uparrow$, $3\!\!\uparrow$, $4\!\!\downarrow$, $3\!\!\downarrow$, $4\!\!\uparrow$. It is easy to see that  any two spin modes of an orbital will at every step of the swap network be at most one qubit apart. The reverse ordering of the spins between right-moving (1,3) and left-moving (2,4) pairs further guarantees that all $J$-terms can be implemented in depth five, as shown using Eq.~\ref{eq:Jterm}.

Finally, we would like to point out that the edge operators in the presented mappings generate a group of stabilizers $S_p$ with weights ranging from 6 to 8, depending on the particular choice of mapping. These, similarly to the mappings in Ref.~\cite{Derby2020}, allow for partial single-qubit error detection, but not for error correction.

\section{Simulating fermionic systems}
\label{sec:fermionicsimulation}
Many different algorithms for studying the properties of fermionic models on quantum computers can be found in literature. The most straightforward example consists of time-evolving a system under a Hamiltonian of the form $H = \textstyle{\sum_{j} h_j}$ using the Suzuki-Trotter decomposition formula~\cite{lloyd1996}:
\begin{align} \label{eq:trotter}
    e^{- i H t} = \lim_{\delta_t \to 0} \prod_k^{t/\delta_t} \prod_j e^{-i h_j \delta_t}
\end{align}
where $t/\delta_t \in \mathbb{Z}$. For any non-zero value of $\delta_t$ this relation becomes approximate and the quality of the result will depend on the smallness of $\delta_t$. Following this approach, there have been proposed additional techniques involving more sophisticated summation formulae \cite{suzuki1976,Nathan2012,Somma2014,Childs2019,Campbell2019}. Many other quantum algorithms developed for fermionic models also require implementing sequential evolution operators \cite{Kitaev1995,Peruzzo2014,Wecker2015,yuan2019theory,Grimsley2019,Somma2019,Bharti2021iqae,bharti2021quantum}, typically equivalent to a single Trotter step or some of it's parts. 

In this section we aim investigate the performance of our four fermion-to-qubit mappings (PPA, PA, PPAA, PAA) used together with the XYZ-decomposition formalism, and evaluated by the depth of the quantum circuits obtained for a single Trotter step ($k=1$ in Eq.~\ref{eq:trotter}). When counting gates (and depth) we consider single-qubit gates as a free resource and fSIM two-qubit gates as native. We study three different system classes with varying complexity: the tight-binding model (TB), the single-band Fermi-Hubbard model (FH) and the multi-orbital Hubbard-Kanamori model (HK). In the case of TB and FH models we consider all eight lattices shown in Fig.~\ref{fig:lattices}. For geometries which are not symmetric with respect to 90 degree rotations, such as the honeycomb and Kagome lattices, we consider both orientations. Note that the Kagome fermionic lattice can also be embedded into a square qubit layout without third-neighbor connections but at the cost of including idling qubits. We therefore consider both options for its embedding and only report the minimal depth found. To obtain a better idea of the performance of our fermion-to-qubit mappings, we benchmark them against the state-of-the-art mapping from Derby and Klassen (DK) \cite{Derby2020} and its later modification for multi-orbital models \cite{Clinton2022}. The details for this mapping and details on how one can embed it into a square qubit layout can be found in Appendix~\ref{appendix:embeddingDK}.

\subsection{Tight-binding model}
\label{subsec:spinlessmodels}

Let us start by considering the TB model. In Fig.~\ref{fig:Picture1} we show a worked-out example for this model on an NNN square lattice using the PA mapping. We investigate a 3-by-3 fermionic NNN square lattice with a single mode per lattice site and open boundary conditions processed using the PA mapping. The total circuit depth, however, is independent of system size. Fig.~\ref{fig:Picture1}.a) shows the fermionic lattice sites together with all existing connections between them. Using the PA mapping we can port this lattice onto a square 18-qubit layout (6-by-3, see Fig.~\ref{fig:Picture1}.b)). We then proceed by defining all vertices, horizontal and vertical edges in Fig.~\ref{fig:Picture1}.c) and, using Eq.~\ref{eq:quadraticsumhc} the operators corresponding to hopping terms of the NNN square lattice combined with their Hermitian conjugates (Fig.~\ref{fig:Picture1}.d)). Diagonal hopping operators can be constructed using the composite rule from a vertex and two edges (one horizontal and one vertical). In the PA mapping, all diagonal hopping terms act on four qubits. For a particular quantum algorithm one can now efficiently decompose these qubit operators using the formalism which we introduced in Sec.~\ref{sec:xyzdecomposition}. The resulting circuit of one step in the Suzuki-Trotter decomposition of the time-evolution for the TB Hamiltonian on this lattice is shown in Fig.~\ref{fig:Picture1}.e).

As previously discussed, we find that the strategy generating the shallowest circuits is to first combine the operator pairs generated from mapping the sum of Hermitian conjugate hopping terms, in the spirit of Eq.~\ref{eq:exhc}, which immediately reduces the total depth by a factor two. In Fig.~\ref{fig:Picture1}.e) we have also rotated (with single-qubit gates) all operators to be of the form $\sigma Z\dots Z \bar{\sigma}$, guaranteeing that all two-qubit operators are of the form $e^{i\alpha (XY+YX)}$, which in this case would suffice as a native gate. Parallel operators acting on distinct sets of qubits can then be collected into layers and applied simultaneously.

The order in which these layers are applied will also affect the circuit depth, as additional compression can be achieved between the layers. In Fig.~\ref{fig:Picture1}.e) we have not performed such compression between layers, but rather indicated, by shaded areas, where this is possible. The two cases allowing additional compression that occur in Fig.~\ref{fig:Picture1}.e) are of one of the two forms (see Appendix \ref{appendix:xyz_derivations} for a detailed derivation):
\begin{equation}\label{eq:xyzdecomp7}
\begin{matrix}
    \raisebox{-.42\height}{\includegraphics[scale=0.08]{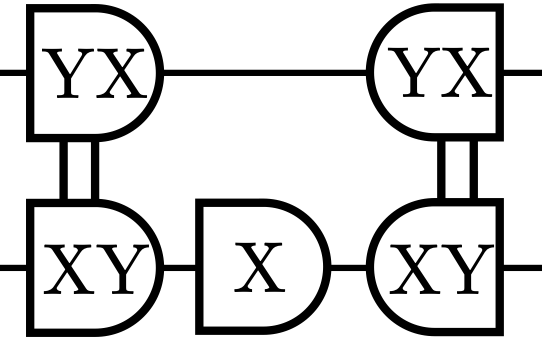}} =
    \raisebox{-.42\height}{\includegraphics[scale=0.08]{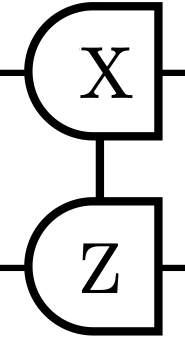}}, \\ \\
    \raisebox{-.42\height}{\includegraphics[scale=0.08]{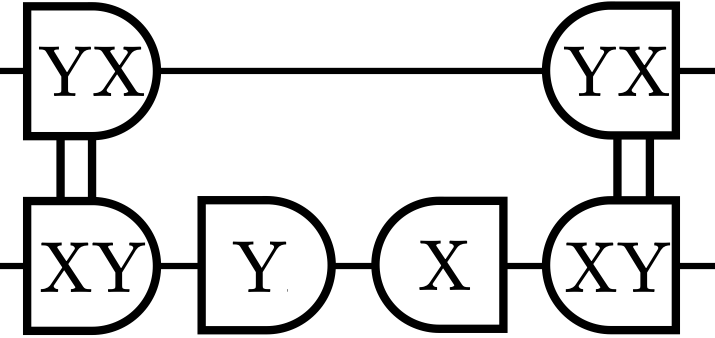}} =
    \raisebox{-.42\height}{\includegraphics[scale=0.08]{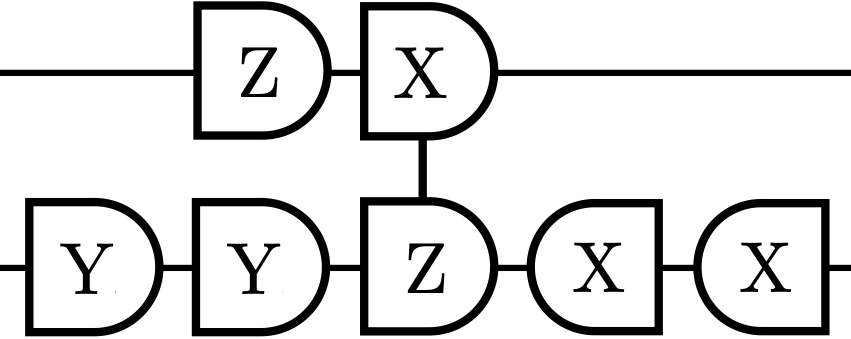}} 
\end{matrix}
\end{equation}
In such cases one will, through the application of Eq.~\ref{eq:xyzdecomp8}, produce two-qubit gates of the form $e^{i\alpha ZZ}$, which same as the $e^{i\alpha (XY+YX)}$ gate can be implemented using a single fSIM gate.

\begin{table*}
\centering
\begin{tabular}{|c|ccccc|}
\hline
Lattice type & PPA & PA & PPAA & PAA & DK \\ \hline
Honeycomb    & 8   & 7    & \textbf{6}    & 8 & 8  \\
Square    & 12  & 9    & \textbf{8}    & 10 & 10 \\
Shastry-Sutherland    & 19  & \textbf{12}   & 16   & 16 & 13 \\
Kagome   & 14  & 14   & \textbf{13}   & \textbf{13} & 17 \\
Triangular    & 26  & \textbf{14}   & 26   & 19 & 23 \\
Checkerboard    & 25  & \textbf{15}   & 24   & 22 & \textbf{15} \\
Tetrakis    & 23  & \textbf{18}   & 20   & 23 & 22 \\
NNN square    & 33  & \textbf{18}   & 31   & 25 & 29 \\ \hline
\hline
 $r$ & 1.5 & 2 & 2 & 3 & 1.5 \\ \hline
\end{tabular}
\caption{Single Trotter step circuit depths for the tight-binding model on different lattice geometries, counted in native two-qubit gates (fSIM), is shown for multiple fermion-to-qubit mappings as well as their fermion-to-qubit ratios $r$. Circuits were obtained using the XYZ formalism. Lowest depths per geometry are indicated in bold.}\label{table:spinless}
\end{table*}

For circuits which consist of multiple Trotter steps one can compress even further between steps. The most straightforward way is by inverting the order of the operators, which is equivalent to the second-order Suzuki-Trotter formula. This allows to merge two parallel layers of hopping operators on the boundary between two Trotter steps (i.e. $e^{-i h_1 \delta_t} e^{-i h_2 \delta_t} e^{-i h_2 \delta_t} e^{-i h_1 \delta_t} = e^{-i h_1 \delta_t} e^{-i 2 h_2 \delta_t} e^{-i h_1 \delta_t}$) and save one such layer for each additional Trotter step. In this case it can pay off to strategically position the most computationally expensive layer at the boundary. 

Our results for the TB model are summarized in Table~\ref{table:spinless} and the prescriptions for obtaining the circuits can be found in Appendix~\ref{app:prescriptions}. We find that a single Trotter step can be simulated with circuit depths ranging from 6 for the honeycomb lattice to 17 for the NNN square lattice which contains all possible next-nearest neighbor hopping terms. For lattices without NNN connectivity (honeycomb and square) we found the PPAA mapping to perform fine, whilst for all the lattices the PA mapping proved optimal. Surprisingly, the state-of-the-art DK mapping performed worse for all geometries we have considered, despite generating lower-weight qubit operators for the vertical hopping operators (see a comparison of operator weights between mappings in Table.~\ref{table:neighbors} of Appendix~\ref{appendix:embeddingDK}). here, for every further Trotter step, one can reduce the circuits by an additional two (honeycomb) to three (NNN square) depth. 

One might be tempted to argue that for small system sizes, the JW mapping performs better than the results we obtained for our local mappings. However, in Appendix~\ref{appendix:jw} we show that even for a 3-by-3 lattice, arguably the smallest truly two-dimensional case, the JW mapping already performs worse than the local mappings considered here.

By itself, the tight-binding Hamiltonian does not present a challenge to classical methods. Other than for benchmarking purposes, it is however also interesting with respect to mapping the Fermi-Hubbard model. Namely, if one would allow for a quantum processor with two superimposed square qubit lattices, one for each spin, all of the hoppings in the FH model could be applied in parallel. Then, one only needs to treat the on-site interaction term between pairs of spins on the same sites which, within this qubit layout, could be immediately implemented with one layer of two-qubit gates (this approach was previously used in Ref.~\cite{Clinton2021}). This means that such circuits for the FH model are always only one deeper than the results from Table~\ref{table:spinless}. We found these circuit depths to always be slightly superior to using mappings involving only one square qubit layout, which will be discussed in the next section. On the other hand, such qubit topologies are rather challenging to be implemented in most quantum computing platforms (with trapped-ion QPUs with all-to-all connectivity being the notable exception).

\subsection{Fermi-Hubbard model}
\label{subsec:spinfulmodels}

Let us now consider how to map the single-band FH model with two spin-types per lattice site to a square qubit layout. For the case of two spin modes per lattice site, we can split all physical qubits $\textit{P}\rightarrow \textit{P}_{\uparrow}\textit{P}_{\downarrow}$. This way, the PA pattern becomes  $\textit{P}_{\downarrow}\textit{P}_{\uparrow}A$ and PAA becomes ${P}_{\uparrow}\textit{P}_{\downarrow}AA$ (we will keep using PA and PAA nomenclature for simplicity). Further, we keep the ordering for vertically neighboring lattice sites the same, but switch it between horizontally neighboring ones, so that a site with pattern $\textit{P}_{\uparrow}\textit{P}_{\downarrow}$ is followed by a site with the pattern $\textit{P}_{\downarrow}\textit{P}_{\uparrow}$. 
This allows for similar modes to interact across neighboring lattice sites. For this model, we will not consider the PPA and PPAA mappings, as we found they generally perform worse  in comparison to PA and PAA due to their lack of vertical connectivity.

The general strategy for all mappings involving multiple modes per lattice site is to devise a fermionic swap network within each lattice site individually and perform all applicable inter-site interactions at each step of the network much as was shown for the tight-binding Hamiltonian. In the case of two modes per lattice site this network consists of a single fSWAP operation. 

\begin{figure}
\centering
  {\includegraphics[width=0.9\linewidth]{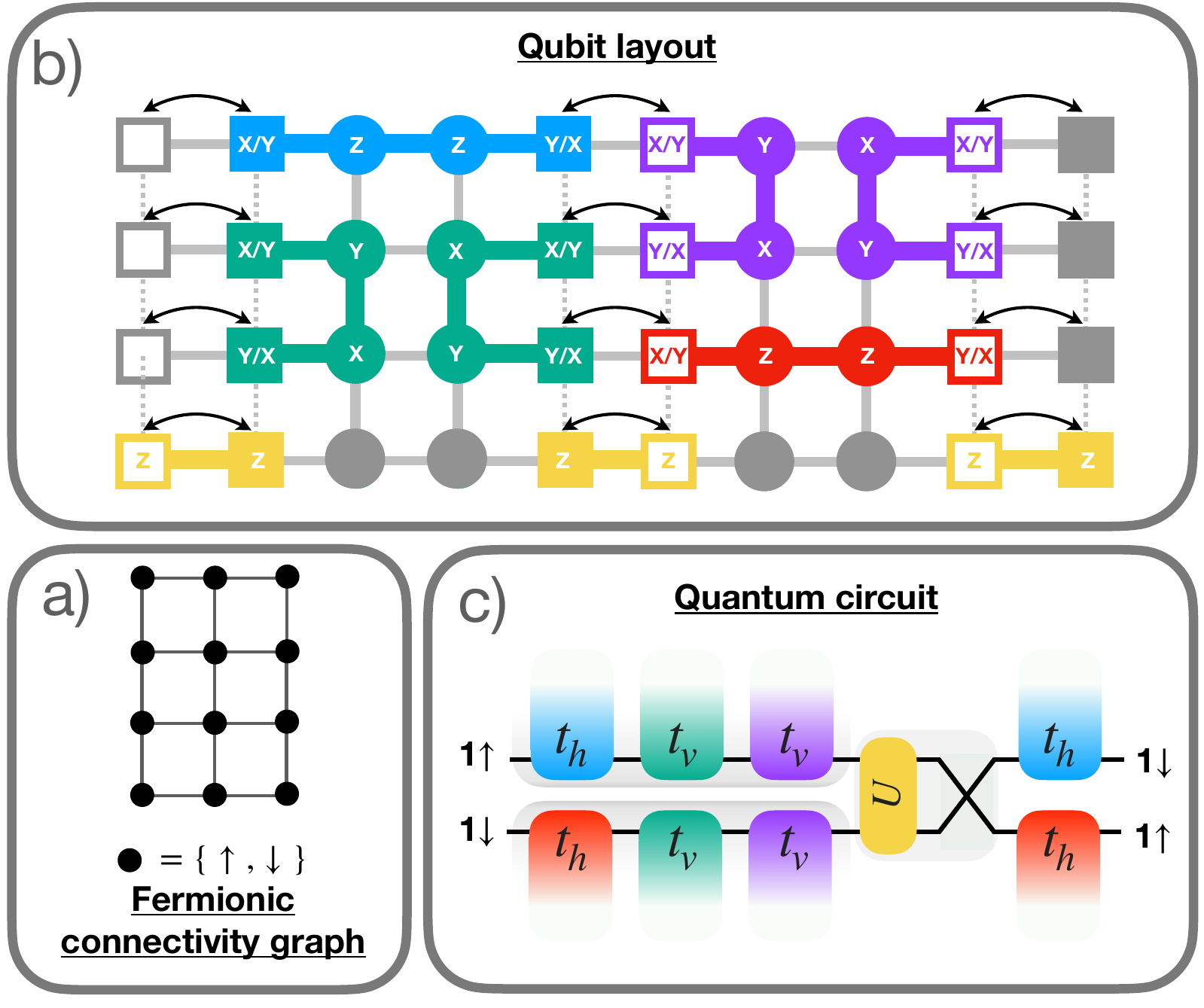}}
\hfill
\caption
{a) Fermionic connectivity graph for the Fermi-Hubbard model on a 3-by-4 square lattice. b) Hopping operators (blue, purple, green and red) and the density-density operator (yellow) for the model using the PAA mapping where circles are ancilla qubits and filled (empty) squares represent spin up (down) physical qubits. The X/Y and Y/X symbolize two distinct terms. Black arrows indicate qubits between which fSWAPs operations are required. c) quantum circuit for a single Trotter step of the model, restricted to a single orbital with two spin modes ($1\!\!\uparrow$, $1\!\!\downarrow$). Color scheme corresponds to a). Further circuit compression can be applied within grey shaded areas of the circuit. Dashed qubit connections are not required. The total circuit depth is 11.
  \label{fig:spinful_fig}
}
\end{figure}

Let us consider the example of the FH model on a NN square lattice and using the PAA mapping, as shown in Fig.~\ref{fig:spinful_fig}. In a) we see the qubit layout containing one spin up and down physical qubit per lattice site. In b) we show the correspond qubit layout together with examples of hopping and density-density interaction operators. Spin-up modes correspond to full squares and spin-down modes to empty squares. In c) we show the full circuit from the point of view of a single fermionic site. Given translational invariance, the circuits for all other lattice sites will be identical. We first apply one horizontal hopping operator pair for each mode and combine it with both of the vertical hoppings pairs in order to reduce their combined depth to seven. Next, we apply the density-density operator, which can be combine with an fSWAP between the two modes into a single fSIM gate. Finally, we can implement the remaining horizontal hopping terms in the second direction  which were not immediately available before we exchanged the two spin modes in depth three. The total circuit depth is therefore 11 per Trotter step. Note, that if only $e^{i\phi (XY+YX)}$ and $e^{i\phi ZZ}$ native gates were available, the depth would only increase by one, due to the combined density-density and fSWAP operators. Generally, this is the highest additional gain from using native fSIM gates instead of $e^{i\phi (XY+YX)}$ and $e^{i\phi ZZ}$ that can be obtained for this model.

In Table~\ref{table:spinful} we provide the single Trotter layer depths for the FH Hamiltonian on various lattices and for the PA and PAA mappings, as well as the DK mapping serving as a benchmark. The prescriptions for obtaining these circuit depths with the proposed mappings in this paper can be found in Appendix~\ref{app:prescriptions}. We use the XYZ formalism for all mappings and optimize over all possible lattice orientations, decompositions and orderings of operators with the Trotter later. The PAA mapping produces the lowest possible depths for any lattice considered, the PA mapping is slightly worse throughout and, surprisingly, the DK mapping produces clearly inferior results with up to 2.5 times longer circuits. We believe that this is due to the lack of parallelism, which is a consequence of the lower number of ancilla qubits. 
The additional parallelism and cancellations within PAA bring the depths for the FH model always below double that of the TB model, which one would naively expect as a lower bound. Remarkably, we find that for the honeycomb and square lattices the resulting depths are 9 and 11, respectively, which is only worse by three compared to the TB model. Additional reductions of up to three can be achieved from cancellations between Trotter layers.

It is common in fermion-to-qubit mapping literature to assume two square layers of qubits (one for each spin) one on top of each other connected one-to-one in the vertical axis as the qubit topology \cite{Clinton2021}. Assumming that topology, we can implement the hoppings for spin up and down at the same time as in our TB model and then the density-density terms in depth one. This implies a depth for the FH model equal to the TB model plus one, i.e. depth $7$ for a honeycomb lattice and $9$ for a square lattice. 

We can compare the latter results with a trivial lower bound for a FHM on a fermionic lattice with a connectivity graph of maximum degree $k$. In that case, the depth of any realization of a Trotter step cannot be less than $k+1$, since it will be limited by the maximally connected site, whose physical qubit has to interact through mutually anticommuting gates with $k$ other physical qubits of the same spin (through hopping operators) and $1$ physical qubit with the opposite spin (through the density-density term). This yields a naive minimal depth of $4$ for the honeycomb lattice and $5$ for the square grid, suggesting that our approach is at least close-to optimal.

\begin{table}
\centering
\begin{tabular}{|c|ccc|}
\hline
Lattice type & PA & PAA & DK \\ \hline
Honeycomb    & 11    & \textbf{9} & 17 \\
Square &  15    & \textbf{11} & 21 \\
Shastry-Sutherland    &  21  & \textbf{19} & 24 \\
Kagome    & 25  & \textbf{22} & 36  \\
Triangular    & 27   & \textbf{22} & 46 \\
Checkerboard    & 27   & \textbf{22} &  30\\
Tetrakis    & 25   & \textbf{23} & 46 \\
NNN square    & 35   & \textbf{30} & 74 \\ \hline \hline
 $r$ & 1.5 & 2 & 1.25\\ \hline
\end{tabular}
\caption{Single Trotter step circuit depths for the Fermi-Hubbard model on different lattice geometries, counted in native two-qubit gates (fSIM), is shown for multiple fermion-to-qubit mappings as well as their fermion-to-qubit ratios $r$. Circuits were obtained using the XYZ formalism. Lowest depths per geometry are indicated in bold.}\label{table:spinful}
\end{table}

\subsection{Hubbard-Kanamori model}
\label{subsec:multiorbital}

Now, we focus on investigating the circuit depths for the Hubbard-Kanamori model on the square lattice and as a function of the number of orbitals $M$ per lattice site. Compared to the FH model, the HK model has additional density-density terms between different orbitals ($U_1$ and $U_2$) and the Hund's coupling terms related spin-exchange and pair-hopping terms ($J$) that need to be implemented. 

We pursue a similar strategy to the FH model and group all physical modes of a given lattice site into horizontal blocks on which we perform an even-odd fermionic swap network \cite{Kivlichan2018}. The modes on a site are ordered, as described earlier, in a way that ensures that all $J$ terms can be implemented within adjacent qubits and decomposed into depth-five circuits. This order is inverted between horizontally subsequent blocks of modes of neighboring lattice sites. We only consider the PAA mapping for the HK model, as it has unambiguously demonstrated the best performance already at the level of the FH model. We also do not consider the model on any other lattice geometries as we do not expect this to yield additional insights. 

\begin{figure*}
    \centering
    \includegraphics[width=\linewidth]{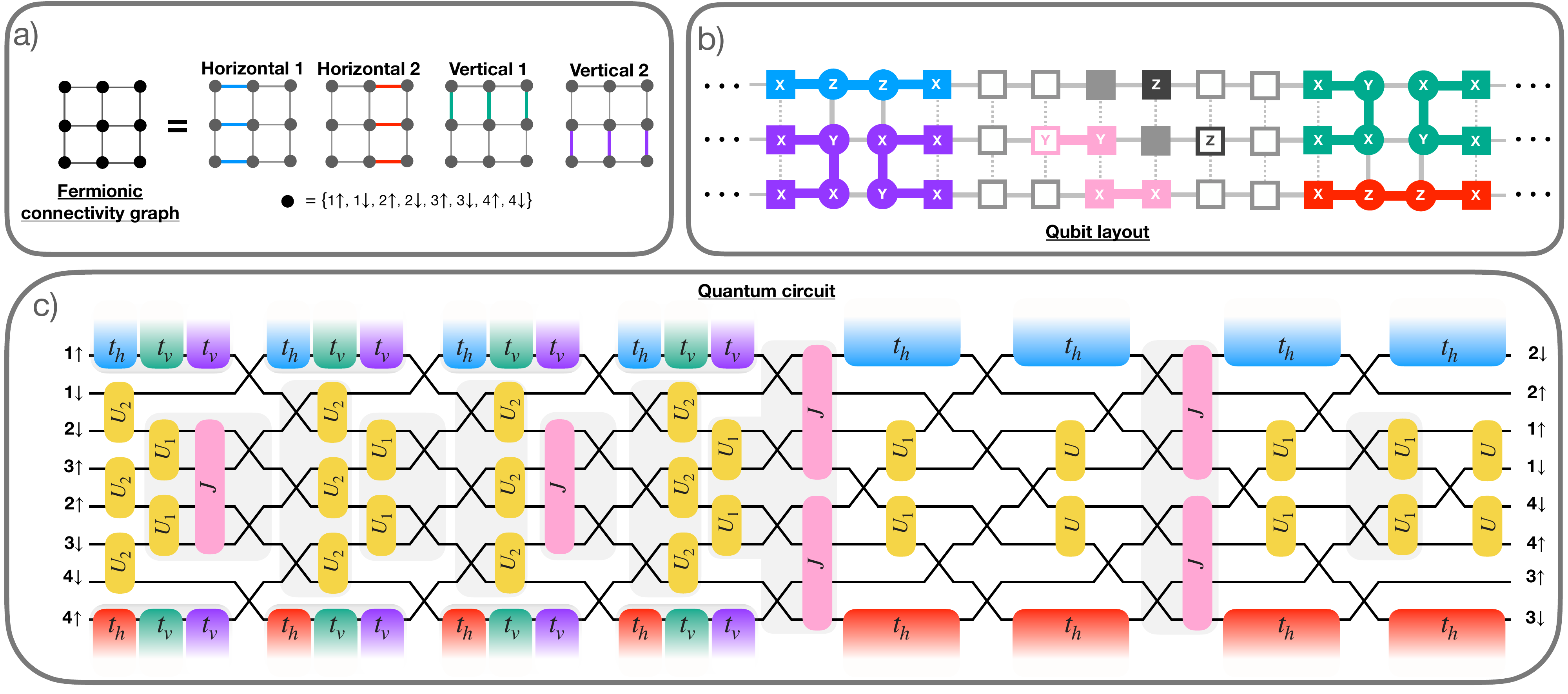} 
    \caption{a) Fermionic connectivity graph for a 3-by-3 square NN lattice with 4 orbitals per lattice site with two spin-modes each. b) Vertex (black) and edge (blue, green, red, purple, pink) operators mapping this fermionic connectivity to a square qubit layout using the PAA mapping. Each lattice site is encoded into a linear chain consisting of four spin-up (full) and four spin-down (empty) physical qubits (squares) connected to the neighboring sites through ancilla qubits (circles). Dashed qubit connections are not required. c) Minimal-depth circuit for the simulation of one Trotter step of the Hubbard-Kanamori model in 55 parallel steps. Size of blocks is not to scale: $t_h$-$t_v$-$t_v$ blocks are depth-seven, $t_h$-terms are depth-three, J terms are depth-five, $U$, $U_1$, $U_2$ terms and fSWAPs can be performed with a single two-qubit gate.
    Further circuit compression can be applied within grey shaded areas of the circuit.
    }\label{fig:hk}
\end{figure*}

In Fig.~\ref{fig:hk} we study the example of the four-band HK model on the square lattice using PAA in more detail (see Fig.~\ref{fig:hk}.a)). In Fig.~\ref{fig:hk}.b) we show a subset of the vertex and edge operators for this mapping. Note that all edges between adjacent modes of the same lattice site are given by either XX or YY (pink). We also differentiate between vertices for the two different spin types (empty vs full squares). In Fig.~\ref{fig:hk}.c) we provide the shallowest, 55-depth, circuit we were able to find for this model. It can be seen that, with the exception of some fSWAP and $J$ layers, the depth is fully governed by hopping operations. Similarly to the FH model we split the hoppings into circuit blocks with depth 7 ($t_h$-$t_v$-$t_v$) and depth 3 ($t_h$). All of the density-density terms, those layers of the fSWAP network and those Hund's terms which don't involve any modes on the external edges of the chain can be effectively parallelized with the hoppings terms. In terms of the swap network, we can omit one fSWAP layer at the start and at the end of the circuit whilst still ensuring that every mode has travelled to both edges. In some cases, we delay fSWAP operations within a layer in order to be able to efficiently implement density-density terms. One can achieve some additional reduction in circuit depth by combining fSWAP operations with Hund's or density-density terms. 

This approach can be generally applied also for the HK model with any other number of orbitals $M$ with only small modifications. In Appendix~\ref{appendix:hk} we show the circuits for $M=2$ and $M=3$, with circuit depths of 27 and 35, respectively. For $M\geq4$ we can provide a general expression for the depth per single Trotter layer, $28\ceil{\frac{M}{2}}-1$. For an even number of orbitals this is obtained by counting $M$ layers of hopping terms with depth of $3+7$, $2M-1$ non-parallelizable layers of fSWAPs of which $\frac{M}{2}$ can be combined with the $J$ terms and $\frac{M}{2}$ non-parallelizable layers of $J$ terms with depth 5. For an odd number of orbitals (with the exception of $M=3$) we did not find any fSWAP network that guarantees that all Hund's terms can be implemented. Our most efficient strategy is to fill up the chain of each lattice site with an additional \emph{dummy} orbital, corresponding to a pair of qubits, and to proceed as in the case of an even number of orbitals (hence the ceiling function in the expression). Similarly to the HK model, we can provide an expression for the depth of a multi-orbital FH model, which is equivalent to removing all $U_1$, $U_2$ and $J$ terms from the circuit. In this case we obtain a depth of $12M-1$ for both even and odd numbers of orbitals $M$.

Some extra terms in a hypothetical Hamiltonian could, in principle, be simulated without any additional resource costs. For example, this includes intra-orbital hoppings, as they can be always be compressed together with fSWAP operations. Another class of terms that may be relatively cheap to add are three-mode terms acting on two orbitals.

\section{Comparison to standard decomposition techniques}
\label{subsec:comparison}

In this section we attempt to disentangle the improvement over state-of-the-art due to our choice of mappings from that coming from the XYZ decomposition and compression techniques. To this end we investigate the circuit depths for four examples: the TB model on an NNN square lattice using the PA mapping, the FH model on an NNN square lattice using PAA and DK mappings, and the HK model on the square lattice using PAA. For these models we attempt to maximaly reduce the circuit depth both using XYZ and standard decomposition techniques (see X-shaped circuit in Eq.~\ref{eq:cnotlad}) and consider either CNOT or the parameterized fSIM gates to be native. Generally, any two-qubit gate can be transformed into at most three CNOTs plus single-qubit gates. However, in the case of  $e^{i\phi Z_1Z_2}$ and $e^{i\phi(X_1Y_2+Y_1X_2)}$, we only need one \cite{nielsen2002quantum} and two CNOTs \cite{Vatan2004}, respectively:
\begin{equation}\label{eq:zzandxytocnot}
    \raisebox{-.40\height}{\includegraphics[scale=0.085]{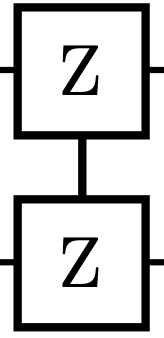}} =
    \raisebox{-.40\height}{\includegraphics[scale=0.085]{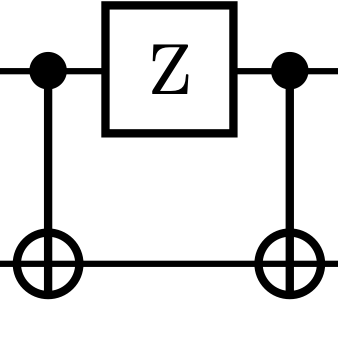}}, \ \
    \raisebox{-.40\height}{\includegraphics[scale=0.085]{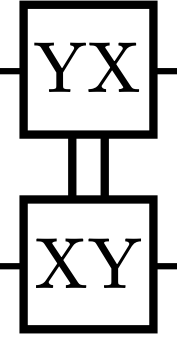}} =
    \raisebox{-.40\height}{\includegraphics[scale=0.085]{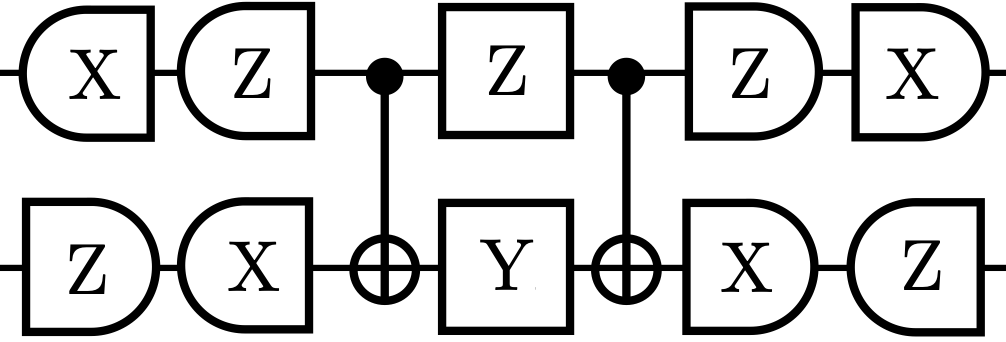}}
\end{equation}
where the central Y-rotation has the opposite angle of the XY+YX rotation. When comparing the XYZ formalism to the standard techniques, it is clear that compression identities also exist for sequences of two-qubit CNOT and single-qubit gates. We give an overview over compression identities we have considered for circuits from the standard decomposition techniques in Appendix~\ref{appendix:app_CNOT}. Equally, we also take into account that it is possible to combine two CNOT gates with a parameterized single-qubit gate between them (as shown in Eq.~\ref{eq:cnotlad}) into a single fSIM gate.

In Table~\ref{table:decompositions} we present the summary of our findings for circuit depth of various configurations. As expected, we find that the combination of XYZ and fSIM gates yields the lowest circuit depths throughout. Further, both the choice of native gate set and the decomposition technique have a strong influence on the result. If we compare the XYZ decomposition with the standard one counted in fSIM gates, we see an improvement of up to a factor $2.3$ in the case of the FH model using PAA. Interestingly, even if we only allow for CNOT gates there is an improvement of up to a factor $1.8$, for the same model and mapping. This holds true for any of the examples in the Table and suggests that using XYZ decomposition generally leads to lower-depth circuits. The number of gates is also reduced to a new minimal bound with respect to previous state-of-the-art when using XYZ-decomposition and DK mapping, since hoppings can be implemented with half of the two-qubit gates needed before.

We can also attempt to quantify the total improvement we have achieved related to the previous state-of-art, which we identify as the combination of using the standard decomposition, parameterized gates and the DK mapping. For the FH model on the NNN square lattice this combination yields a circuit depth of 97 per single Trotter layer, compared to 30 for using XYZ together with the PAA mapping. This corresponds to an improvement of 70 \%. 

\begin{table*}
\begin{center}
\begin{tabular}{|c|c|cccc|}
\hline
\multirow{2}{*}{Decomp.} &
\multirow{2}{*}{\begin{tabular}[c]{@{}c@{}}Native\\ TQGs\end{tabular}} & \multirow{2}{*}{\begin{tabular}[c]{@{}c@{}}TB NNN \, \\ PA\end{tabular}} & \multirow{2}{*}{\begin{tabular}[c]{@{}c@{}}FH NNN \, \\ PAA\end{tabular}} & \multirow{2}{*}{\begin{tabular}[c]{@{}c@{}}FH NNN \, \\ DK\end{tabular}} & 
\multirow{2}{*}{\begin{tabular}[c]{@{}c@{}}HK NN \\ PAA\end{tabular}} \\
 &  &  &  &  &  \\ \hline
XYZ & fSIM & 17 & 30 & 74 & 55 \\
Standard & fSIM & 31 & 68 & 97 & 84 \\
XYZ & CNOT & 30 & 53 & 125 & 100 \\
Standard & CNOT & 47 & 93 & 130 & 132 \\ \hline
\end{tabular}
\caption{Depths of a single Trotter step for different models using either the XYZ or the standard decomposition and expressed in terms of different native two-qubit gates.}\label{table:decompositions}
\end{center}
\end{table*}

\section{Discussion}
\label{sec:discussion}

We have presented an efficient operator decomposition and circuit compression formalism. Whilst generally leading to improvements in circuit depth for the three fermionic systems studied here, it can, in principle, be applied to any other systems of interest, including non-fermionic ones \cite{xyzdecomposition}. We found that the XYZ formalism is more powerful when native fSIM gates are available, however, it can still lead to improvements over standard decomposition techniques when parameterized gates are not available since it can optimally combine the fermionic operators acting on each pair of modes.
We have also introduced a set of local fermion-to-qubit mappings which are restricted to a realistic, square, qubit connectivity graph. For systems with multiple modes per lattice site we have chosen the strategy of using the composite edge rule to implement inter-site operators. At the same time we have placed modes belonging to the same site onto local Jordan-Wigner chains and performed a fermionic swap network to allow for all inter-site operators to be implemented. This idea is much in the spirit of Ref.~\cite{Clinton2022}.

Especially the PAA mapping allows for an increased degree of parallelism which leads to significantly lower circuit depths for scientifically relevant systems compared to previously introduced mappings~\cite{Ball2005, Verstraete2005, Derby2020}. In particular, we found one single Trotter step of the FH model on a honeycomb lattice, the FH model on an NNN square lattice and the three-band HK model on a NN square lattice can be simulated on a square QPU in depths 9, 30 and 35, respectively. These, to the best of our knowledge, correspond to either the first or the lowest number reported for these examples to date. It is important to state that the chosen FH and HK models constitute unsolved actively researched strongly correlated systems in condensed matter physics with actively debated phase diagrams~\cite{Huang2022} and direct relevance to the physics of cuprates, nickelates and transitions-metal oxides. Our formalism could also be applied to ab-initio models from quantum chemistry, especially when the degree of geometric locality allows to neglect transfer integrals below a given threshold. Since the here developed mappings allow for low-weight interactions between far-neighbors in one lattice dimension (see Table~\ref{table:neighbors}), they might be particularly useful for materials with weak couplings in the second and third dimensions \cite{helmholz1947crystal, amsler2023quantum}. Concluding, we believe that the circuit depths reported in this work certainly improve the chances of fermionic simulation algorithms successfully being implemented on future NISQ quantum hardware.

We note that the parallelism of the PAA mapping comes at the expense of introducing additional ancilla qubits which don't necessarily have to be readily available in current processors. The use of techniques meant to reduce the number of qubits by using the symmetries of the Hamiltonian~\cite{Ralli2022} could potentially improve this issue. Equally, if the goal is to optimize for the minimal total gate count, rather than the circuit depth, we expect the DK mapping to yield favourable results compared to the other mappings introduced here. 

We would like to stress that in most cases one single Trotter layer is not sufficient to reliably simulate a system and one would potentially need (polynomially) many such layers instead. This is due to the Trotter error introduced from approximating the limit in Eq.~\ref{eq:trotter} with a finite number of steps. It would be of interest to study the actual necessary number of Trotter steps for these fermionic model in various regimes, for various system sizes, as well as the impact on the Trotter error that the ordering of operators in the decomposition has. It has been shown that this effect can be significant \cite{Tranter2019} and could therefore easily out-weight the minimal additional cancellations achieved by our optimal orderings of operators. On the other hand, it was also shown in Ref.~\cite{Tranter2019} that most orderings produce close to optimal Trotter errors, whereas finding the optimal ordering is a factorially hard problem. 

We note that the optimal circuit depths derived in this work do not take into account state preparation or measurement and we point the interested reader to Refs~\cite{Cleve1997, bharti2021noisy, higgott2021optimal} for more details on these subjects. Our results can, however, potentially also be useful for other quantum algorithms, e.g. the Variational Quantum Eigensolver (VQE) algorithm with a Hamiltonian Variational Ansatz (HVA)~\cite{tilly2022variational}. We have also not considered the effect of errors on our circuits, as has been done in Ref.~\cite{Clinton2021}. The results of such analyses are, however, highly error-model dependent and we leave this topic to future studies.

The structure of the XYZ decomposition can naturally be translated into a QEC framework, as all the external legs with semicircular shapes are Clifford gates, while only the square gate at the center is a non-Clifford gate. The latter can be further decomposed until a single qubit non-Clifford gate is obtained. This can, in turn, be approximated with bounded error by using standard approaches in error correction such as the Ross-Selinger algorithm \cite{ross2016}. When using fault-tolerant quantum algorithms like qubitization \cite{Low2019}, however, fermionic evolution operators may not preserve the required structure which is necessary for taking advantage of the XYZ decomposition.

\section*{Acknowledgments} The authors would like to thank S.~Cheylan, S.~Ikäläinen, F.~R.~Fernandes Pereira, I.~de Vega and K.~Q.~Dog for useful discussions.

\bibliography{Bibliography}
\bibliographystyle{apsrev4-2}

% \newpage
\appendix

\section{CNOT identities}
\label{appendix:app_CNOT}
In order to obtain the circuit depths in  Table~\ref{table:decompositions}, some CNOT identities were used with the aim of compressing the circuit and making a fair comparison of the XYZ formalism against the standard decomposition techniques. Most of these identities have been applied on the boundaries between different fermionic operators. The identities, obtained from Qiskit \cite{Qiskit}, use CNOTs, H as Hadamard gate, S as phase gate and any Pauli rotation $R_\sigma(\alpha)=e^{-i\frac{\alpha}{2}\sigma}$ where $\sigma \in \{X,Y,Z\}$, as well as their Hermitian conjugates:
\begin{equation}\label{eq:qiskit_1}
\begin{matrix}
    \raisebox{-.40\height}{\includegraphics[scale=0.400]{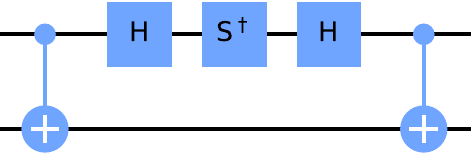}} =
    \raisebox{-.45\height}{\includegraphics[scale=0.400]{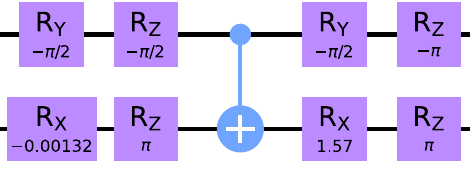}} , \\ \\
    \raisebox{-.40\height}{\includegraphics[scale=0.400]{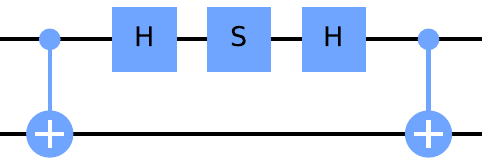}} =
    \raisebox{-.43\height}{\includegraphics[scale=0.400]{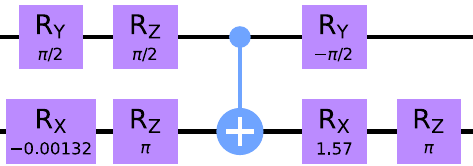}} 
\end{matrix}
\end{equation}
\begin{equation}\label{eq:qiskit_2}
\begin{matrix}
    \raisebox{-.40\height}{\includegraphics[scale=0.400]{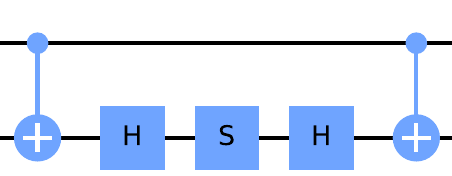}} =
    \raisebox{-.47\height}{\includegraphics[scale=0.400]{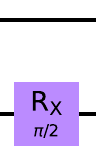}} ,  \\ \\
    \raisebox{-.40\height}{\includegraphics[scale=0.400]{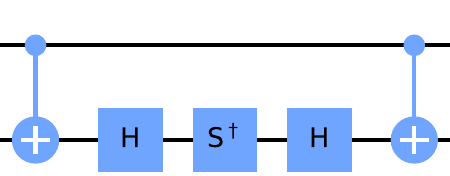}} =
    \raisebox{-.45\height}{\includegraphics[scale=0.400]{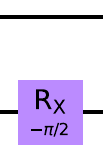}} 
\end{matrix}
\end{equation}
\begin{equation}\label{eq:qiskit_3}
\begin{matrix}
    \raisebox{-.40\height}{\includegraphics[scale=0.400]{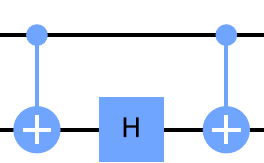}} =
    \raisebox{-.40\height}{\includegraphics[scale=0.400]{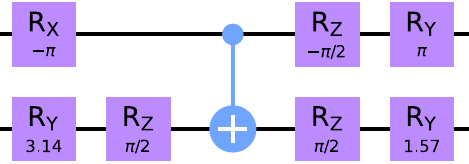}}
\end{matrix}
\end{equation}
\begin{equation}\label{eq:qiskit_4}
\begin{matrix}
    \raisebox{-.40\height}{\includegraphics[scale=0.400]{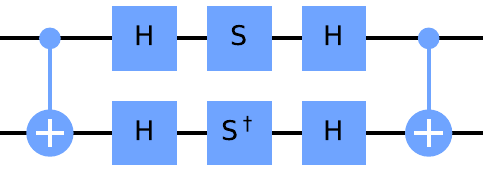}} =
    \raisebox{-.40\height}{\includegraphics[scale=0.400]{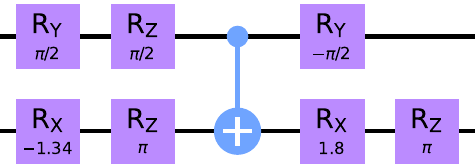}} ,\\ \\
    \raisebox{-.40\height}{\includegraphics[scale=0.400]{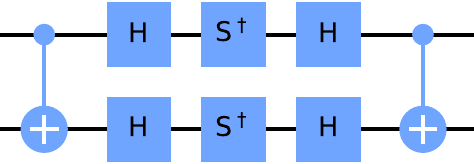}} =
    \raisebox{-.40\height}{\includegraphics[scale=0.400]{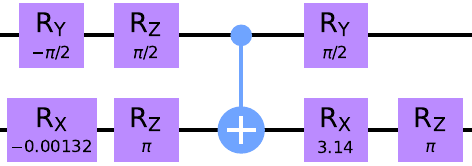}}
\end{matrix}
\end{equation}

We note that, in principle, CNOT (and XYZ) decompositions exist that yield a logarithmic depth at the cost of having an all-to-all connectivity structure \cite{Clinton2022}. However, they would not improve the results obtained in this paper, since for operators with weight lower than 6 using a binary tree does not make ant difference compared to using the standard approach. Hence, higher connectivities will not give extra advantage for the cases considered using our mappings.

\section{XYZ derivations}
\label{appendix:xyz_derivations}

Proofs for some of the equations from the main text are provided in this section. For Eq.~\ref{eq:sqgcompression}, we use Eq.~\ref{eq:xyzdecomp} to obtain:
\begin{align}
    e^{i\alpha \sigma^a_1 \rho_2}e^{\pm i\frac{\pi}{4}\sigma^b_1}&=e^{\pm i\frac{\pi}{4}\sigma^b_1}e^{i\alpha \bar{\sigma}_1 \rho_2}e^{\mp i\frac{\pi}{4}\sigma^b_1}e^{\pm i\frac{\pi}{4}\sigma^b_1}=\nonumber\\&=e^{\pm i\frac{\pi}{4}\sigma^b_1}e^{i\alpha \bar{\sigma}_1 \rho_2}
\end{align}
where $\bar{\sigma}_1=\pm \frac{i}{2}[\sigma^a_1,\sigma^b_1]$. However, if $\sigma^a_1=\sigma^b_1$, Eq.~\ref{eq:xyzdecomp} does not hold anymore. In order to extend its validity to this case an extra term has to be introduced $\bar{\sigma}_1=\pm \frac{i}{2}[\sigma^a_1,\sigma^b_1]+\delta_{ab}\sigma^a_1$, which also treats the aforementioned exception.

In order to obtain Eq.~\ref{eq:xyzdecomp8} we equally apply Eq.~\ref{eq:xyzdecomp}:
\begin{align}
    e^{i\alpha \sigma^a_1\rho_2}e^{i\frac{\pi}{4}\sigma^b_1\rho_2} = &e^{i\frac{\pi}{4}\sigma^b_1\rho_2}e^{i\alpha \bar{\sigma}_1}e^{-i\frac{\pi}{4}\sigma^b_1\rho_2}e^{i\frac{\pi}{4}\sigma^b_1\rho_2} = \nonumber \\& = e^{i\frac{\pi}{4}\sigma^b_1\rho_2}e^{i\alpha \bar{\sigma}_1}
\end{align}
where $\sigma^a_1= \frac{i}{2}[\bar{\sigma}_1,\sigma^b_1]$. For Eq.~\ref{eq:xyzdecomp7}.a:
\begin{align}
    &e^{-i\frac{\pi}{4}(X_1Y_2+Y_1X_2)}e^{-i\frac{\pi}{4}X_2} e^{i\frac{\pi}{4}(X_1Y_2+Y_1X_2)}=\nonumber\\&= e^{-i\frac{\pi}{4}Y_1X_2}e^{-i\frac{\pi}{4}X_1Y_2}e^{-i\frac{\pi}{4}X_2} e^{i\frac{\pi}{4}X_1Y_2}e^{i\frac{\pi}{4}Y_1X_2}=\nonumber\\&= e^{-i\frac{\pi}{4}Y_1X_2}e^{i\frac{\pi}{4}X_1Z_2}e^{i\frac{\pi}{4}Y_1X_2}=e^{i\frac{\pi}{4}X_1Z_2}
\end{align}

While for Eq.~\ref{eq:xyzdecomp7}.b:

\begin{align}
    &e^{-i\frac{\pi}{4}(X_1Y_2+Y_1X_2)}e^{-i\frac{\pi}{4}Y_2}e^{i\frac{\pi}{4}X_2} e^{i\frac{\pi}{4}(X_1Y_2+Y_1X_2)}\nonumber=\\&=e^{-i\frac{\pi}{4}Y_1X_2}e^{-i\frac{\pi}{4}Y_2}e^{i\frac{\pi}{4}Y_1X_2}e^{i\frac{\pi}{4}X_2}\nonumber\\& e^{-i\frac{\pi}{4}X_1Y_2}e^{i\frac{\pi}{4}X_2}e^{i\frac{\pi}{4}X_1Y_2}=e^{-i\frac{\pi}{4}Y_1Z_2}e^{-i\frac{\pi}{4}X_1Z_2}\nonumber=\\&=e^{-i\frac{\pi}{4}X_1Z_2}e^{i\frac{\pi}{4}Z_1}e^{i\frac{\pi}{4}X_1Z_2}e^{-i\frac{\pi}{4}X_1Z_2}=\nonumber\\&=e^{-i\frac{\pi}{4}X_1Z_2}e^{i\frac{\pi}{4}Z_1}
\end{align}

\section{Embedding the DK mapping into a square qubit layout}
\label{appendix:embeddingDK}
For the sake of completeness, we give a brief overview on the DK mapping proposed in Ref.~\cite{Derby2020}. It is a local fermion-to-qubit mapping constructed through the definition of edge and vertex operators, much like the mappings presented in the main text. It has a fermion-to-qubit ratio of $r=1.5$ and weight operators $w=3$.

\begin{figure}
  \centering
  \begin{minipage}[c]{.49\linewidth}
    \subfloat
      {\includegraphics[width=\linewidth]{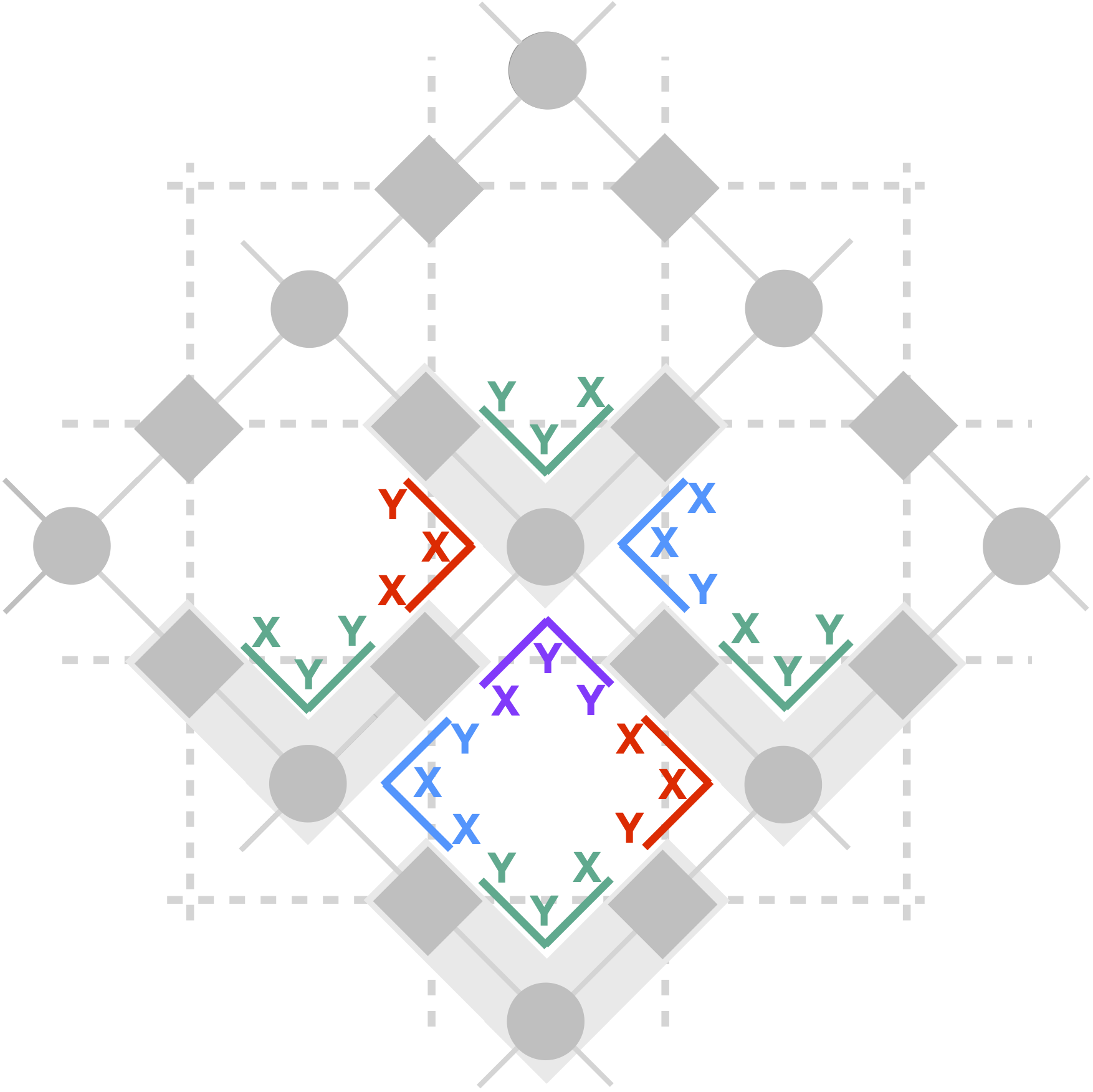}}
  \end{minipage}
  \hfill
  \begin{minipage}[c]{.49\linewidth}
    \subfloat
      {\includegraphics[width=\linewidth]{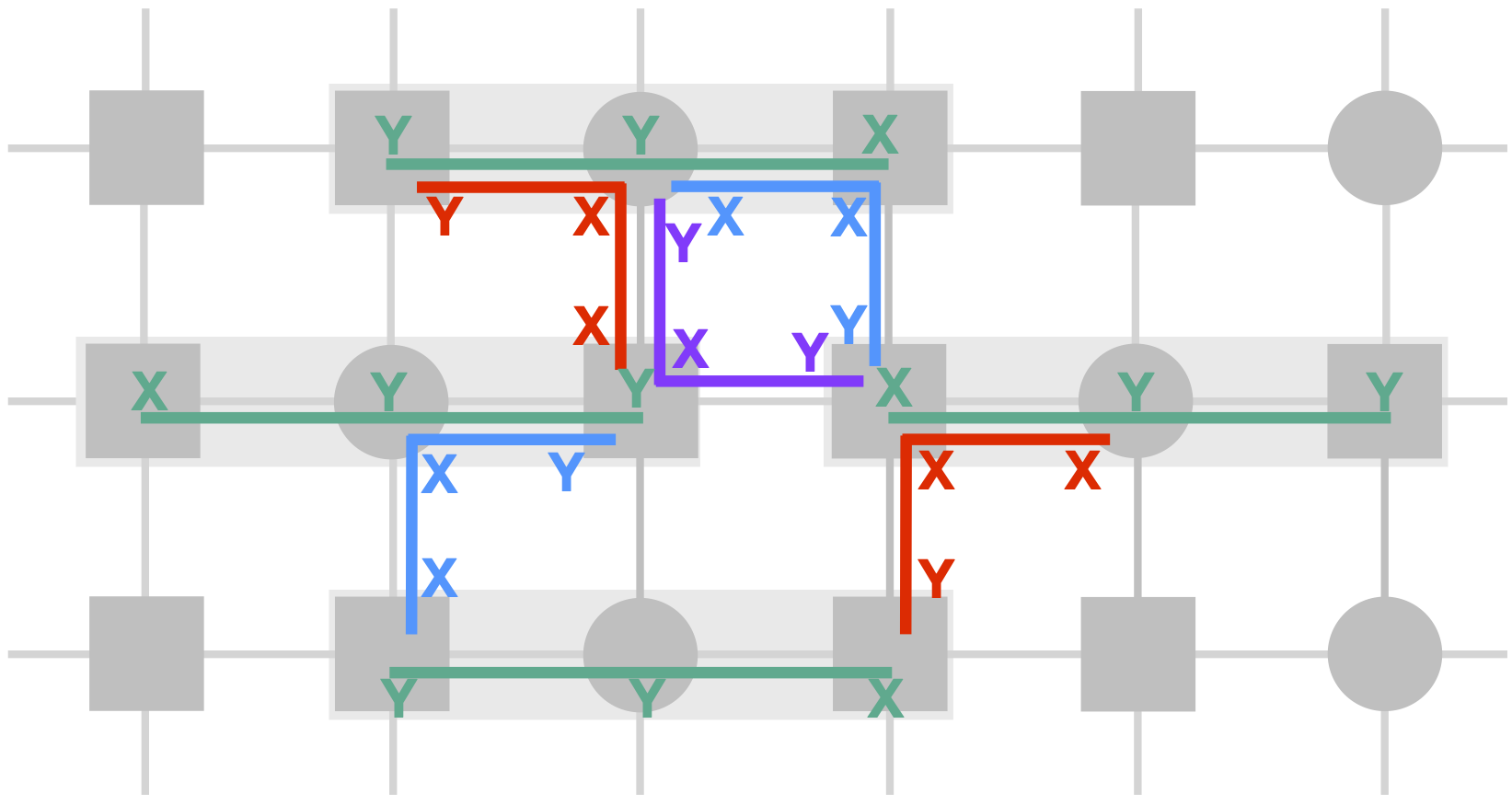}}
  \end{minipage}
  \caption
    {Edge operators for a) the original DK mapping \cite{Derby2020} with one spin and for b) the square grid-embedded version. Squares represent physical qubits and circles are ancilla qubits. The dashed lines show the edge operators connectivity while the solid lines are qubit connections.
      \label{fig:spinlessembeddingDK}
    }
\end{figure}

The original topology can be understood as a square grid of physical qubits with ancilla qubits placed on the center of alternating faces of the lattice, connected to the four closest qubits. See the dashed lattice from Fig.~\ref{fig:spinlessembeddingDK}.a), in which some faces have no ancilla qubits in the middle. Each edge $(i,j)$ on a face with no ancilla qubits is assigned an orientation following a clockwise or counterclockwise direction over the face edges, alternating on every row of faces. The vertex and edge operators are:
\begin{equation}
    V_i \equiv Z_i
\end{equation}
{\small{
\begin{equation}
E_{i,j} \equiv \left\{\begin{array}{cl}X_i Y_j X_{f(i,j)} &   \mathrm{if}\ (i,j)\ \textrm{is oriented downwards} \\-X_iY_jX_{f(i,j)} &    \mathrm{if}\ (i,j)\ \textrm{is oriented upwards} \\X_iY_j Y_{f(i,j)} &   \mathrm{if}\ (i,j)\ \textrm{is horizontal}\end{array}\right.
\end{equation}}}
where $i,j$ are physical qubits and $f$ is an ancilla qubit on the face corresponding to (i,j). This mapping yields weight-3 hopping operators, weight-2 density-density operators and weight-8 stabilizers.

\begin{figure}
  \centering
  \begin{minipage}[c]{.59\linewidth}
    \subfloat
      {\includegraphics[width=\linewidth]{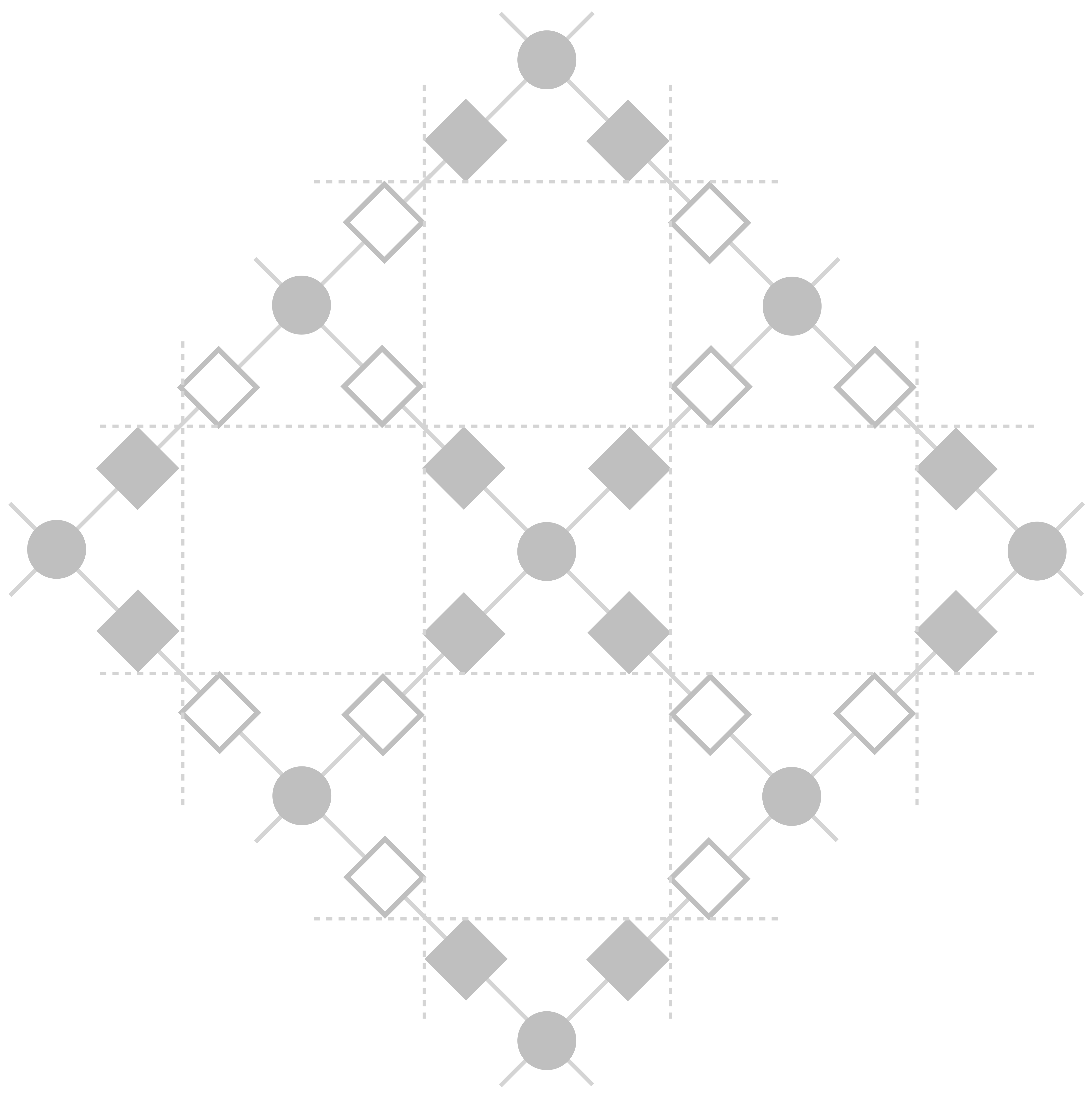}}
  \end{minipage}
  \hfill
  \begin{minipage}[c]{.39\linewidth}
    \subfloat
      {\includegraphics[width=.9\linewidth]{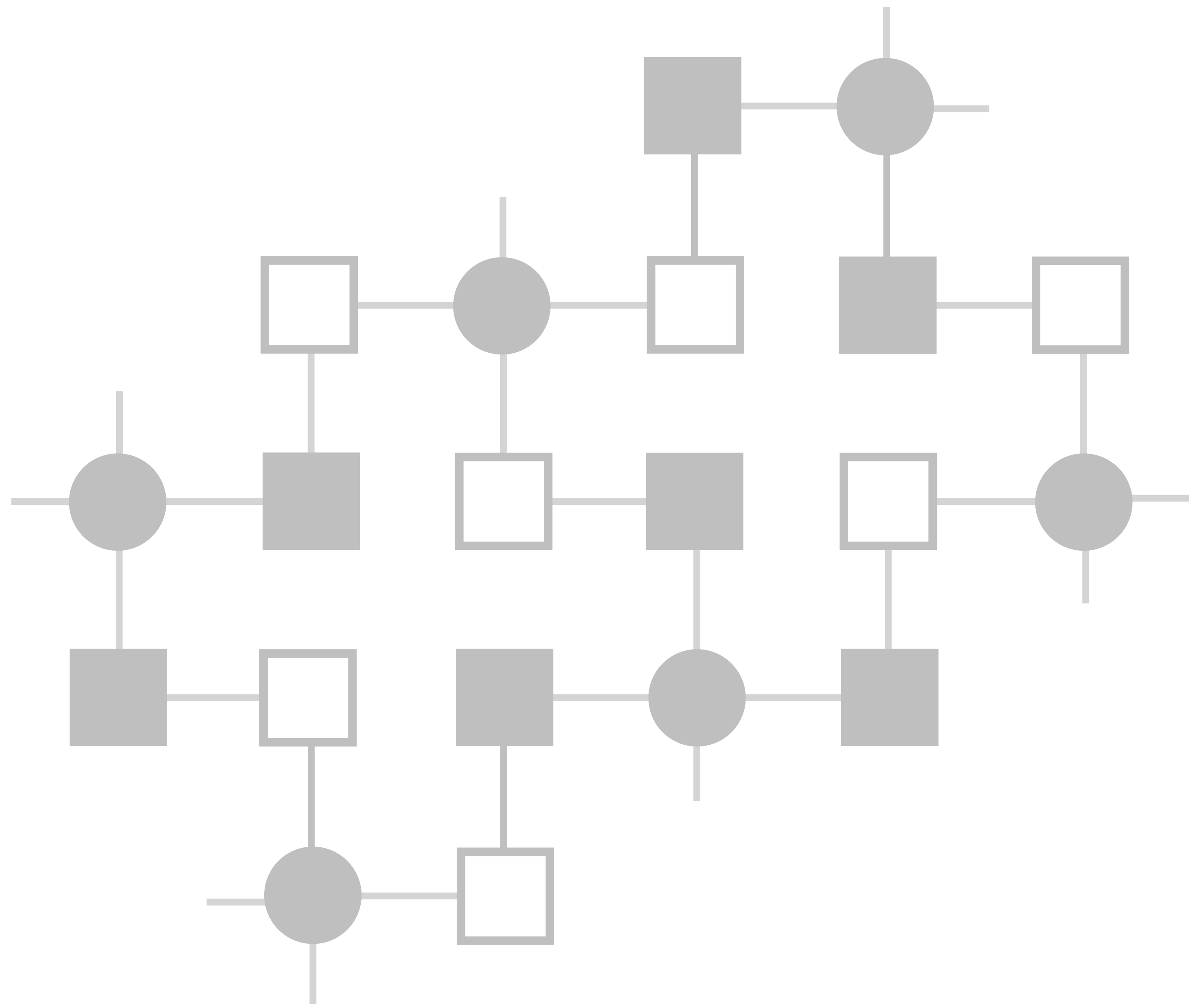}}
  \end{minipage}
  \caption
    {a) Spinful DK mapping from \cite{Clinton2022} and b) its embedding on a square grid topology. Full (empty) squares represent physical qubits with spin up (down) qubits and circles are ancilla qubits. The dashed lines show the edge operators connectivity while the solid lines are qubit connections.
      \label{fig:spinfulembeddingDK}
e    }
\end{figure}

\begin{table*}
\begin{center}
\begin{tabular}{|c|ccc|}
\hline
Axis & Weight (DK) & Weight (PA) & Weight (PAA)  \\ \hline
$\shortarrow{0}$    & $2+s+M(s-1)$   & $2+s+M(s-1)$  & $2+2s+M(s-1)$   \\
$\shortarrow{2}$    & $2+s+M(s-1)$  & $3+s$ & $3+s$    \\
$\shortarrow{1}$   & $2+2s+Ms$  & $2+2s+M(s-1)$ & $2+3s+M(s-1)$  \\ 
$\shortarrow{3}$    & $2+2s+Ms$  & $2+2s+M(s-1)$ & $2+3s+M(s-1)$  \\ \hline
\end{tabular}
\caption{Worst-case operator weight of the edge operators between lattice sites separated by $s\geq 1$ steps on the respective axis for a fermionic model with $M$ modes.}\label{table:neighbors}
\end{center}
\end{table*}

While it is obvious that the DK mapping can be embedded into a 4-connected square grid topology by adding idling qubits in the faces that have no ancilla qubits, an embedding without idling qubits can also be found as shown in Fig.~\ref{fig:spinlessembeddingDK}.b). We found that the circuit depths obtained in Table~\ref{table:spinless} do not change if we use this implementation of the DK mapping.

An extension to multi-orbital models was proposed in \cite{Clinton2022}, substituting every physical qubit from the DK mapping by a chain of physical qubits representing different fermionic modes on each site and using JW chains for defining the edges, identical to our treatment in the main text. The resulting topology for two spins can be seen in Fig.~\ref{fig:spinfulembeddingDK}.a) and its (more straight-forward) embedding into a square grid topology is shown in Fig.~\ref{fig:spinfulembeddingDK}.b).

The interactions between fermions on non-neighboring sites can also be included in the DK mapping, however the operator weights will scale with the distance and number of modes per lattice site. In Table~\ref{table:neighbors} we show the edge operator weights of the DK, PA and PAA mappings for a model with $M$ modes per lattice site between far-neighbors along different axes, and separated by $s \geq 1$ steps. For the DK mapping, the unit cell consists of two lattice sites for which this edge operator weight in some cases scales differently. We consider the worst-case operator weight in such case. While in DK, the edges along all axes scale with $M$, in the case of PA and PAA lattice sites can be connected vertically through ancilla qubits, and thus do not scale with $M$. For some multi-orbital fermionic systems with higher-connected geometries this scaling can yield an advantage in the operator weights and hence also the circuit depths. Since the linear dependence on $M$ is usually translated into quadratic dependence of the circuit depth, we can therefore expect to reduce the depth from quadratic to linear in some cases.

\section{Circuit prescriptions
}\label{app:prescriptions}
\begin{figure}
\centering
  {\includegraphics[width=0.95\linewidth]{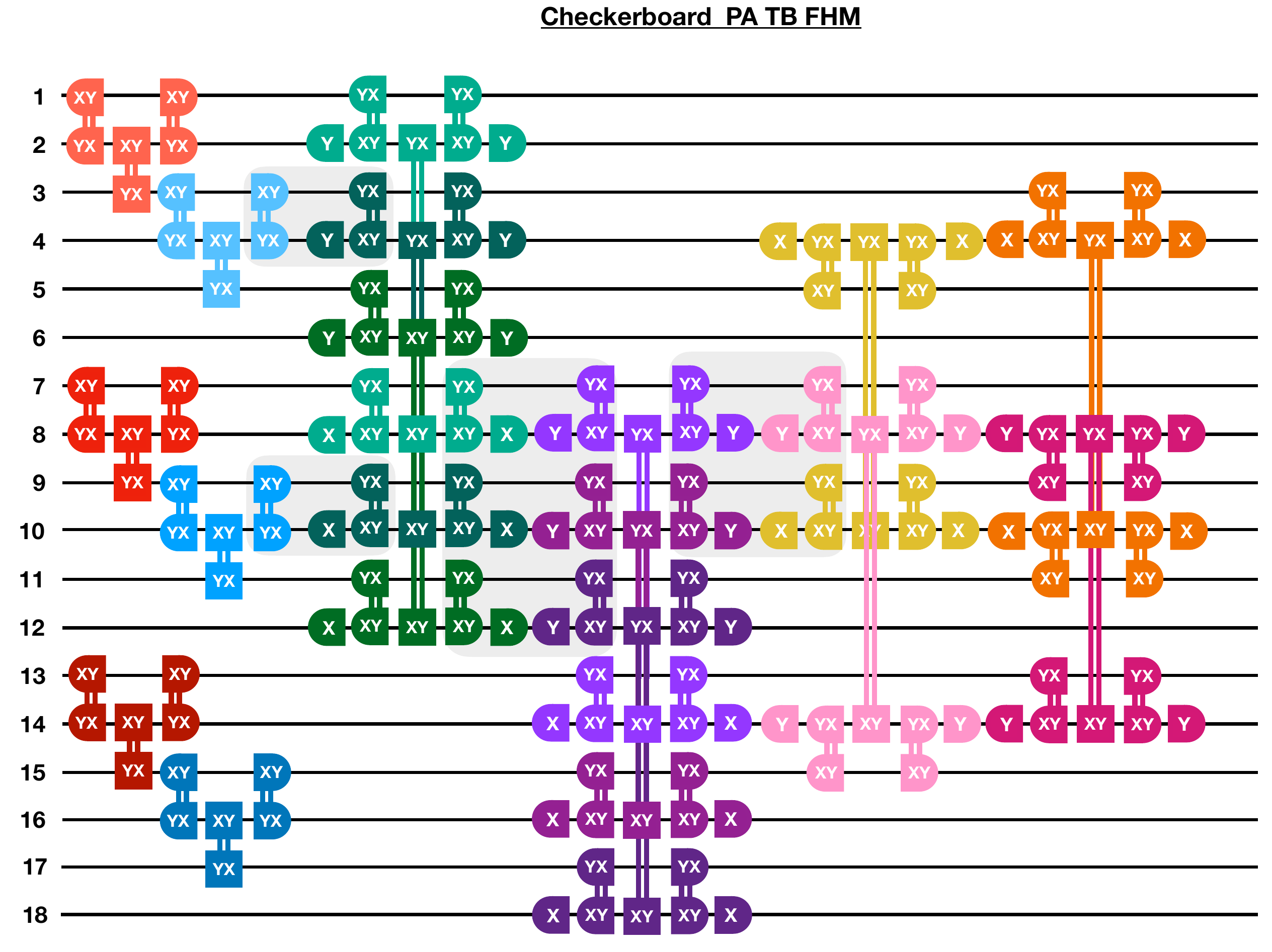}}
\hfill
\caption
{ Quantum circuit for simulating one Trotter step of the checkerboard TB FHM with the PA mapping.
}\label{fig:checkerboard_PA}
\end{figure}

\begin{figure}
\centering
  {\includegraphics[width=0.95\linewidth]{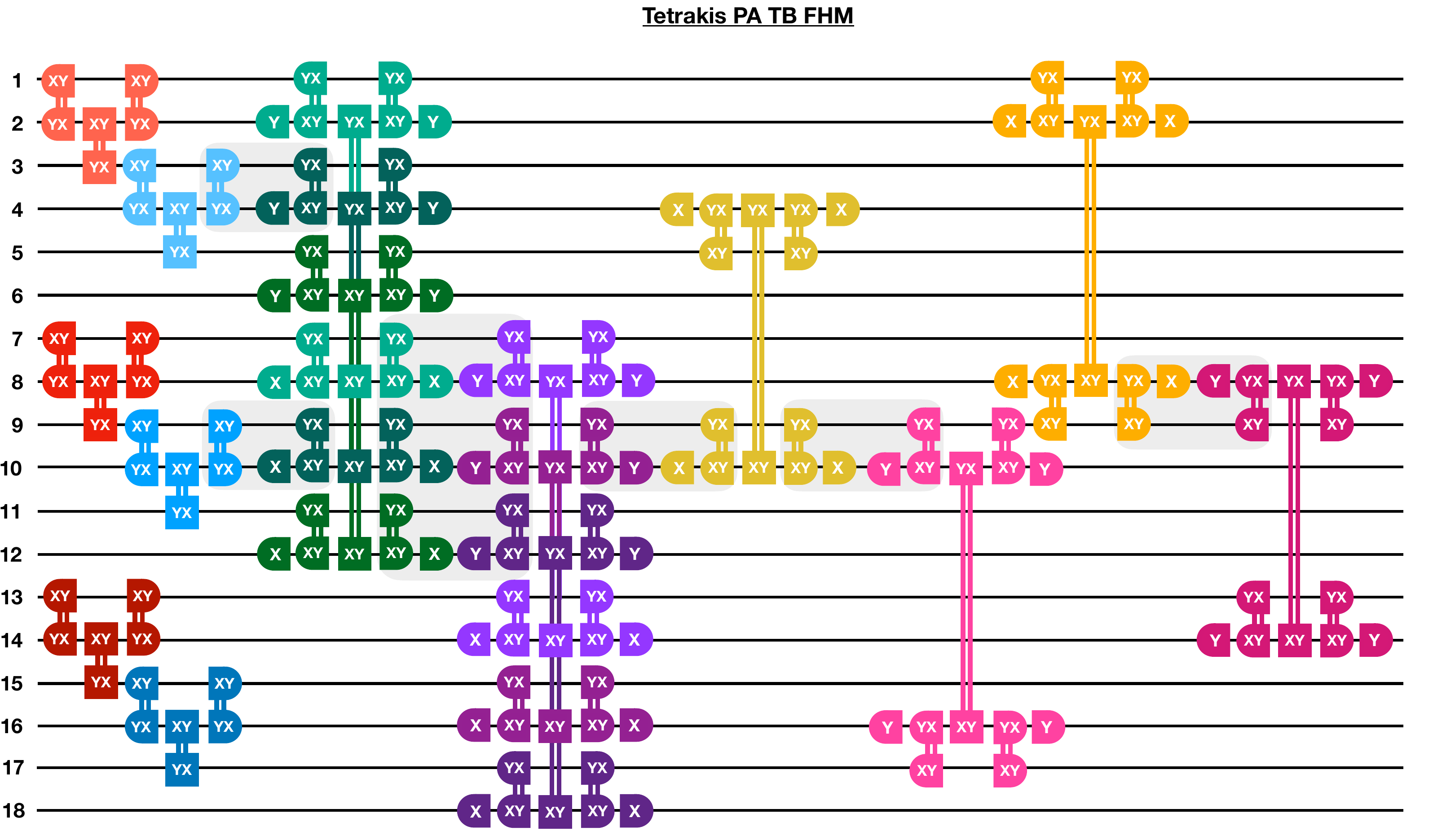}}
\hfill
\caption
{Quantum circuit for simulating one Trotter step of the tetrakis TB FHM with the PA mapping.
}\label{fig:tetrakis_PA}
\end{figure}

\begin{figure}
\centering
  {\includegraphics[width=0.75\linewidth]{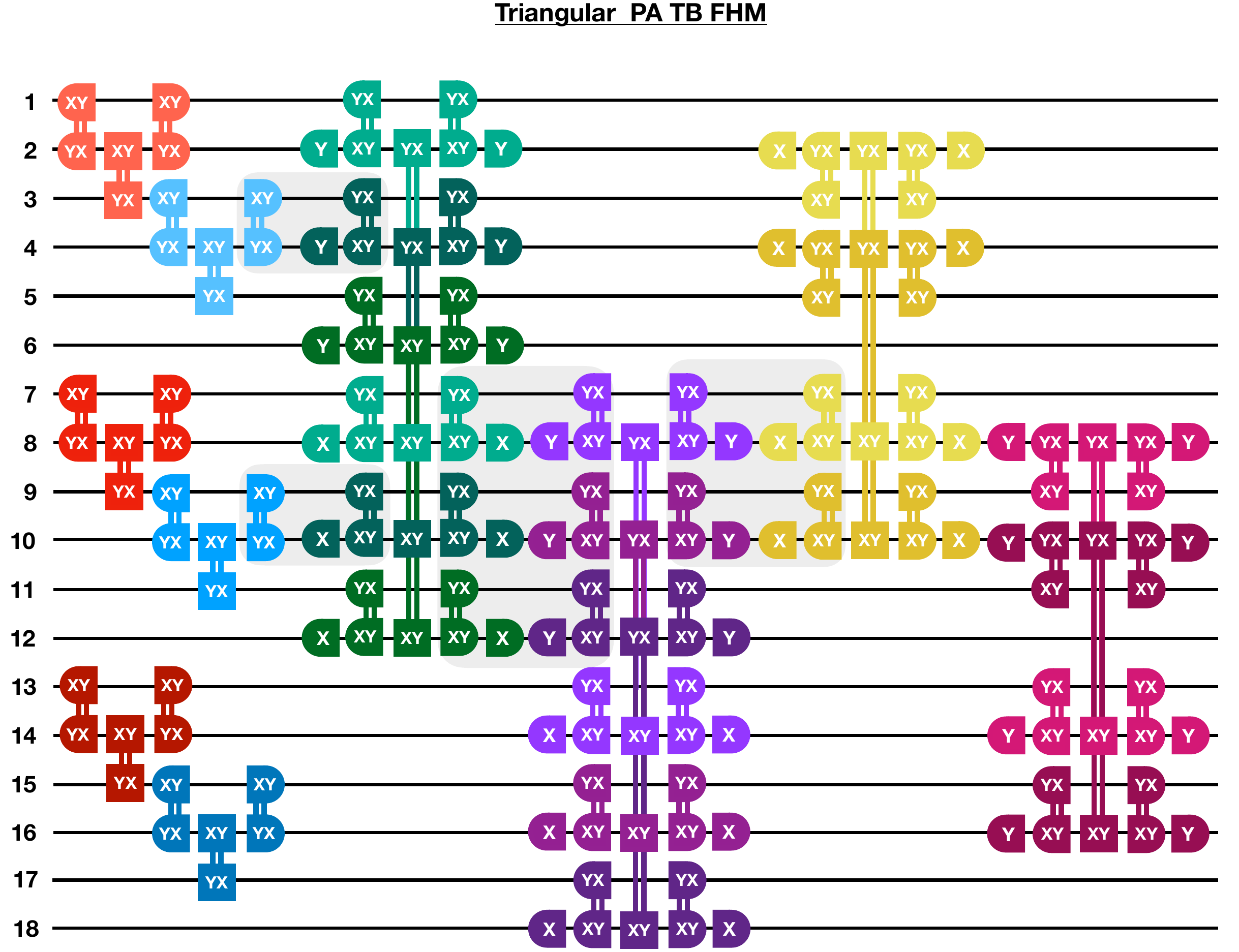}}
\hfill
\caption
{Quantum circuit for simulating one Trotter step of the triangular TB FHM with the PA mapping.
}\label{fig:triangular_PA}
\end{figure}

\begin{figure*}[h!btp]
\centering
\hbox{}{
\includegraphics[scale=0.3]{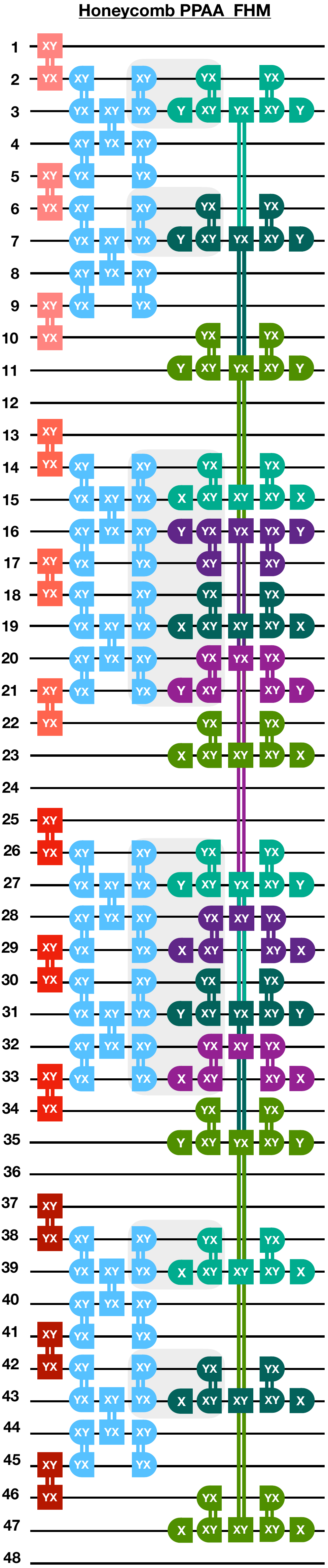}
\includegraphics[scale=0.3]{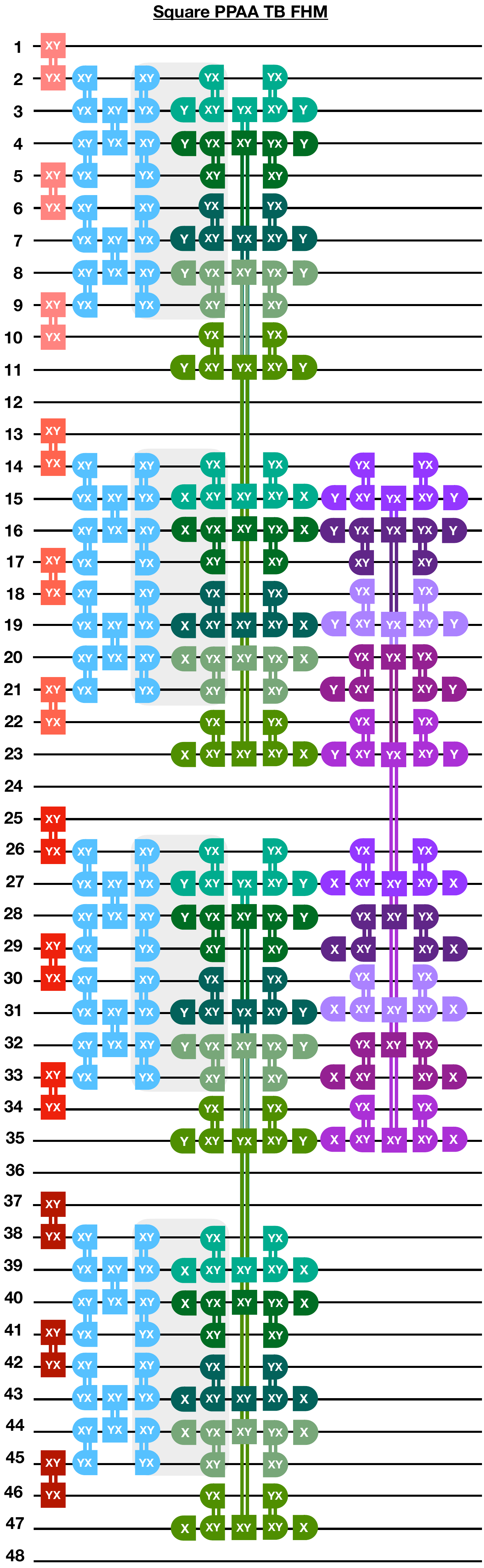}}
\caption{Quantum circuit for simulating one Trotter step of the honeycomb and square TB FHM with the PPAA mapping.}\label{fig:honeycomb_and_square_PPAA}
\end{figure*}

\begin{figure}
\centering
  {\includegraphics[width=0.65\linewidth]{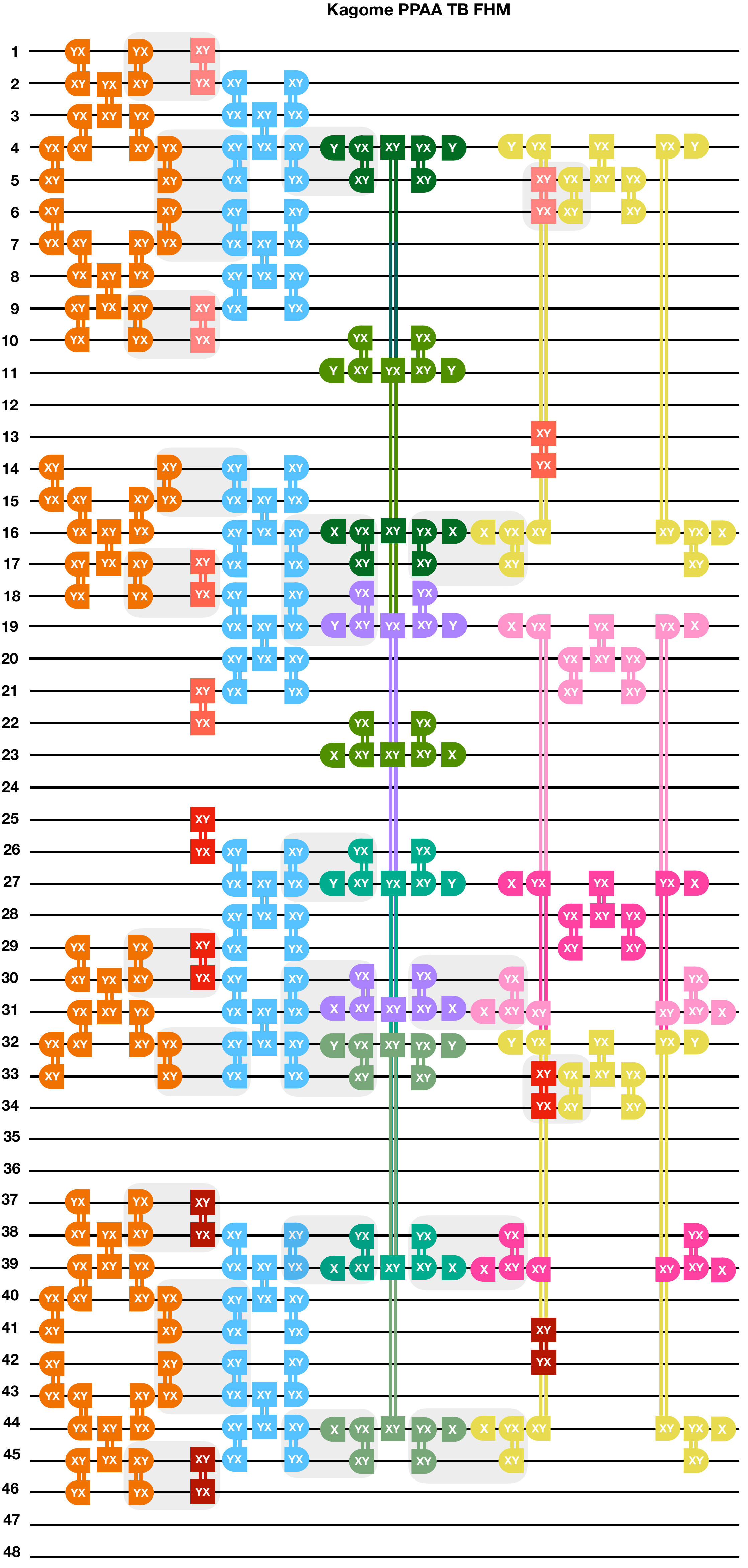}}
\hfill
\caption
{Quantum circuit for simulating one Trotter step of the kagome TB FHM with the PPAA mapping.
}\label{fig:kagome_PPAA}
\end{figure}

\begin{figure}
\centering
  {\includegraphics[width=0.95\linewidth]{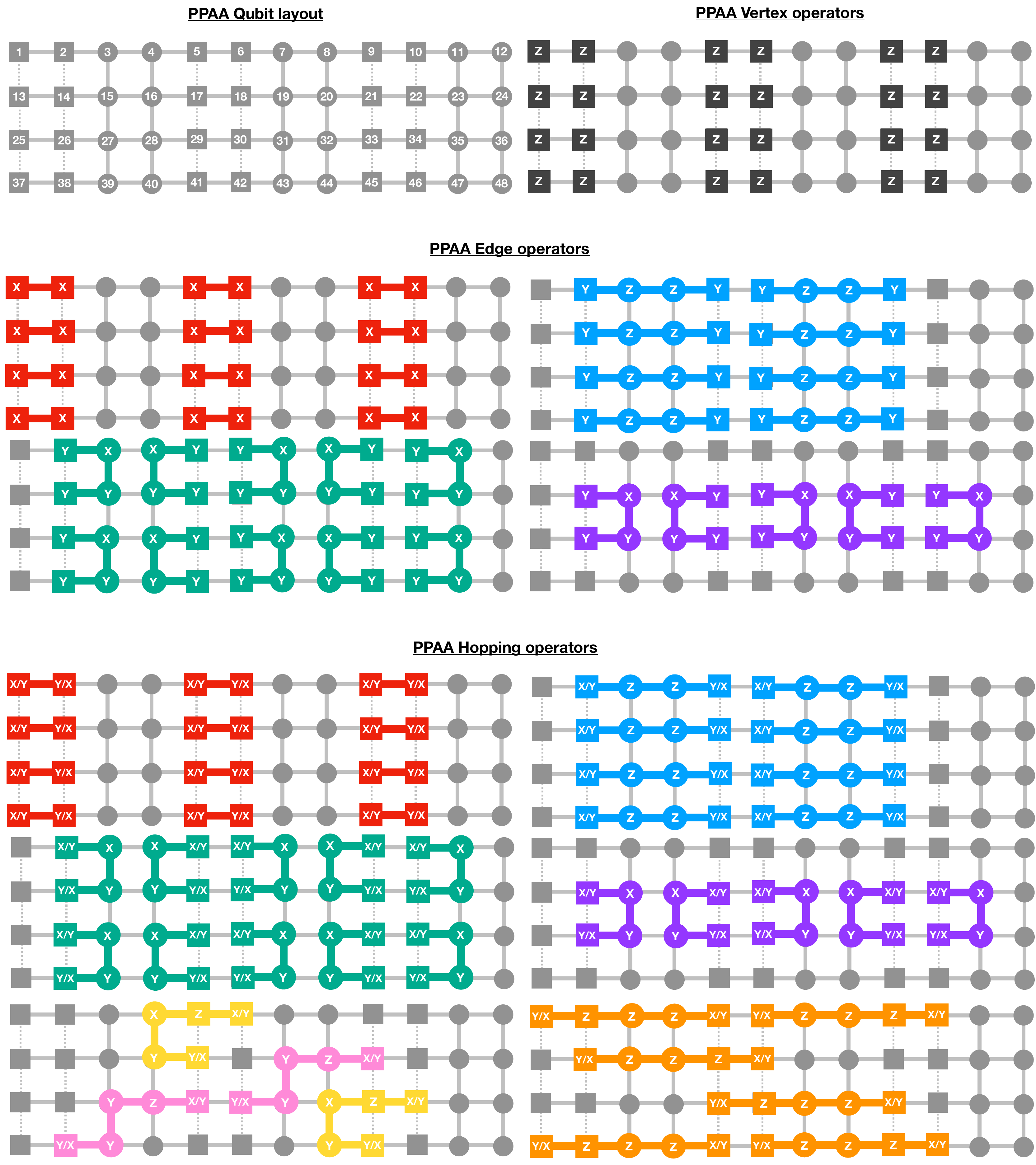}}
\hfill
\caption
{ Required information for building quantum circuits using the PPAA mapping.
}\label{fig:PPAA_info}
\end{figure}

For obtaining the reported depths in Tables~\ref{table:spinless} and \ref{table:spinful}, the employed ordering of the operators in the quantum circuits is shown in Tables~\ref{table:spinless_prescr_1} and \ref{table:spinless_prescr_2} for the TB model and in Tables~\ref{table:spinful_prescr_1} and \ref{table:spinful_prescr_2} for the FH model. For the TB model the explicit quantum circuits with the best depth for each 3-by-3 lattice can be 
found in Figs.~\ref{fig:checkerboard_PA}, \ref{fig:tetrakis_PA}, \ref{fig:triangular_PA}, \ref{fig:honeycomb_and_square_PPAA} and \ref{fig:kagome_PPAA}. Additional information for the two last figures using the PPAA mapping can be found in Fig.~\ref{fig:PPAA_info}.

\onecolumn

\begin{table}[]
\begin{adjustwidth}{-1.8cm}{}
\begin{tabular}{|c|c|c|}
\hline
Mapping              & Model                      & Prescription        \\ \hline
\multirow{20}{*}{PPA} & \multirow{2}{*}{TB Honeycomb}     & $H^{2,1}_{x+1,y\rightarrow x+2,y}H^{2,1}_{x,y\rightarrow x+1,y}$ \\
                     &                            &    $V^{4,2}_{x,y\rightarrow x,y+1}V^{4,2}_{x+3,y\rightarrow x+3,y+1}V^{4,2}_{x+1,y\rightarrow x+1,y+1}V^{4,2}_{x+2,y+1\rightarrow x+2,y+2}$                 \\ \cline{2-3} 
                     & \multirow{2}{*}{TB Square} & $H^{2,1}_{x+1,y\rightarrow x+2,y}H^{2,1}_{x,y\rightarrow x+1,y}V^{4,2}_{x,y+1\rightarrow x,y+2}V^{4,2}_{x+3,y+1\rightarrow x+3,y+2}V^{4,2}_{x,y\rightarrow x,y+1}$               \\
                     &                            & $V^{4,2}_{x+3,y\rightarrow x+3,y+1}V^{4,2}_{x+1,y+1\rightarrow x+1,y+2}V^{4,2}_{x+2,y+1\rightarrow x+2,y+2}V^{4,2}_{x+1,y\rightarrow x+1,y+1}V^{4,2}_{x+2,y\rightarrow x+2,y+1}$\\ \cline{2-3} 
                     & \multirow{2}{*}{TB Shastry-Sutherland} & $H^{2,1}_{x+1,y\rightarrow x+2,y}D^{2,2}_{x+1,y+2\rightarrow x+2,y+1}H^{2,1}_{x,y\rightarrow x+1,y}V^{4,2}_{x,y+1\rightarrow x,y+2}V^{4,2}_{x+3,y+1\rightarrow x+3,y+2}$               \\
                     &                            & $V^{4,2}_{x,y\rightarrow x,y+1}V^{4,2}_{x+3,y\rightarrow x+3,y+1}V^{4,2}_{x+1,y+1\rightarrow x+2,y+2}V^{4,2}_{x+3,y+1\rightarrow x+3,y+2}D^{2,2}_{x,y\rightarrow x+1,y+1}$\\ \cline{2-3} 
                     & \multirow{2}{*}{TB Kagome} & $H^{2,2}_{x,y\rightarrow x+1,y}V^{2,2}_{x+1,y\rightarrow x+1,y+1}V^{2,2}_{x+1,y+1\rightarrow x+1,y+2}D^{2,4}_{x,y \rightarrow x+1,y+1}D^{2,4}_{x+2,y+2 \rightarrow x+1,y+3}$               \\
                     &                            & $D^{2,4}_{x+1,y+1 \rightarrow x,y+2}D^{2,4}_{x+1,y+3 \rightarrow x+2,y+4}H^{2,2}_{x+1,y\rightarrow x+2,y}$\\ \cline{2-3} 
                     & \multirow{2}{*}{TB Triangular} & $H^{2,1}_{x+1,y\rightarrow x+2,y}H^{2,1}_{x,y\rightarrow x+1,y}V^{4,2}_{x,y+1\rightarrow x,y+2}V^{4,2}_{x+3,y+1\rightarrow x+3,y+2}V^{4,2}_{x,y\rightarrow x,y+1}$               \\
                     &                            & $V^{4,2}_{x+3,y\rightarrow x+3,y+1}D^{2,2}_{x,y \rightarrow x+1,y+1}D^{2,2}_{x,y+1 \rightarrow x+1,y+2}D^{2,2}_{x+1,y+1 \rightarrow x+2,y+2}D^{2,2}_{x+1,y \rightarrow x+2,y+1}$\\ \cline{2-3} 
                     & \multirow{3}{*}{TB Checkerboard} & $H^{2,1}_{x+1,y\rightarrow x+2,y}H^{2,1}_{x,y\rightarrow x+1,y}V^{4,2}_{x,y+1\rightarrow x,y+2}V^{4,2}_{x+3,y+1\rightarrow x+3,y+2}V^{4,2}_{x,y\rightarrow x,y+1}$               \\
                     &                            & $V^{4,2}_{x+3,y\rightarrow x+3,y+1}V^{4,2}_{x+1,y+1\rightarrow x+1,y+2}V^{4,2}_{x+2,y+1\rightarrow x+2,y+2}V^{4,2}_{x+1,y\rightarrow x+1,y+1}$\\
                     &                            & $V^{4,2}_{x+2,y\rightarrow x+2,y+1}D^{2,2}_{x,y+2 \rightarrow x+1,y+1}D^{2,2}_{x,y+1 \rightarrow x+1,y+2}D^{2,2}_{x+1,y \rightarrow x+2,y+1}D^{2,2}_{x+1,y+1 \rightarrow x+2,y}$\\ \cline{2-3} 
                     & \multirow{3}{*}{TB Tetrakis} & $H^{2,1}_{x+1,y\rightarrow x+2,y}D^{2,2}_{x+1,y+2\rightarrow x+2,y+1}D^{2,2}_{x+1,y\rightarrow x+2,y+1}H^{2,1}_{x,y\rightarrow x+1,y}V^{4,2}_{x,y+1\rightarrow x,y+2} $               \\
                     &                            & $V^{4,2}_{x+3,y+1\rightarrow x+3,y+2}V^{4,2}_{x,y\rightarrow x,y+1}V^{4,2}_{x+3,y\rightarrow x+3,y+1}V^{4,2}_{x+1,y+1\rightarrow x+1,y+2}V^{4,2}_{x+2,y+1\rightarrow x+2,y+2}$\\  
                     &                            & $V^{4,2}_{x+1,y\rightarrow x+1,y+1}V^{4,2}_{x+2,y\rightarrow x+2,y+1}D^{2,2}_{x,y+1\rightarrow x+1,y}D^{2,2}_{x,y+1\rightarrow x+1,y+2}$\\ \cline{2-3} 
                     & \multirow{4}{*}{TB NNN Square} & $H^{2,1}_{x+1,y\rightarrow x+2,y}H^{2,1}_{x,y\rightarrow x+1,y}V^{4,2}_{x,y+1\rightarrow x,y+2}V^{4,2}_{x+3,y+1\rightarrow x+3,y+2}V^{4,2}_{x,y\rightarrow x,y+1}$               \\
                     &                            & $V^{4,2}_{x+3,y\rightarrow x+3,y+1}V^{4,2}_{x+1,y+1\rightarrow x+1,y+2}V^{4,2}_{x+2,y+1\rightarrow x+2,y+2}V^{4,2}_{x+1,y\rightarrow x+1,y+1}$\\  
                     &                            & $V^{4,2}_{x+2,y\rightarrow x+2,y+1}D^{2,2}_{x,y+2\rightarrow x+1,y+1}D^{2,2}_{x,y\rightarrow x+1,y+1}D^{2,2}_{x,y+1\rightarrow x+1,y}$\\  
                     &                            & $D^{2,2}_{x,y+1\rightarrow x+1,y+2}D^{2,2}_{x+1,y+2\rightarrow x+2,y+1}D^{2,2}_{x+1,y\rightarrow x+2,y+1}D^{2,2}_{x+1,y+1\rightarrow x+2,y+2}D^{2,2}_{x+1,y+1\rightarrow x+2,y}$\\    \hline

\multirow{12}{*}{PA} & TB Honeycomb     & $H^{2,1}_{x,y\rightarrow x+1,y}H^{2,1}_{x+1,y\rightarrow x+2,y}V^{2,2}_{x,y+1\rightarrow x,y+2}V^{2,2}_{x,y\rightarrow x,y+1}$ \\ \cline{2-3} 
                     & TB Square & $H^{2,1}_{x,y\rightarrow x+1,y}H^{2,1}_{x+1,y\rightarrow x+2,y}V^{1,2}_{x,y+1\rightarrow x,y+2}V^{1,2}_{x,y\rightarrow x,y+1}$               \\ \cline{2-3} 
                     & TB Shastry-Sutherland & $H^{2,1}_{x,y\rightarrow x+1,y}H^{2,1}_{x+1,y\rightarrow x+2,y}V^{1,2}_{x,y+1\rightarrow x,y+2}V^{1,2}_{x,y\rightarrow x,y+1}D^{2,2}_{x,y+1\rightarrow x+1,y+2}D^{2,2}_{x+1,y+1\rightarrow x+2,y}$               \\ \cline{2-3} 
                     & \multirow{2}{*}{TB Kagome} & $H^{2,2}_{x,y\rightarrow x+1,y}H^{2,2}_{x+1,y\rightarrow x+2,y}V^{2,2}_{x,y+1\rightarrow x,y+2}V^{2,2}_{x,y\rightarrow x,y+1}$               \\
                     &                            & $D^{2,4}_{x+1,y \rightarrow x,y+1}D^{2,4}_{x,y+1 \rightarrow x+1,y+2}D^{2,4}_{x+1,y+4 \rightarrow x+2,y+3}D^{2,4}_{x+1,y+2 \rightarrow x+2,y+3}$\\ \cline{2-3} 
                     & TB Triangular & $H^{2,1}_{x,y\rightarrow x+1,y}H^{2,1}_{x+1,y\rightarrow x+2,y}V^{1,2}_{x,y+1\rightarrow x,y+2}V^{1,2}_{x,y\rightarrow x,y+1}D^{1,2}_{x,y+1\rightarrow x+1,y+2}D^{1,2}_{x,y\rightarrow x+1,y+1}$              \\ \cline{2-3} 
                     & \multirow{2}{*}{TB Checkerboard} & $H^{2,1}_{x,y\rightarrow x+1,y}H^{2,1}_{x+1,y\rightarrow x+2,y}V^{1,2}_{x,y+1\rightarrow x,y+2}V^{1,2}_{x,y\rightarrow x,y+1}$               \\
                     &                            & $D^{2,2}_{x+1,y+1\rightarrow x+2,y+2}D^{2,2}_{x,y+1\rightarrow x+1,y}D^{2,2}_{x+1,y+2\rightarrow x+2,y+1}D^{2,2}_{x+1,y\rightarrow x+2,y+1}$\\ \cline{2-3} 
                     & \multirow{3}{*}{TB Tetrakis} & $H^{2,1}_{x,y\rightarrow x+1,y}H^{2,1}_{x+1,y\rightarrow x+2,y}V^{1,2}_{x,y+1\rightarrow x,y+2}V^{1,2}_{x,y\rightarrow x,y+1}$               \\
                     &                            & $D^{2,2}_{x+1,y+1\rightarrow x+2,y+2}D^{2,2}_{x+1,y+1\rightarrow x+2,y}D^{2,2}_{x,y+2\rightarrow x+1,y+1}D^{2,2}_{x,y\rightarrow x+1,y+1}$\\ \cline{2-3} 
                     & \multirow{2}{*}{TB NNN Square} & $H^{2,1}_{x,y\rightarrow x+1,y}H^{2,1}_{x+1,y\rightarrow x+2,y}V^{1,2}_{x,y+1\rightarrow x,y+2}V^{1,2}_{x,y\rightarrow x,y+1}$               \\
                     &                            & $D^{1,2}_{x,y+1\rightarrow x+1,y+2}D^{1,2}_{x,y+1\rightarrow x+1,y}D^{1,2}_{x,y+2\rightarrow x+1,y+1}D^{1,2}_{x,y\rightarrow x+1,y+1}$\\   \hline
\end{tabular}
\end{adjustwidth}
\caption{Ordering prescription for the application of horizontal ($H$), vertical ($V$) and diagonal ($D$) hoppings in the quantum circuit. The superindices $m,n$ represent the regularity of the applied hoppings in each direction, so they will have to be applied every $m$ sites in the horizontal axis and every $n$ sites in the vertical axis. Note that the Kagome lattice used here is the traditional one with idling qubits.}\label{table:spinless_prescr_1}
\end{table}

\begin{table}[]
\begin{adjustwidth}{-1.8cm}{}
\begin{tabular}{|c|c|c|}
\hline
Mapping              & Model                      & Prescription        \\ \hline
\multirow{20}{*}{PPAA} & TB Honeycomb     & $H^{2,1}_{x,y\rightarrow x+1,y}H^{2,1}_{x+1,y\rightarrow x+2,y}V^{2,2}_{x+1,y\rightarrow x+1,y+1}V^{2,2}_{x+3,y+1\rightarrow x+3,y+2}$            \\ \cline{2-3} 
                     & \multirow{2}{*}{TB Square} & $H^{2,1}_{x,y\rightarrow x+1,y}H^{2,1}_{x+1,y\rightarrow x+2,y}V^{2,2}_{x+1,y+1\rightarrow x+1,y+2}V^{2,2}_{x+2,y\rightarrow x+2,y+1}$               \\
                     &                            & $V^{2,2}_{x+1,y\rightarrow x+1,y+1}V^{2,2}_{x+2,y+1\rightarrow x+2,y+2}$ \\ \cline{2-3} 
                     & TB Shastry-Sutherland & $V^{1,2}_{x,y+1\rightarrow x,y+2}V^{1,2}_{x,y\rightarrow x,y+1}H^{2,1}_{x+1,y \rightarrow x+2,y}D^{2,2}_{x+1,y+1\rightarrow x+2,y+2}H^{2,1}_{x,y\rightarrow x+1,y}D^{2,2}_{x,y+1\rightarrow x+1,y}$\\ \cline{2-3} 
                     & \multirow{4}{*}{TB Kagome} & $\tilde{H}^{3,3}_{x,y\rightarrow x+2,y}\tilde{H}^{3,3}_{x+2,y+1\rightarrow x+4,y+1}\tilde{H}^{3,3}_{x+1,y+2\rightarrow x+3,y+2}H^{4,1}_{x,y\rightarrow x+1,y}H^{2,1}_{x+1,y\rightarrow  x+2,y}$               \\
                     &                            & $V^{6,3}_{x+1,y\rightarrow x+1,y+1}V^{6,3}_{x,y+1\rightarrow x,y+2}V^{6,3}_{x+3,y+1\rightarrow x+3,y+2}V^{6,3}_{x+2,y+2\rightarrow x+2,y+3}$ \\
                     &                            & $V^{6,3}_{x+5,y+2\rightarrow x+5,y+3}V^{6,3}_{x+4,y\rightarrow x+4,y+1}H^{4,1}_{x+2,y\rightarrow x+3,y}D^{3,3}_{x+1,y\rightarrow x+2,y+1}$ \\
                     &                            & $D^{3,3}_{x,y+1\rightarrow x+1,y+2}D^{3,3}_{x+2,y+2\rightarrow x+3,y+3}$ \\ \cline{2-3} 
                     & \multirow{2}{*}{TB Triangular} & $H^{2,1}_{x,y\rightarrow x+1,y}V^{4,2}_{x+1,y+1\rightarrow x+1,y+2}V^{4,2}_{x,y\rightarrow x,y+1}H^{2,1}_{x+1,y\rightarrow x+2,y}V^{4,2}_{x+1,y\rightarrow x+1,y+1}V^{4,2}_{x,y+1\rightarrow x,y+2}$               \\
                     &                            & $D^{2,2}_{x+1,y+1 \rightarrow x+2,y+2}D^{2,2}_{x+1,y \rightarrow x+2,y+1}D^{2,2}_{x,y+1 \rightarrow x+1,y+2}D^{2,2}_{x,y \rightarrow x+1,y+1}$\\ \cline{2-3} 
                     & \multirow{2}{*}{TB Checkerboard} & $V^{1,2}_{x,y\rightarrow x,y+1}V^{1,2}_{x,y+1\rightarrow x,y+2}H^{2,1}_{x+1,y\rightarrow x+2,y}D^{2,2}_{x+1,y+1 \rightarrow x+2,y+2}D^{2,2}_{x,y \rightarrow x+1,y+1}$               \\
                     &                            & $H^{2,1}_{x,y\rightarrow x+1,y}D^{2,2}_{x,y+1 \rightarrow x+1,y}D^{2,2}_{x+1,y+2 \rightarrow x+2,y+1}$\\ \cline{2-3} 
                     & \multirow{2}{*}{TB Tetrakis} & $H^{2,1}_{x,y\rightarrow x+1,y}H^{2,1}_{x+1,y\rightarrow x+2,y}V^{1,2}_{x,y\rightarrow x,y+1}V^{1,2}_{x,y+1\rightarrow x,y+2}$               \\
                     &                            & $D^{2,2}_{x+1,y+1\rightarrow x+2,y+2}D^{2,2}_{x+1,y+1\rightarrow x+2,y}D^{2,2}_{x,y+2\rightarrow x+1,y+1}D^{2,2}_{x,y\rightarrow x+1,y+1}$\\ \cline{2-3} 
                     & \multirow{3}{*}{TB NNN Square} & $H^{2,1}_{x,y\rightarrow x+1,y}H^{2,1}_{x+1,y\rightarrow x+2,y}V^{1,2}_{x,y\rightarrow x,y+1}V^{1,2}_{x,y+1\rightarrow x,y+2} D^{2,2}_{x+1,y+1 \rightarrow x+2,y+2}$               \\
                     &                            & $D^{2,2}_{x+2,y+2 \rightarrow x+1,y+1}D^{2,2}_{x+1,y+1\rightarrow x+2,y}D^{2,2}_{x,y+2\rightarrow x+1,y+1}D^{2,2}_{x,y\rightarrow x+1,y+1}$\\  
                     &                            & $D^{2,2}_{x,y+1 \rightarrow x+1,y+2}D^{2,2}_{x,y+1 \rightarrow x+1,y}D^{2,2}_{x+1,y+2 \rightarrow x+2,y+1}D^{2,2}_{x+1,y \rightarrow x+2,y+1}$\\    \hline
\multirow{13}{*}{PAA} & TB Honeycomb     & $V^{2,2}_{x,y+1\rightarrow x,y+2}V^{2,2}_{x+1,y\rightarrow x+1,y+1}H^{2,1}_{x+1,y\rightarrow x+2,y}H^{2,1}_{x,y\rightarrow x+1,y}$ \\ \cline{2-3} 
                     & TB Square & $H^{2,1}_{x,y\rightarrow x+1,y}H^{2,1}_{x+1,y\rightarrow x+2,y}V^{1,2}_{x,y\rightarrow x,y+1}V^{1,2}_{x,y+1\rightarrow x,y+2}$               \\ \cline{2-3} 
                     & TB Shastry-Sutherland & $H^{2,1}_{x,y\rightarrow x+1,y}H^{2,1}_{x+1,y\rightarrow x+2,y}V^{1,2}_{x,y\rightarrow x,y+1}V^{1,2}_{x,y+1\rightarrow x,y+2}D^{2,2}_{x,y+1\rightarrow x+1,y+2}D^{2,2}_{x+1,y+1\rightarrow x+2,y}$               \\ \cline{2-3} 
                     & \multirow{3}{*}{TB Kagome} & $V^{1,2}_{x,y\rightarrow x,y+1}V^{1,2}_{x,y+1\rightarrow x,y+2}D^{3,3}_{x+1,y+1\rightarrow x+2,y}D^{3,3}_{x,y+3\rightarrow x+1,y+2}D^{3,3}_{x+2,y+2\rightarrow x+3,y+1}$               \\
                     &                            & $H^{3,3}_{x,y\rightarrow x+1,y}H^{3,3}_{x+1,y+1\rightarrow x+2,y+1}H^{3,3}_{x+2,y+2\rightarrow x+3,y+2}$ \\
                     &                            & $\tilde{V}^{3,3}_{x+1,y\rightarrow x+1,y+2}\tilde{V}^{3,3}_{x+2,y+1\rightarrow x+2,y+3}\tilde{V}^{3,3}_{x,y+2\rightarrow x,y+4}$ \\ \cline{2-3} 
                     & TB Triangular & $H^{2,1}_{x,y\rightarrow x+1,y}H^{2,1}_{x+1,y\rightarrow x+2,y}V^{1,2}_{x,y\rightarrow x,y+1}V^{1,2}_{x,y+1\rightarrow x,y+2}D^{1,2}_{x,y+1\rightarrow x+1,y+2}D^{1,2}_{x,y\rightarrow x+1,y+1}$              \\ \cline{2-3} 
                     & \multirow{2}{*}{TB Checkerboard} & $H^{2,1}_{x,y\rightarrow x+1,y}H^{2,1}_{x+1,y\rightarrow x+2,y}V^{1,2}_{x,y\rightarrow x,y+1}V^{1,2}_{x,y+1\rightarrow x,y+2}$               \\
                     &                            & $D^{2,2}_{x,y+1\rightarrow x+1,y}D^{2,2}_{x+1,y+1\rightarrow x+1,y+2}D^{2,2}_{x,y\rightarrow x+1,y+1}D^{2,2}_{x+1,y+2\rightarrow x+2,y+1}$\\ \cline{2-3} 
                     & \multirow{2}{*}{TB Tetrakis} & $V^{1,2}_{x,y\rightarrow x,y+1}V^{1,2}_{x,y+1\rightarrow x,y+2}H^{2,1}_{x,y\rightarrow x+1,y}D^{2,2}_{x,y+1\rightarrow x+1,y}D^{2,2}_{x,y+1\rightarrow x+1,y+2}$               \\
                     &                            & $D^{2,2}_{x+1,y+2\rightarrow x+2,y+1}D^{2,2}_{x+1,y\rightarrow x+2,y+1}H^{2,1}_{x+1,y\rightarrow x+2,y}$\\ \cline{2-3} 
                     & \multirow{2}{*}{TB NNN Square} & $H^{2,1}_{x,y\rightarrow x+1,y}H^{2,1}_{x+1,y\rightarrow x+2,y}V^{1,2}_{x,y\rightarrow x,y+1}V^{1,2}_{x,y+1\rightarrow x,y+2}D^{1,2}_{x,y+1\rightarrow x+1,y}D^{1,2}_{x,y+1\rightarrow x+1,y+2}$               \\
                     &                            & $D^{1,2}_{x,y+2\rightarrow x+1,y+1}D^{1,2}_{x,y\rightarrow x+1,y+1}$\\   \hline
                     
\end{tabular}
\end{adjustwidth}
\caption{Ordering prescription for the application of horizontal ($H$), vertical ($V$) and diagonal ($D$) hoppings in the quantum circuit. The superindices $m,n$ represent the regularity of the applied hoppings in each direction, so they will have to be applied every $m$ sites in the $x$-direction and every $n$ sites in the $y$-direction. Note that the Kagome lattice used for PPAA and PAA is the modified version described in \ref{fig:lattices} without idling qubits with a 90\textdegree rotation in the PAA case.}\label{table:spinless_prescr_2}
\end{table}

\begin{table}[]
\begin{adjustwidth}{-1.8cm}{}
\begin{tabular}{|c|c|c|}
\hline
Mapping              & Model                      & Prescription        \\ \hline
\multirow{27}{*}{PA} & \multirow{2}{*}{FH Honeycomb} & $H^{2,1}_{x+1,y\rightarrow x+2,y}V^{4,2}_{x+1,y+1\rightarrow x+1,y+2}V^{4,2}_{x+3,y\rightarrow x+3,y+1}U^{2,1}_{x,y\rightarrow x+1,y}\text{fSWAP}^{2,1}_{x,y\rightarrow x+1,y}$               \\
                     &                            & $H^{2,1}_{x+1,y\rightarrow x+2,y}V^{4,2}_{x+1,y+1\rightarrow x+1,y+2}V^{4,2}_{x+3,y\rightarrow x+3,y+1}$ \\ \cline{2-3} 
                     & \multirow{2}{*}{FH Square} & $H^{2,1}_{x+1,y\rightarrow x+2,y}V^{2,2}_{x+1,y\rightarrow x+1,y+1}V^{2,2}_{x+1,y+1\rightarrow x+1,y+2}U^{2,1}_{x,y\rightarrow x+1,y}\text{fSWAP}^{2,1}_{x,y\rightarrow x+1,y}$               \\
                     &                            & $H^{2,1}_{x+1,y\rightarrow x+2,y}V^{2,2}_{x+1,y\rightarrow x+1,y+1}V^{2,2}_{x+1,y+1\rightarrow x+1,y+2}$ \\ \cline{2-3} 
                     & \multirow{3}{*}{FH Shastry-Sutherland} & $H^{2,1}_{x+1,y\rightarrow x+2,y}V^{2,2}_{x+1,y\rightarrow x+1,y+1}V^{2,2}_{x+1,y+1\rightarrow x+1,y+2}D^{4,2}_{x+1,y\rightarrow x+2,y+1}D^{4,2}_{x+3,y+2\rightarrow x+4,y+1}$ \\
                     &                            & $U^{2,1}_{x,y\rightarrow x+1,y}\text{fSWAP}^{2,1}_{x,y\rightarrow x+1,y} H^{2,1}_{x+1,y\rightarrow x+2,y}V^{2,2}_{x+1,y\rightarrow x+1,y+1}V^{2,2}_{x+1,y+1\rightarrow x+1,y+2}$ \\
                     &                            & $D^{4,2}_{x+1,y\rightarrow x+2,y+1}D^{4,2}_{x+3,y+2\rightarrow x+4,y+1}$  \\ \cline{2-3} 
                     & \multirow{5}{*}{FH Kagome} & $H^{2,2}_{x+1,y+1\rightarrow x+2,y+1}V^{4,2}_{x+1,y\rightarrow x+1,y+1}V^{4,2}_{x+1,y+1\rightarrow x+1,y+2}D^{4,4}_{x+1,y+2\rightarrow x+2,y+3}$               \\
                     &                            & $D^{4,4}_{x+1,y+2\rightarrow x+2,y+1}D^{4,4}_{x+3,y+1\rightarrow x+4,y}D^{4,4}_{x+3,y+3\rightarrow x+4,y+4}U^{2,2}_{x,y+1\rightarrow x+1,y+1}$ \\
                     &                            & $U^{4,2}_{x,y\rightarrow x+1,y}\text{fSWAP}^{2,2}_{x,y+1\rightarrow x+1,y+1}\text{fSWAP}^{4,2}_{x,y\rightarrow x+1,y}H^{2,2}_{x+1,y+1\rightarrow x+2,y+1}V^{4,2}_{x+1,y\rightarrow x+1,y+1}$ \\
                     &                            & $V^{4,2}_{x+1,y+1\rightarrow x+1,y+2}D^{4,4}_{x+1,y+2\rightarrow x+2,y+3}D^{4,4}_{x+1,y+2\rightarrow x+2,y+1}$ \\
                     &                            & $D^{4,4}_{x+3,y+1\rightarrow x+4,y}D^{4,4}_{x+3,y+3\rightarrow x+4,y+4}$  \\ \cline{2-3} 
                     & \multirow{3}{*}{FH Triangular} & $H^{2,1}_{x+1,y\rightarrow x+2,y}V^{2,2}_{x+1,y\rightarrow x+1,y+1}V^{2,2}_{x+1,y+1\rightarrow x+1,y+2}D^{4,2}_{x+1,y\rightarrow x+2,y+1}$               \\
                     &                            & $D^{2,2}_{x+1,y+1\rightarrow x+2,y+2}U^{2,1}_{x,y\rightarrow x+1,y}\text{fSWAP}^{2,1}_{x,y\rightarrow x+1,y}$ \\
                     &                            & $H^{2,1}_{x+1,y\rightarrow x+2,y}V^{2,2}_{x+1,y\rightarrow x+1,y+1}V^{2,2}_{x+1,y+1\rightarrow x+1,y+2}D^{4,2}_{x+1,y\rightarrow x+2,y+1}D^{2,2}_{x+1,y+1\rightarrow x+2,y+2}$ \\ \cline{2-3} 
                     & \multirow{4}{*}{FH Checkerboard} & $H^{2,1}_{x+1,y\rightarrow x+2,y}V^{2,2}_{x+1,y\rightarrow x+1,y+1}V^{2,2}_{x+1,y+1\rightarrow x+1,y+2}D^{4,2}_{x+1,y+2\rightarrow x+2,y+1}D^{4,2}_{x+3,y\rightarrow x+4,y+1}$               \\
                     &                            & $D^{4,2}_{x+1,y+1\rightarrow x+2,y+2}D^{4,2}_{x+3,y+1\rightarrow x+4,y}U^{2,1}_{x,y\rightarrow x+1,y}\text{fSWAP}^{2,1}_{x,y\rightarrow x+1,y}H^{2,1}_{x+1,y\rightarrow x+2,y}$ \\
                     &                            & $V^{2,2}_{x+1,y\rightarrow x+1,y+1}V^{2,2}_{x+1,y+1\rightarrow x+1,y+2}D^{4,2}_{x+1,y+2\rightarrow x+2,y+1}$ \\
                     &                            & $D^{4,2}_{x+3,y\rightarrow x+4,y+1}D^{4,2}_{x+1,y+1\rightarrow x+2,y+2}D^{4,2}_{x+3,y+1\rightarrow x+4,y}$ \\ \cline{2-3} 
                     & \multirow{4}{*}{FH Tetrakis} & $H^{2,1}_{x+1,y\rightarrow x+2,y}V^{2,2}_{x+1,y\rightarrow x+1,y+1}V^{2,2}_{x+1,y+1\rightarrow x+1,y+2}D^{4,2}_{x+3,y\rightarrow x+4,y+1}D^{4,2}_{x+1,y+1\rightarrow x+2,y+2}$               \\
                     &                            & $D^{4,2}_{x+1,y+1\rightarrow x+2,y}D^{4,2}_{x+3,y+2\rightarrow x+4,y+1}U^{2,1}_{x,y\rightarrow x+1,y}\text{fSWAP}^{2,1}_{x,y\rightarrow x+1,y}H^{2,1}_{x+1,y\rightarrow x+2,y}$ \\
                     &                            & $V^{2,2}_{x+1,y\rightarrow x+1,y+1}V^{2,2}_{x+1,y+1\rightarrow x+1,y+2}D^{4,2}_{x+3,y\rightarrow x+4,y+1}$ \\
                     &                            & $D^{4,2}_{x+1,y+1\rightarrow x+2,y+2}D^{4,2}_{x+1,y+1\rightarrow x+2,y}D^{4,2}_{x+3,y+2\rightarrow x+4,y+1}$ \\ \cline{2-3} 
                     & \multirow{4}{*}{FH NNN Square} & $H^{2,1}_{x+1,y\rightarrow x+2,y}V^{2,2}_{x+1,y\rightarrow x+1,y+1}V^{2,2}_{x+1,y+1\rightarrow x+1,y+2}D^{2,2}_{x+1,y\rightarrow x+2,y+1}D^{2,2}_{x+1,y+2\rightarrow x+2,y+1}$               \\
                     &                            & $D^{2,2}_{x+1,y+1\rightarrow x+2,y+2}D^{2,2}_{x+1,y+1\rightarrow x+2,y}U^{2,1}_{x,y\rightarrow x+1,y}\text{fSWAP}^{2,1}_{x,y\rightarrow x+1,y}H^{2,1}_{x+1,y\rightarrow x+2,y}$\\  
                     &                            & $V^{2,2}_{x+1,y\rightarrow x+1,y+1}V^{2,2}_{x+1,y+1\rightarrow x+1,y+2}D^{2,2}_{x+1,y\rightarrow x+2,y+1}D^{2,2}_{x+1,y+2\rightarrow x+2,y+1}$ \\  
                     &                            & $D^{2,2}_{x+1,y+1\rightarrow x+2,y+2}D^{2,2}_{x+1,y+1\rightarrow x+2,y}$ \\    \hline
    
\end{tabular}
\end{adjustwidth}
\caption{Ordering prescription for the application of on-site interaction ($U$), horizontal ($H$), vertical ($V$), diagonal ($D$), next-nearest neighbor horizontal ($\tilde{H}$) and next-nearest neighbor vertical ($\tilde{V}$) hoppings in the quantum circuit. The superindices $m,n$ represent the regularity of the applied hoppings in each direction, so they will have to be applied every $m$ sites in the $x$-direction and every $n$ sites in the $y$-direction. Note that the Kagome lattice used for PA is the version with idling qubits. and PAA is the modified version described in \ref{fig:lattices} without idling qubits with a 90\textdegree  rotation in the PAA case.}\label{table:spinful_prescr_1}
\end{table}

\begin{table}[]
\begin{adjustwidth}{-1.8cm}{}
\begin{tabular}{|c|c|c|}
\hline
Mapping              & Model                      & Prescription        \\ \hline
\multirow{25}{*}{PAA} & \multirow{2}{*}{FH Honeycomb} & $H^{2,1}_{x+1,y\rightarrow x+2,y}V^{4,2}_{x,y\rightarrow x,y+1}V^{4,2}_{x+3,y\rightarrow x+3,y+1}V^{4,2}_{x+1,y+1\rightarrow x+1,y+2}$               \\
                     &                            & $V^{4,2}_{x+2,y+1\rightarrow x+2,y+2}U^{2,1}_{x,y\rightarrow x+1,y}\text{fSWAP}^{2,1}_{x,y\rightarrow x+1,y}H^{2,1}_{x+1,y\rightarrow x+2,y}$ \\ \cline{2-3} 
                     & FH Square & $H^{2,1}_{x+1,y\rightarrow x+2,y}V^{1,2}_{x,y\rightarrow x,y+1}V^{1,2}_{x,y+1\rightarrow x,y+2}U^{2,1}_{x,y\rightarrow x+1,y}\text{fSWAP}^{2,1}_{x,y\rightarrow x+1,y}H^{2,1}_{x+1,y\rightarrow x+2,y}$               \\ \cline{2-3} 
                     & \multirow{2}{*}{FH Shastry-Sutherland} & $V^{1,2}_{x,y\rightarrow x,y+1}V^{1,2}_{x,y+1\rightarrow x,y+2}D^{4,2}_{x+1,y\rightarrow x+2,y+1}D^{4,2}_{x+3,y+2\rightarrow x+4,y+1}H^{2,1}_{x+1,y\rightarrow x+2,y}$ \\
                     &                            & $U^{2,1}_{x,y\rightarrow x+1,y}\text{fSWAP}^{2,1}_{x,y\rightarrow x+1,y}D^{4,2}_{x+1,y\rightarrow x+2,y+1}D^{4,2}_{x+3,y+2\rightarrow x+4,y+1}H^{2,1}_{x+1,y\rightarrow x+2,y}$\\ \cline{2-3} 
                     & \multirow{5}{*}{FH Kagome} & $\tilde{V}^{6,3}_{x,y\rightarrow x,y+2}\tilde{V}^{6,3}_{x+1,y\rightarrow x+1,y+2}\tilde{V}^{6,3}_{x+2,y+1\rightarrow x+2,y+3}\tilde{V}^{6,3}_{x+3,y+1\rightarrow x+3,y+3}$               \\
                     &                            & $\tilde{V}^{6,3}_{x+4,y+2\rightarrow x+4,y+4}\tilde{V}^{6,3}_{x+5,y+2\rightarrow x+5,y+4}V^{1,2}_{x,y\rightarrow x,y+1}H^{6,3}_{x+1,y+1\rightarrow x+2,y+1}$ \\
                     &                            & $H^{6,3}_{x+3,y+2\rightarrow x+4,y+2}H^{6,3}_{x+5,y+3\rightarrow x+6,y+3}V^{1,2}_{x,y+1\rightarrow x,y+2}D^{6,3}_{x+1,y+1\rightarrow x+2,y}D^{6,3}_{x+3,y+2\rightarrow x+4,y+1}$ \\
                     &                            & $D^{6,3}_{x+5,y+3\rightarrow x+6,y+2}U^{2,1}_{x,y\rightarrow x+1,y}\text{fSWAP}^{2,1}_{x,y\rightarrow x+1,y}H^{6,3}_{x+1,y+1\rightarrow x+2,y+1}H^{6,3}_{x+3,y+2\rightarrow x+4,y+2}$ \\
                     &                            & $H^{6,3}_{x+5,y+3\rightarrow x+6,y+3}D^{6,3}_{x+1,y+1\rightarrow x+2,y}D^{6,3}_{x+3,y+2\rightarrow x+4,y+1}D^{6,3}_{x+5,y+3\rightarrow x+6,y+2}$  \\ \cline{2-3} 
                     & \multirow{3}{*}{FH Triangular} & $V^{1,2}_{x,y\rightarrow x,y+1}V^{1,2}_{x,y+1\rightarrow x,y+2}D^{4,2}_{x+1,y+1\rightarrow x+2,y+2}D^{4,2}_{x+3,y\rightarrow x+4,y+1}$               \\
                     &                            & $H^{2,1}_{x+1,y\rightarrow x+2,y}D^{4,2}_{x+1,y \rightarrow x+2,y+1}D^{4,2}_{x+3,y+1 \rightarrow x+4,y+2}U^{2,1}_{x,y\rightarrow x+1,y}\text{fSWAP}^{2,1}_{x,y\rightarrow x+1,y}$ \\
                     &                            & $D^{4,2}_{x+1,y+1\rightarrow x+2,y+2}D^{4,2}_{x+3,y\rightarrow x+4,y+1}H^{2,1}_{x+1,y\rightarrow x+2,y}D^{4,2}_{x+1,y \rightarrow x+2,y+1}D^{4,2}_{x+3,y+1 \rightarrow x+4,y+2}$ \\ \cline{2-3} 
                     & \multirow{4}{*}{FH Checkerboard} & $V^{1,2}_{x,y\rightarrow x,y+1}V^{1,2}_{x,y+1\rightarrow x,y+2}D^{4,2}_{x+1,y+1\rightarrow x+2,y+2}D^{4,2}_{x+3,y\rightarrow x+4,y+1}$               \\
                     &                            & $H^{2,1}_{x+1,y\rightarrow x+2,y}D^{4,2}_{x+1,y+2\rightarrow x+2,y+1}D^{4,2}_{x+3,y+1\rightarrow x+4,y}U^{2,1}_{x,y\rightarrow x+1,y}\text{fSWAP}^{2,1}_{x,y\rightarrow x+1,y}$ \\
                     &                            & $D^{4,2}_{x+1,y+1\rightarrow x+2,y+2}D^{4,2}_{x+3,y\rightarrow x+4,y+1}H^{2,1}_{x+1,y\rightarrow x+2,y}D^{4,2}_{x+1,y+2\rightarrow x+2,y+1}D^{4,2}_{x+3,y+1\rightarrow x+4,y}$ \\
                     &                            & $D^{4,2}_{x+3,y\rightarrow x+4,y+1}D^{4,2}_{x+1,y+1\rightarrow x+2,y+2}D^{4,2}_{x+3,y+1\rightarrow x+4,y}$ \\ \cline{2-3} 
                     & \multirow{3}{*}{FH Tetrakis} & $V^{1,2}_{x,y\rightarrow x,y+1}V^{1,2}_{x,y+1\rightarrow x,y+2}D^{4,2}_{x+1,y+1\rightarrow x+2,y}D^{4,2}_{x+3,y\rightarrow x+4,y+1}$               \\
                     &                            & $D^{4,2}_{x+1,y+1\rightarrow x+2,y+2}D^{4,2}_{x+3,y+2\rightarrow x+4,y+1}H^{2,1}_{x+1,y\rightarrow x+2,y}U^{2,1}_{x,y\rightarrow x+1,y}\text{fSWAP}^{2,1}_{x,y\rightarrow x+1,y}$ \\
                     &                            & $D^{4,2}_{x+1,y+1\rightarrow x+2,y}D^{4,2}_{x+3,y\rightarrow x+4,y+1}D^{4,2}_{x+1,y+1\rightarrow x+2,y+2}D^{4,2}_{x+3,y+2\rightarrow x+4,y+1}H^{2,1}_{x+1,y\rightarrow x+2,y}$\\ \cline{2-3} 
                     & \multirow{5}{*}{FH NNN Square} & $V^{1,2}_{x,y\rightarrow x,y+1}V^{1,2}_{x,y+1\rightarrow x,y+2}D^{4,2}_{x+1,y+1\rightarrow x+2,y}D^{4,2}_{x+3,y\rightarrow x+4,y+1}D^{4,2}_{x+1,y+1\rightarrow x+2,y+2}$               \\
                     &                            & $D^{4,2}_{x+3,y+2\rightarrow x+4,y+1}H^{2,1}_{x+1,y\rightarrow x+2,y}D^{4,2}_{x+1,y+2\rightarrow x+2,y+1}D^{4,2}_{x+3,y+1\rightarrow x+4,y+2}$\\  
                     &                            & $D^{4,2}_{x+1,y\rightarrow x+2,y+1}D^{4,2}_{x+3,y+1\rightarrow x+4,y}U^{2,1}_{x,y\rightarrow x+1,y}\text{fSWAP}^{2,1}_{x,y\rightarrow x+1,y}D^{4,2}_{x+1,y+1\rightarrow x+2,y}$ \\  
                     &                            & $D^{4,2}_{x+3,y\rightarrow x+4,y+1}D^{4,2}_{x+1,y+1\rightarrow x+2,y+2}D^{4,2}_{x+3,y+2\rightarrow x+4,y+1}H^{2,1}_{x+1,y\rightarrow x+2,y}$ \\  
                     &                            & $D^{4,2}_{x+1,y+2\rightarrow x+2,y+1}D^{4,2}_{x+3,y+1\rightarrow x+4,y+2}D^{4,2}_{x+1,y\rightarrow x+2,y+1}D^{4,2}_{x+3,y+1\rightarrow x+4,y}$ \\    \hline
    
\end{tabular}
\end{adjustwidth}
\caption{Ordering prescription for the application of on-site interaction ($U$), horizontal ($H$), vertical ($V$), diagonal ($D$) and next-nearest neighbor vertical ($\tilde{V}$) hoppings in the quantum circuit. The superindices $m,n$ represent the regularity of the applied hoppings in each direction, so they will have to be applied every $m$ sites in the $x$-direction and every $n$ sites in the $y$-direction. Note that the Kagome lattice used for PAA is the version with idling qubits and rotated 90\textdegree.}\label{table:spinful_prescr_2}
\end{table}


\clearpage
\twocolumn
\section{Jordan-Wigner vs PA on a 3-by-3 lattice}
\label{appendix:jw}
\begin{figure}
\centering
  {\includegraphics[width=0.95\linewidth]{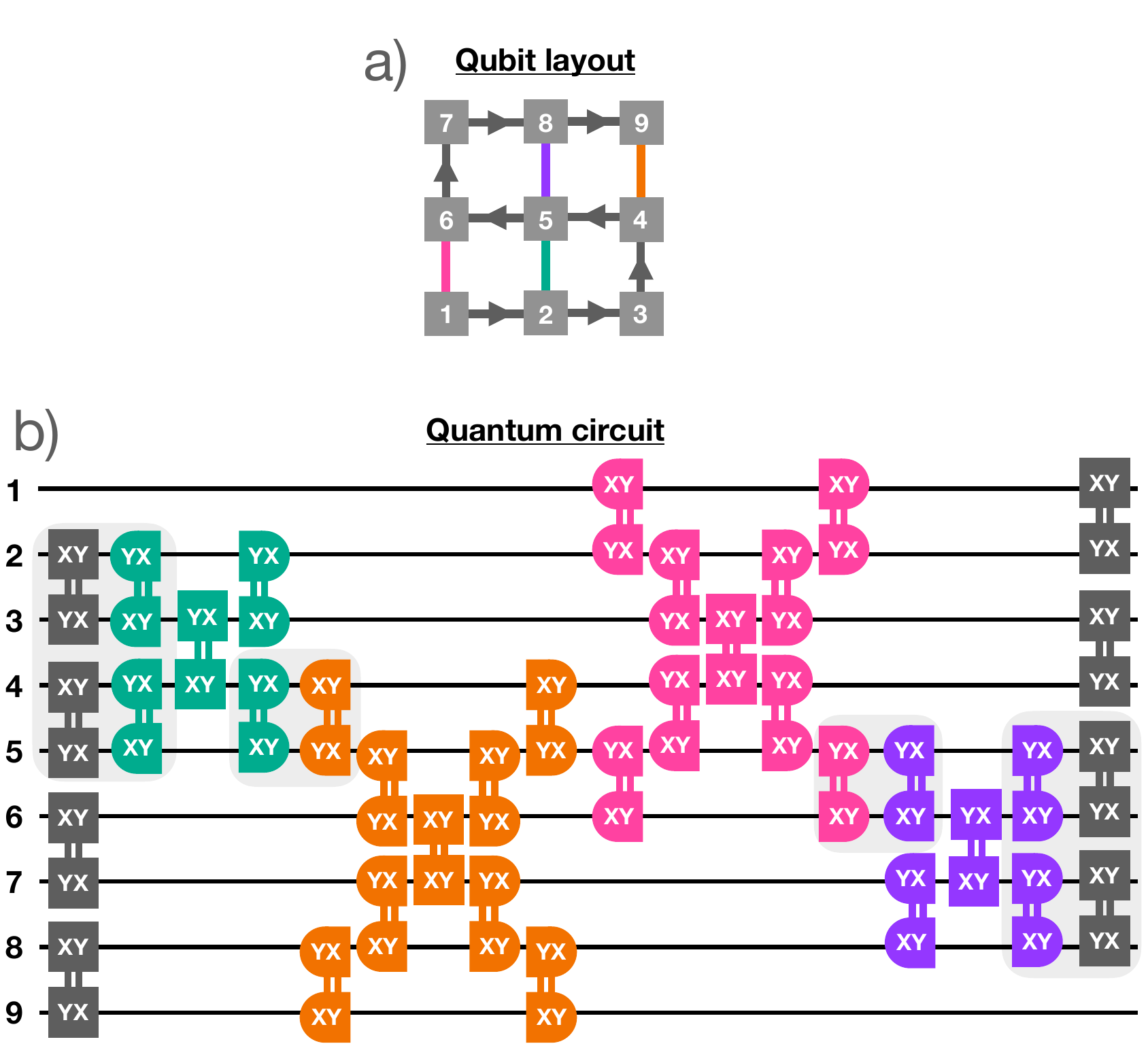}}
\hfill
\caption
{ a) Qubit layout representing the JW snake-like pattern of a TB model on a 3-by-3 fermionic lattice and b) quantum circuit of the corresponding Trotter step, showing in grey the combinations of gates and through red lines the cancellations of gates. The two-qubit gates represent fSIM gates.
  \label{fig:jw3by3}
}
\end{figure}

Due to the relatively local nature of hopping operators when using the Jordan Wigner mapping (JW) on small lattices, one could argue that this will yield smaller circuit depths than local mappings. To address this question, we investigate in Fig.~\ref{fig:jw3by3} the minimal JW example of a TB model on a 3-by-3 square grid and obtain a circuit depth of 10, which is worse than the depth 9 obtained for PA or depth 8 for PPAA in Table~\ref{table:spinless}. Using the same approach, one can also show that the number of two-qubit gates for this system size is exactly the same between the DK mapping and JW. This result leads us to the conclusion that it is never favourable to use JW for the simulation of such fermionic models on two-dimensional lattices with a system size larger than 3-by-3.

It should, however, be noted, that we have not included any overhead due to state preparation for local mappings into this calculation. This overhead can be either negligible, when stabilizers can be measured at the start of the circuit, or otherwise add a one-time overhead at the beginning of the circuit, which scales linearly with the system size. In the second case, it is expected that local mappings still become favourable to the JW mapping, albeit at somewhat larger system sizes.

\section{Hubbard-Kanamori simulation}
\label{appendix:hk}

The simulation of a HK model for 4 orbitals has been shown in the main text. In this appendix, we additionally provide the circuits for the HK model with two and three orbitals in  Fig.~\ref{fig:hk}. 

In the case of two orbitals (Fig.~\ref{fig:hk}.a)) we follow the standard fSWAP network strategy described in the main text. In order to minimize the circuit depth we combine fSWAP operators with interaction terms from the Hamiltonian. Note that in one case an fSWAP operator can be sandwiched between a $U_1$ and a $J$ interaction term. The final depth for this combination of operators is therefore still five. As a consequence, we can implement the circuit in a total depth of 27.  

In the case of the three-orbital model we pursue a somewhat different ordering of modes within a site: $\{1\!\!\uparrow$,$2\!\!\uparrow$,$3\!\!\downarrow$,$2\!\!\downarrow$,$3\!\!\uparrow$,$1\!\!\downarrow\}$ together with a modified fSWAP network (see the circuit in Fig.~\ref{fig:hk}.b)), which allows for fSWAP operators to be merged with $J$-terms from both the left and the right sides. The total circuit depth for this model is 35, remarkably only 8 steps longer than for the two-orbital model. This is mainly due to the fact that we can implement all quartic $J$-terms in parallel with the hoppings between lattice sites. This system is also of scientific interest as it is considered to be the minimal number of orbitals that is necessary in order for Hund's physics to appear in the HK model \cite{Georges2013}.

\begin{figure*}
  \centering
%   \begin{minipage}[c]{\linewidth}
    \subfloat
      {\includegraphics[width=.65\linewidth]{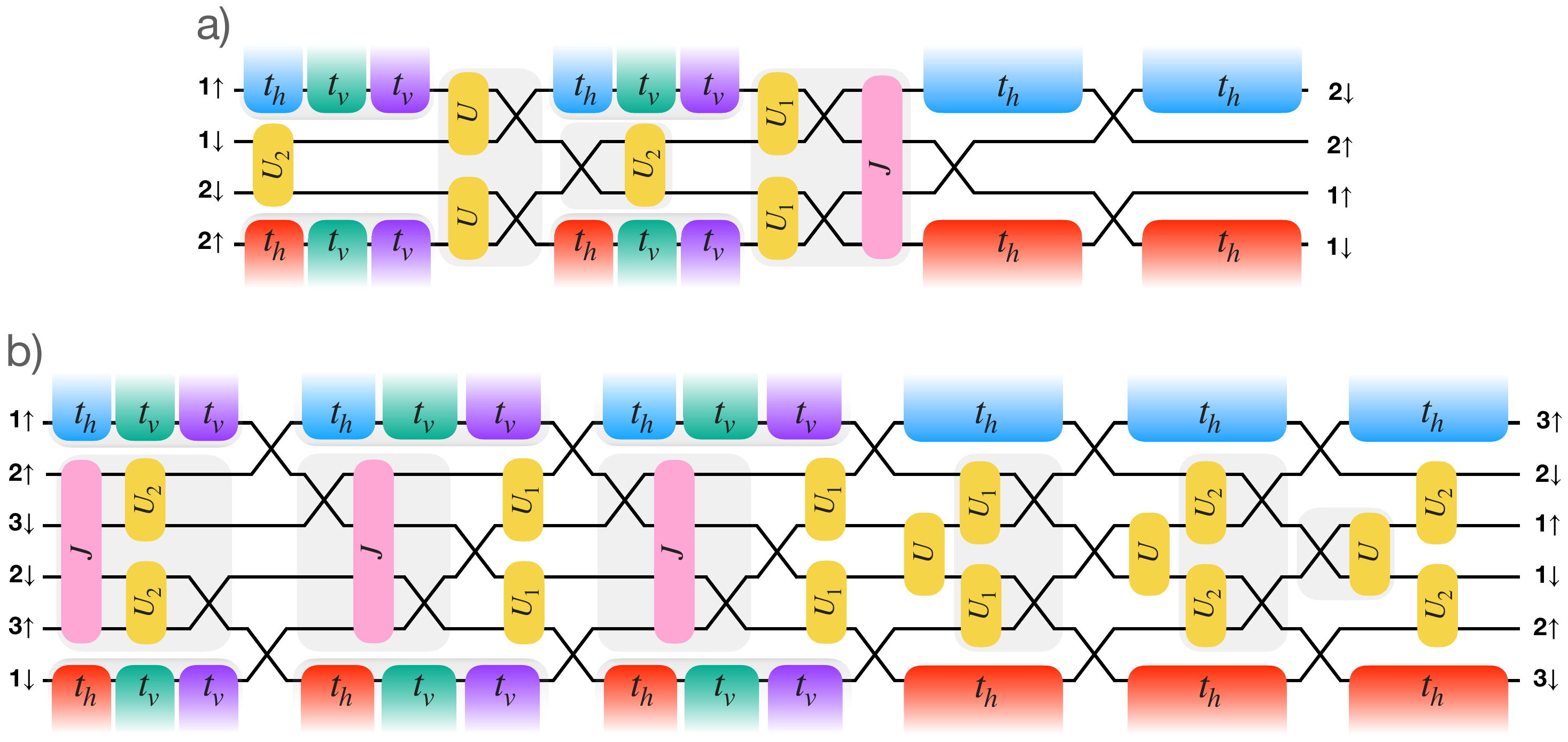}}
%   \end{minipage}
   \hfill
  \caption
    {Single-Trotter step simulation circuits for Hubbard-Kanamori model with a) 2 orbitals and b) 3 orbitals. Size of blocks is not to scale: $t_h$-$t_v$-$t_v$ blocks are depth-seven, $t_h$-terms are depth-three, J terms are depth-five, $U$, $U_1$, $U_2$ terms and fSWAPs can be performed with a single two-qubit gate.
    Further circuit compression can be applied within grey shaded areas of the circuit.
      \label{fig:hk_appendix}
    }
\end{figure*}

\end{document}